\documentclass{article}
\usepackage{emulateapj,apjfonts}
\usepackage{lscape}

\newcommand{\etal}{et~al.\ }

\def\ts{\thinspace}

\def\omit#1{\empty}

\newdimen\sa  \def\sd{\sa=.1em \ifmmode $\rlap{.}$''$\kern -\sa$
                               \else \rlap{.}$''$\kern -\sa\fi}

\reversemarginpar

\begin{document}

\lefthead{The Challenge of Pure-Disk Galaxies}

\righthead{Kormendy et al.}

\centerline{\null}\vskip -25pt

\title{Bulgeless Giant Galaxies Challenge our Picture of Galaxy Formation\\
       by Hierarchical Clustering\altaffilmark{1,2}}

\author{
John Kormendy\altaffilmark{3,4,5},
Niv Drory\altaffilmark{5},
Ralf Bender\altaffilmark{4,5}, and
Mark E.~Cornell\altaffilmark{3}
}

\altaffiltext{1}{Based on observations obtained with the Hobby-Eberly Telescope,
                 which is a joint project of 
                 the University of Texas at Austin, 
                 the Pennsylvania State University, 
                 Stanford University, 
                 Ludwig-Maximilians-Universit\"at M\"unchen, and 
                 Georg-August-Universit\"at G\"ottingen.} 

\altaffiltext{2}{Based on observations made with the NASA/ESA {\it Hubble Space Telescope}, 
                 obtained from the Data Archive at STScI, which is operated by AURA, Inc., under 
                 NASA contract NAS 5-26555.  These observations are associated with program numbers
                 7330,  
                 7919,  
                 8591,  
                 8597,  
                 8599,   
                 9293,  
                 9360,  
                 9490,  
                 9788,   
                 and
                 11080.} 

\altaffiltext{3}{Department of Astronomy, The University of Texas at Austin,
                 1 University Station C1400, Austin,
                 Texas 78712-0259, USA; kormendy@astro.as.utexas.edu; \hbox{cornell@astro.as.utexas.edu}}

\altaffiltext{4}{Universit\"ats-Sternwarte, Scheinerstrasse 1,
                 M\"unchen D-81679, Germany}

\altaffiltext{5}{Max-Planck-Institut f\"ur Extraterrestrische Physik,
                 Giessenbachstrasse, D-85748 Garching-bei-M\"unchen, Germany; 
                 bender@mpe.mpg.de; drory@mpe.mpg.de}

\pretolerance=15000  \tolerance=15000

\begin{abstract} 

      To better understand the prevalence of bulgeless galaxies in the nearby field, we dissect giant 
Sc -- Scd galaxies with {\it Hubble Space Telescope\/} (HST) photometry
and Hobby-Eberly Telescope (HET) spectroscopy.  We use the HET High Resolution Spectrograph 
(resolution $R \equiv \lambda / {\rm FWHM} \simeq 15,000$) to measure stellar velocity dispersions 
in the nuclear star clusters and (pseudo)bulges of the pure-disk galaxies 
M{\ts}33,
M{\ts}101,
NGC 3338, 
NGC 3810,
NGC 6503, and
NGC 6946.  The dispersions range from $20 \pm 1$ km s$^{-1}$ in the nucleus of M{\ts}33 to
$78 \pm 2$ km s$^{-1}$ in the pseudobulge of NGC 3338.  We use HST archive 
images to measure the brightness profiles of the nuclei and (pseudo)bulges in M{\ts}101, NGC 6503, and 
NGC 6946 and hence to estimate their masses.  The results imply small mass-to-light ratios consistent with 
young stellar populations.  These observations lead to two conclusions: 

\lineskip=-4pt \lineskiplimit=-4pt

      (1) Upper limits on the masses of any supermassive black holes (BHs) are $M_\bullet \lesssim
(2.6 \pm 0.5) \times 10^6$ $M_\odot$ in M{\ts}101 and $M_\bullet \lesssim (2.0 \pm 0.6) \times 10^6$ $M_\odot$
in NGC 6503. 

      (2) We show that the above galaxies contain only tiny pseudobulges that make up $\lesssim 3$\ts\% of the 
stellar mass.  This provides the strongest constraints to date on the lack of classical bulges in the biggest 
pure-disk galaxies.  We inventory the galaxies in a sphere of radius 8 Mpc centered on our Galaxy to see 
whether giant, pure-disk galaxies are common or rare.  We find that at least 11 of 19 galaxies with 
$V_{\rm circ} > 150$ km~s$^{-1}$, including M{\ts}101, NGC 6946, IC 342, and our Galaxy, show no evidence 
for a classical bulge.  Four may contain small classical bulges that contribute 5\ts--12\ts\%
of the light of the galaxy.  Only four of the 19 giant galaxies are ellipticals or have classical bulges that
contribute $\sim$ 1/3 of the galaxy light.  We conclude that pure-disk galaxies are far~from~rare.  
It is hard to understand how bulgeless galaxies could form as the quiescent tail of a distribution of merger 
histories.  Recognition of pseudobulges makes the biggest problem with cold dark matter galaxy formation more 
acute: How can hierarchical clustering make so many giant, pure-disk galaxies with 
no evidence for merger-built bulges?  Finally, we emphasize that this problem is a strong 
function of environment: the Virgo cluster is not a puzzle, because more than 2/3 of its stellar mass is in 
merger remnants.

\end{abstract}


\section{Introduction}

\pretolerance=15000  \tolerance=15000

      This paper has two aims.  First,~we~derive upper limits on the masses $M_\bullet$ of any 
supermassive BHs in two giant, pure-disk galaxies.  This provides data for a study 
(Kormendy \etal 2010) of the lack of correlation (Kormendy \& Gebhardt 2001) between BHs and 
galaxy disks.  Second, we inventory disks, pseudobulges, and classical bulges in the nearby 
universe and show that giant, pure-disk galaxies are not rare.  This highlights the biggest 
problem with our mostly well supported picture of hierarchical clustering: How can so many pure-disk 
galaxies form, given so much merger violence?  Both studies need the same observations: photometry 
to measure structure and spectroscopy to measure velocity dispersions and masses.

\subsection{A Practical Guide to Readers}

      In~\S\thinspace2, we measure stellar velocity dispersions in high-mass, Sc{\ts}--{\ts}Scd 
galaxies that contain only nuclei or extremely small pseudobulges.  In \S\ts3, we derive HST- and 
ground-based surface photometry of the most useful subset of our galaxies to see whether they contain 
small classical bulges, pseudobulges, or nuclei and to measure nuclear masses and  $M_\bullet$ limits.  
Sections\ts2 and 3 are long.  Readers who need $\sigma$ and $M_\bullet$ results can find
them in Figure 1 and Table~1.  Readers who are interested in the smallest pseudobulges
can concentrate on \S\thinspace3.  Readers whose interest is the challenge that pure-disk galaxies 
present for our picture of galaxy formation should skip directly to \S\thinspace 4. 

\subsection{Introduction to the Velocity Dispersion Measurements}

      Nuclei are expected to have velocity dispersions that range from those of globular
clusters, $\sigma \sim 10$ km s$^{-1}$, to values similar to those in the smallest 
classical bulges and ellipticals (e.{\ts}g., M{\ts}32: $\sigma \simeq 60$ km s$^{-1}$; Tonry 1984;
1987; Dressler \& Richstone 1988; van der Marel \etal 1994a, b; Bender \etal 1996).  But nuclei 
are faint and embedded in bright disks.  The $\sigma$ constraint 
implies that we need high dispersion, and the faintness implies that we need a large telescope.  
As a result, few $\sigma$ measurements of nuclei are available.  The best 
object\thinspace--{\thinspace}now very well measured{\thinspace}--{\thinspace}is M{\ts}33, whose 
exceptionally well defined nucleus has a velocity dispersion of $\sigma \simeq 21 \pm 3$ km s$^{-1}$ 
(Kormendy \& McClure 1993; Gebhardt \etal 2001). We use it as a test case for our observations.
The ``gold standard'' of nuclear dispersion measurements is Walcher \etal (2005); they used the 
Ultraviolet and Visual Echelle Spectrograph on the Very Large Telescope to measure 9 nuclei of 
generally modest-sized galaxies at a resolution of $R = 35,000$. 

      This paper reports $R = 15,000$ measurements of $\sigma$~in~M{\ts}33,
NGC{\ts}3338, 
NGC\ts3810,
NGC\ts5457,
NGC\ts6503, and
NGC\ts6946.
In choosing targets, we favored the largest pure-disk galaxies that have the smallest possible
pseudobulges (Kormendy \& Kennicutt 2004) and the smallest possible distances.  The most important 
galaxies for our purposes are NGC 5457 = M{\ts}101 and NGC 6946.  A closely similar object is
IC 342, for which B\"oker \etal (1999) measured a nuclear dispersion of $\sigma = 33 \pm 3$ 
km s$^{-1}$ at a spectral resolution of $R = 21,500$.  All three are Scd galaxies with extremely small 
pseudobulges or nuclear star clusters but essentially the largest possible asymptotic rotation 
velocities $\sim 200$ km s$^{-1}$ consistent with our requirement that they contain no classical bulges. 

      Late in our data reduction, we were scooped by Ho \etal (2009), who measured $\sigma$ 
and collected published $\sigma$ data for 428 galaxies.  All of our objects are included in their paper.  
However, their instrumental velocity dispersion is $\sigma_{\rm instr} = 42$ km s$^{-1}$, whereas ours 
is 8 km s$^{-1}$.  Our measurements therefore provide important confirmation.  Most of their measurements 
prove to be remarkably accurate, even when $\sigma < \sigma_{\rm instr}$.  We disagree on two values.  
Also, our measurements have estimated errors that are a factor of $\sim 4$ smaller than theirs.
Confidence in our understanding of the smallest central velocity dispersions in the biggest
pure-disk galaxies is correspondingly increased.

\section{Observations and Data Reduction}

\subsection{Observations}

      The spectra were obtained with the High Resolution Spectrograph (HRS: Tull 1988) and the
9 m Hobby-Eberly Telescope (HET: Ramsey \etal 1998).  The queue-scheduled observations were 
made between 2006 October 17 and 2007 April 21.  HRS is fed by optical fibers; the image scrambling
provided by the 34 m long fibers guarantees that different seeing conditions, source light distributions,
and object centering accuracies do not affect the wavelength resolution.  We used 3$^{\prime\prime}$
fibers.  A central fiber was postitioned on the galaxy nucleus and two bracketing ``sky fibers'' measured 
the night sky and galaxy disk at radii of 10$^{\prime\prime}$ immediately outside the nucleus.  
We confirmed the nominal resolution of $R = 15,000$ by measuring night sky emission
lines.  The corresponding instrumental dispersion is $\sigma_{\rm instr} \simeq 8.2$ km s$^{-1}$ 
at the Ca infrared triplet lines, 8498 \AA, 8542 \AA, and 8662 \AA.  The instrument is an echelle; 
the above lines were positioned in three orders that were combined into a single spectrum as described
in \S\thinspace2.2.

      We obtained 4, 900 s exposures of NGC 6946.  Exposure times were 600 s per spectrum for the
other galaxies; we obtained two such spectra for M{\ts}33, three for NGC 6503, four for NGC 3338, 
five for NGC 5457, and seven for NGC 3810.  In a few cases, seeing or transparency was 
poor -- the latter is judged by signal level and the former is judged by the contrast of the nucleus 
against the disk, that is, by the ratio of the flux from the galaxy to that from the sky plus disk. 
Low-quality spectra were not reduced.  Each spectrum was taken in 2{\ts}--{\ts}3 
subexposures to allow correction for cosmic ray hits.

      We also obtained 1 s to 50 s exposures of five velocity standard stars, HD 117176 (G4 V),
HR 1327 (G4 III), $\eta$ Cyg (K0 III), $\gamma$ Tau (K0 III), and $\delta$ And (K3 III).  We have
used the K0 -- K3 stars in many previous papers; they reliably fit the spectra of low- to 
moderate-dispersion galaxies very well.  In any case, (1) the Ca infrared triplet region is relatively
insensitive to template mismatch (Dressler 1984), and (2) the FCQ program that provides our 
final dispersion values is specifically engineered to minimize problems with template mismatch 
(Bender 1990).

\subsection{Preprocessing of Spectra}    

      Spectral reductions were carried out using the interactive image processing system
IRAF\footnote{IRAF is distributed by the National Optical Astronomy Observatories, which are 
              operated by the Association of Universities for Research in Astronomy, Inc., 
              under cooperative agreement with the National Science Foundation.} (Tody 1993).
The {\tt echelle} software package was used to remove 
instrumental signatures from the data. For each night, we created a bias frame, a continuum flat, 
and a Thorium-Argon arc-lamp spectrum from calibration data taken before and after the observations. 
Science frames were first bias and overscan corrected.  Bad pixels were flagged.  Next, we removed
cosmic ray hits using the spectroscopy-optimized version of L.A.COSMIC (van Dokkum 2001), and we
coadded the three spectra obtained for each galaxy.  We traced and fitted the spectral orders in the 
continuum flats using 3$^{\rm rd}$-order Legendre polynomials.  We removed the spectral signature of 
the continuum lamp from the flat fields by fitting the continuum using 7$^{\rm th}$-order Legendre 
polynomials along the dispersion direction and divided the flat field frames by the fits. Next, we 
extracted the science spectra (galaxies and velocity standard stars) using these normalized flat
fields to obtain one-dimensional spectra for each order.  This provided one, multi-order spectrum of 
the galaxy nucleus from the object fiber plus two ``sky'' spectra at galactocentric distances of 
10$^{\prime\prime}$ bracketing the nucleus.  Finally, all spectra were wavelength calibrated using 
the Th-Ar lamp spectra to an accuracy of $\sim$\ts0.02 pixel $\simeq$\ts0.003 \AA~$\simeq$\ts0.1 km s$^{-1}$.

      The remaining tasks are sky subtraction and the combination of separate orders into a single 
final spectrum rewritten on a $\log{\lambda}$ scale.  These steps were carried out using a combination 
of IRAF and VISTA (Lauer 1985, Stover 1988).  Two aspects of the reduction are tricky and require 
special care:

      The first is sky subtraction.  The good news is that sky spectra are taken simultaneously 
with the object spectra; this is important because sky lines vary on short time scales.  The bad
news is that sky subtraction is more difficult than it is with long-slit spectrographs, because 
we cannot average many spatial elements to get high signal-to-noise ratios $S/N$.  There are only
two sky fibers.  Sky subtraction contributes significant noise.  Moreover, the sky fibers have 
slightly different throughputs than the object fiber, so the two sky spectra must be scaled 
(differently) to the object spectrum.  Sky subtraction is more difficult for some objects than
for others, because the Ca triplet lines are badly positioned with respect to night sky lines
for some galaxy redshifts and benignly positioned for others.

      The second tricky problem is the polynomial fit to the continuum that must be divided
into each spectral order to remove the blaze efficiency function.  These functions are 
approximately $\bigcap$-shaped.  They are not vertical at the ends of orders, but the
signal is low there and hence multiplied upward by the continuum division.  When Ca lines fall well
away from the ends of orders, then continuum fitting is easy and results are robustly reliable.
But when Ca triplet lines fall near the ends of orders, then (1) the fit becomes difficult
because there is little continuum to fit, and (2) small fitting errors matter a lot 
because the line profile is divided by the low blaze efficiency.

      Both problems are differently severe for different galaxies. As a result, the spectral lines 
that produce the most reliable results are different for different galaxies.  We therefore discuss 
the consequences for each galaxy separately.

      After continuum division, we kept parts of three orders that contain the Ca triplet lines.   
Two wavelength regions were used, 8450 -- 8750 \AA~in 1024 pixels for the Fourier quotient program 
(FQ: Sargent \etal 1977) and 8450 -- 8730 \AA~in 2048 pixels for the Fourier correlation quotient 
program (FCQ: Bender 1990).  The final reduced spectra and the best fit of a standard star spectrum 
are illustrated in Figure 1.  Our velocity dispersion measurements are discussed in \S\ts2.3 and
listed in Table 1.

\subsection{Velocity Dispersion Measurements}

\subsubsection{M{\ts}33}

     We included M{\ts}33 to check our ability to measure small $\sigma$.  We obtained two, 
600{\ts}s exposures on different nights.  They presented no problems.~The nucleus is very compact and
high in surface brightness and very distinct from the surrounding~disk (see Figure 1 in \S\ts3.1 here
and Kormendy \& McClure 1993).  By a factor of 2, it provides the highest flux of any of our galaxies. 
Sky lines fall in the red wings of the 8498 \AA~and 8542 \AA~lines and in both wings of the 8662 \AA~line. 
But the galaxy signal is 11 and 9 times larger than the sky signal in the two exposures.  We measured 
the factor by which to scale the sky spectra to the galaxy spectra to 2\ts\% accuracy using
56 and 61 emission lines in the two spectra.  This results in excellent sky subtraction.  The continuum 
fit is relatively easy at the heliocentric velocity of $-179$ km s$^{-1}$ (NED).  It also helps that
the absorption lines are so narrow that they clobber few continuum pixels.  Combining orders is easy and
the dispersion measurements are reliable.

      FQ and FCQ give consistent results.  The K3 III standard star $\delta$ And gives consistently larger 
scatter in $\sigma$ values than the other stars; we omit it from our averages.  Then the average velocity
dispersion given by the other 4 standard stars is 
$\sigma = 21.2 \pm 0.6$ km s$^{-1}$ using FQ and 
$\sigma = 19.8 \pm 0.7$ km s$^{-1}$ using FCQ.  Quoted errors are for any one standard star.  Adding
in quadrature the estimated error in the mean from averaging results for 4 stars, we get
$\sigma = 19.8 \pm 0.8$ km s$^{-1}$ using FCQ.
We adopt this result. 

      We tested it further by reducing six subregions of the spectra.  They isolate individual Ca 
lines or combinations of weaker lines between the Ca lines.   FCQ finds that the average intrinsic 
widths of the template lines in the wavelength regions tested are $\sigma_* = 6$ km s$^{-1}$ to 
38 km s$^{-1}$.  Signal-to-noise ratios are lower and estimated $\sigma$ errors are bigger when subregions
are~used.  But results for different subregions are~consistent.  In particular, we found no significant 
dependence of the measured $\sigma$ on $\sigma_*$ (see also \S\ts2.3.5).   There is a hint that $\sigma$ 
may be 1.0 km s$^{-1}$ smaller in M{\ts}33 than we derive.  It is not significant, so we retain the result 
from the whole spectrum.  Similar wavelength region tests gave similar results for all galaxies.  {\it We 
always tried at least one wavelength region that contains only a single Ca triplet line.  This guarantees that 
template mismatch cannot be a problem.}

      Our result agrees very well with $\sigma = 21 \pm 3$ km s$^{-1}$ measured with 
$\sigma_{\rm instr} = 20$ km s$^{-1}$ by Kormendy \& McClure (1993).  
Gebhardt \etal (2001) derived an integrated velocity dispersion for the whole nucleus
of $\sigma = 24.0 \pm 1.2$ km s$^{-1}$ using HST~STIS.  Merritt \etal (2001), also with
STIS, found a central dispersion of $\sigma = 24 \pm 3$ km s$^{-1}$ but a rapid rise in 
$\sigma$ to $\sim 35 \pm 5$ km s$^{-1}$ at $\pm 0\farcs3$ radius; Ferrarese (2002) quoted an 
integrated velocity dispersion of $\sigma = 27 \pm 7$ km s$^{-1}$ from these data.  Most recently, 
Ho \etal (2009) got $\sigma = 20.0 \pm 8.5$ km s$^{-1}$ in their catalog of 428 $\sigma$ 
measurements; this is remarkably accurate given that $\sigma < \sigma_{\rm instr} = 42$ km s$^{-1}$.   
We adopt our measurement of $\sigma = 19.8 \pm 0.8$ km s$^{-1}$; it was obtained with the highest 
wavelength resolution.

\subsubsection{NGC 3338 and NGC 3810}

      Despite having the largest $\sigma$ values in our sample -- which means that there are
fewer continuum pixels to fit -- these are the easiest galaxies to reduce.  The reason is that the
heliocentric velocities ($V_\odot = 1302$ and 993 km s$^{-1}$, respectively) put the Ca triplet 
lines far from night sky lines and far from the ends of orders.  Sky subtraction and continuum
fitting are both easy.  Wavelength range tests show that all lines give reliable results.

\lineskip=-4pt \lineskiplimit=-4pt

      For NGC 3810, we obtained four, 600 s exposures with good seeing and nucleus centering,
all taken on different nights.  The flux in the nuclear spectra ranged from 3.8 to 4.7 times
that of the sky spectra in the four exposures.  Sky apertures were located 
at $r \simeq 10^{\prime\prime}$ on each side of the galaxy major axis.  They clearly contained
galaxy absorption lines.  Because of disk rotation, their wavelengths differed from those of
the corresponding lines in the nuclear spectrum.  It was necessary to be exceptionally 
careful to subtract disk absorption and sky emission lines correctly from the 
nuclear spectrum.  This was done by cleaning the sky lines out of each sky spectrum to leave 
only the galaxy lines, smoothing the result slightly, and subtracting this from the sky
spectrum to leave only sky lines.  The cleaned sky spectrum was clipped to zero at low
intensities to reduce noise and leave behind only the significant sky emission.  The result
was subtracted from the nuclear~spectrum.  This was done separately for each of the four nucleus
spectra.

      FCQ gave a velocity dispersion of $\sigma = 62.3 \pm 1.7$ km s$^{-1}$
for the nucleus.  This is the mean of the $\sigma$ values for the four spectra.  Each
$\sigma$ for one spectrum is an average over four standard stars.  The quoted error is the 
sum in quadrature of the estimated error given by FCQ for one spectrum reduced with one 
star and the error in the mean for four standard stars.

      Our $\sigma$ measurement can be compared with $\sigma = 73 \pm 16$ km s$^{-1}$ measured with 
an instrumental $\sigma_{\rm instr} = 78$ km s$^{-1}$ by H\'eraudeau \etal (1999) and
$\sigma = 58 \pm 12$ km s$^{-1}$ measured with $\sigma_{\rm instr} = 65$ km s$^{-1}$ 
by Vega Beltr\'an \etal (2001).  Ho \etal (2009) adopt $\sigma = 64.6 \pm 8.7$ km s$^{-1}$ 
credited to HyperLeda (Paturel \etal 2003) who averaged the above values.

      We also reduced the absorption-line spectrum obtained through the ``sky'' aperture and got
$\sigma = 56.1 \pm 2.3$ km s$^{-1}$ at $r \sim 10^{\prime\prime}$ along the major axis of
the disk of NGC 3810.  This is not significantly different from the nuclear dispersion.

      For NGC 3338, we also obtained four, 600 s exposures on different nights.  The galaxy flux
again was $\sim 5$ times that in the sky spectra, but this time, the sky spectra showed no 
significant galaxy lines.  Because the overall $S/N$ was also a factor of two lower than for
NGC 3810 and because all galaxy lines are benignly positioned with respect to night sky lines,
we reduced only the sum of the four nuclear spectra.  Sky subtraction was easy and FCQ gave
$\sigma = 77.5 \pm 1.5$ km s$^{-1}$.  

      For comparison, H\'eraudeau \etal (1999) got $\sigma = 91 \pm 18$ km s$^{-1}$ at 
$\sigma_{\rm instr} = 78$ km s$^{-1}$, and Ho \etal (2009) got 
$\sigma = 120.6 \pm 9.6$ km s$^{-1}$ at $\sigma_{\rm instr} = 42$ km s$^{-1}$.
Our two largest and most easily measured $\sigma$ values are the only ones that 
disagree with the Ho \etal (2009) measurements.  For NGC 3810, Ho adopted
a smaller $\sigma$ from HyperLeda.  For NGC 3338, Fig.~1 shows~that $\sigma \leq 89$ km s$^{-1}$, 
the near-central value in M{\ts}32.  We are confident 
in our result.  Note:~our science conclusions are based on low-$\sigma$ galaxies.  No conclusions 
depend on NGC 3338.  It is included to anchor our measurements at the high-$\sigma$ end in Fig.~1.

\begin{figure*}[ht]

\vskip 9.3truein

\includegraphics{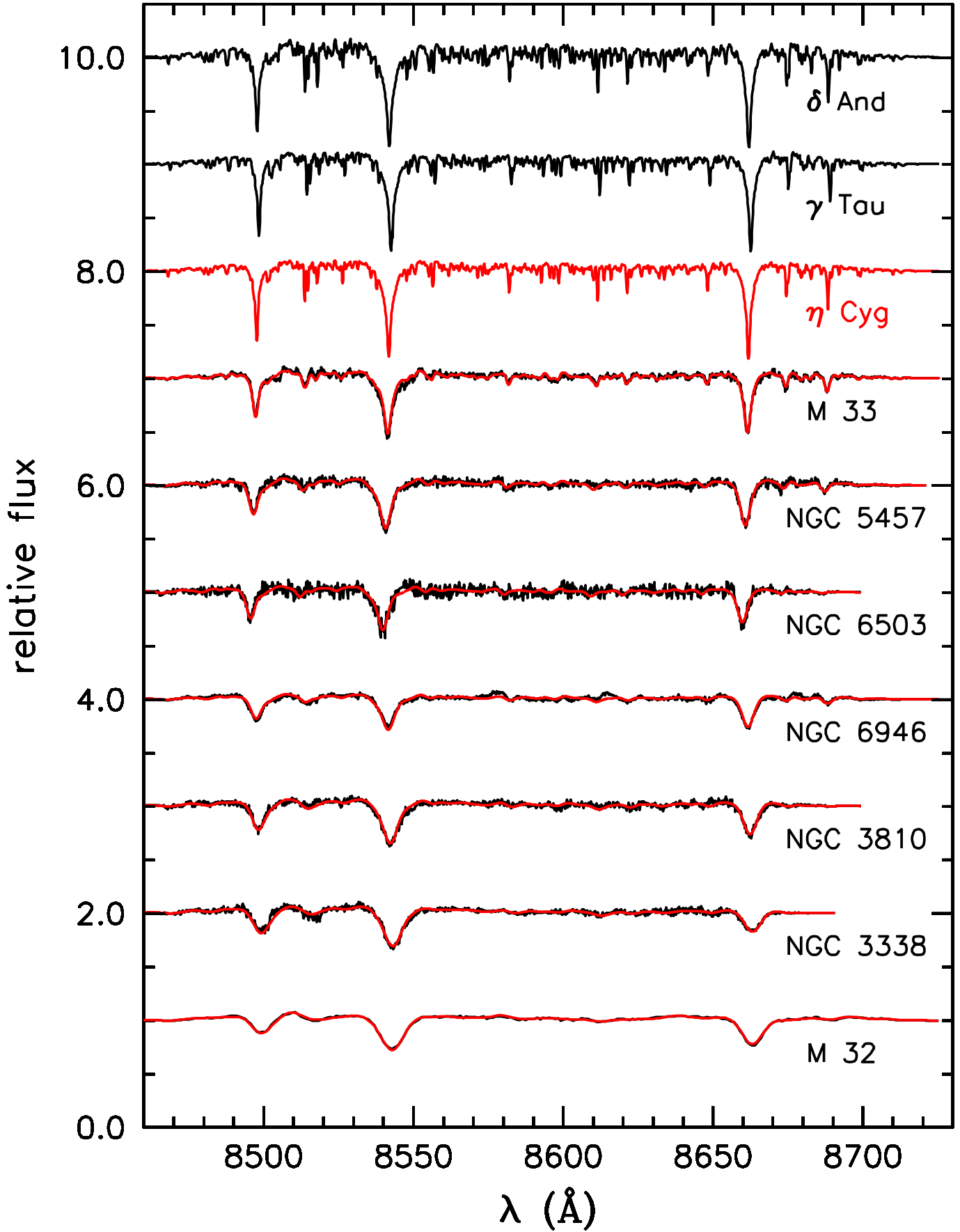}

\figcaption[]
{HET HRS spectra of standard stars $\delta$ And (K3 III), $\gamma$ Tau (K0 III) and $\eta$ Cyg (K0 III), 
our galaxies, and M{\ts}32.  The galaxies are ordered from top to bottom by increasing velocity 
dispersion from $\sigma = 20$ km s$^{-1}$ in M{\ts}33 to $\sigma = 89$ km s$^{-1}$ in a central spectrum
of M{\ts}32.  The spectrum of standard star $\eta$ Cyg broadened to each galaxy's line-of-sight velocity 
distribution is superposed in red on the galaxy's spectrum.
\lineskip=-4pt \lineskiplimit=-4pt
}
\end{figure*}

\vfill\clearpage

\subsubsection{NGC 6503}

      NGC 6503 and NGC 6946 (\S\ts2.3.4) are more difficult than M{\ts}33.  At heliocentric 
velocities of $\sim$\thinspace36 km s$^{-1}$ and 46 km{\thinspace}s$^{-1}$ (HyperLeda), their 
Ca triplet 8498 \AA~and 8542 \AA~lines are bracketed by sky lines where they merge into the continuum.  
Most of each line profile is safe, but special care is required in sky subtraction.  The 8662 
\AA~line falls in a ``picket fence'' of night sky lines.  They subtract well, but it is 
necessary to interpret results from this line with caution.  Sky scaling factors were again 
determined from several dozen night sky lines.  

      These galaxies are also difficult because the nuclei are faint.  The $S/N$
was therefore improved by subtracting sky only above a level of about 1.5 times the RMS fluctuations
in brightness.  That is, significant sky lines were subtracted, but the sky was not subtracted 
between the lines where it consists mostly of noise.  This is consistent with advice in the
online HRS manual 
{{\footnotesize\tt http://www.as.utexas.edu/mcdonald/het/het.html}}, links on the {\tt HRS} and 
thence on {\tt HRS Data Reduction Tips}.

      The less-than-fortuitous redshifts also put the 8542 \AA~and 8662 \AA~lines near the ends 
of spectral orders.  Continuum fitting is more uncertain than normal for these lines.  Therefore
the most reliable velocity dispersion comes from the 8498 \AA~line.  

      For NGC 6503, we took two, 600 s exposures on different nights.  The spectra of the nucleus 
are only 2.5~and~2.1 times brighter than the corresponding sky spectra.  Therefore (i) the $S/N$ 
is relatively low and (ii) the seeing or the centering -- more specifically: the 
degree to which the nuclear spectrum isolates light from the nucleus -- is significantly better in 
one exposure than in the other.  We reduce only the better exposure to measure the velocity dispersion 
of the nucleus.  We reduce the sum of the sky exposures to get the velocity dispersion -- albeit with 
still lower $S/N$ -- in the disk.  The sky apertures were positioned at $\Delta${\thinspace}PA 
$\simeq 53^\circ$ from the major axis; that is, relatively near the minor axis of the $i = 74^\circ$, 
highly inclined disk of NGC 6503.

      The nuclear spectrum was sky-subtracted using a median of the associated ``sky''
exposures.  We assume that the inner exponential profile of the galaxy disk -- 
i.{\thinspace}e., the outermost part of the profile that is fitted with an exponential 
in Figure 16 -- extends to the center.  Then ``sky subtraction'' removes both 
the night sky emission lines and the contribution to the nuclear spectrum from the underlying disk.  
This leaves us with a spectrum (Figure 1) of the combination of the nucleus and pseudobulge (see 
Figure 16, where $r^{1/4} = 1.11$ is the radius of the spectral aperture).  FCQ gives 
$\sigma = 40.5 \pm 2.0$ km s$^{-1}$ for this spectrum.  As noted above, the $\lambda = 8498$ \AA~line 
alone is more reliable than the other Ca triplet lines, given the HRS configuration and galaxy redshift.  
Using only this line, FCQ gives $\sigma = 39.9 \pm 2.0$ km s$^{-1}$.  We adopt the latter value.

      The sky exposures are noisy, but they show absorption lines.  We therefore
reduced the sum of the sky exposures to give us an estimate of the velocity dispersion in the galaxy
disk.  Subtracting sky emission lines without affecting the galaxy absorption lines was tricky,
because we could not afford sky exposures taken far from the galaxy.  Fortunately, we had one
spectrum each of NGC 3810 and NGC 5457 taken on the same night as one of the NGC 6503 spectra and 
one more spectrum of each of these galaxies taken a few nights later.  Their redshifts are
different enough so that galaxy lines in their ``sky'' spectra do not overlap.  We therefore
scaled the above four sky spectra of two galaxies together in intensity and medianed them,
rejecting one low value.  This very effectively removed absorption lines from the median sky
spectrum.  The median was scaled to the emission-line strengths in the summed NGC 6503 sky
spectrum and subtracted.  The result was a noisy but relatively clean spectrum of the disk of
NGC 6503 at radius $r = 10^{\prime\prime}$ at PA = 53$^\circ$ from the major axis.  This corresponds
to a true radius of 30$^{\prime\prime}$ along the major axis.  For this spectrum, FCQ gave 
$\sigma = 35.6 \pm 1.4$ km s$^{-1}$.  Using only the $\lambda$ 8498 \AA~line,
FCQ gave $\sigma = 31 \pm 4$ km s$^{-1}$.  We adopt the latter value.

      In \S\thinspace3.4, we use $\sigma$ to constrain $M_\bullet$.  However, a problem
is revealed when we compare our central $\sigma$ with published results:

      NGC 6503 is a well known galaxy; it has an extended flat rotation curve and 
one of the best rotation curve decompositions into visible and dark matter (Bottema 
1997).  And it was one of the first galaxies in which a drop in $\sigma$ toward the galaxy
center was reported.  Bottema (1989) measured the dispersion profile shown in Figure 2.
He found a maximum $\sigma = 45$ km s$^{-1}$ at $r \simeq 12^{\prime\prime}$ and then a drop at
larger radii to $\sim$ 15~km~s$^{-1}$.  The outward drop in $\sigma$ is no surprise.  But at 
$r < 10^{\prime\prime}$, Bottema observed a highly significant drop in $\sigma$ to 25 km s$^{-1}$ 
in two independent central radial bins.  This was unexpected at the time, but it has 
since become a common observation (e.{\thinspace}g., Emsellem \etal 2001; see especially the 
extensive results on ``$\sigma$  drops'' from the SAURON group: 
Ganda \etal 2006;
Falc\'on-Barroso \etal 2006;
Peletier \etal 2007a, b).
Small central velocity dispersions are now known to be a defining signature of pseudobulges
that are believed to be grown by secular evolution of isolated galaxy disks 
(Kormendy 1993;
Emsellem \etal 2001;
M\'arquez \etal 2003;
Kormendy \& Kennicutt 2004;
Chung \& Bureau 2004;
Peletier \etal 2006, 2007a, b;
Kormendy \& Fisher 2008).  We will conclude in \S\thinspace3.4 that NGC 6503 contains a small 
pseudobulge, based on other classification criteria.  But we do not confirm the central
$\sigma$ drop.

      Instead, our measurement of $\sigma = 40 \pm 2$ km s$^{-1}$ agrees with $\sigma = 46 \pm 3$
km s$^{-1}$ observed by Barth \etal (2002) using the Ca triplet lines at $\sigma_{\rm instr} = 
25$ km s$^{-1}$.  Both values agree with the dispersion peak observed by Bottema (1989).  Who is correct?
Is there a central $\sigma$ drop?

\vfill

\includegraphics{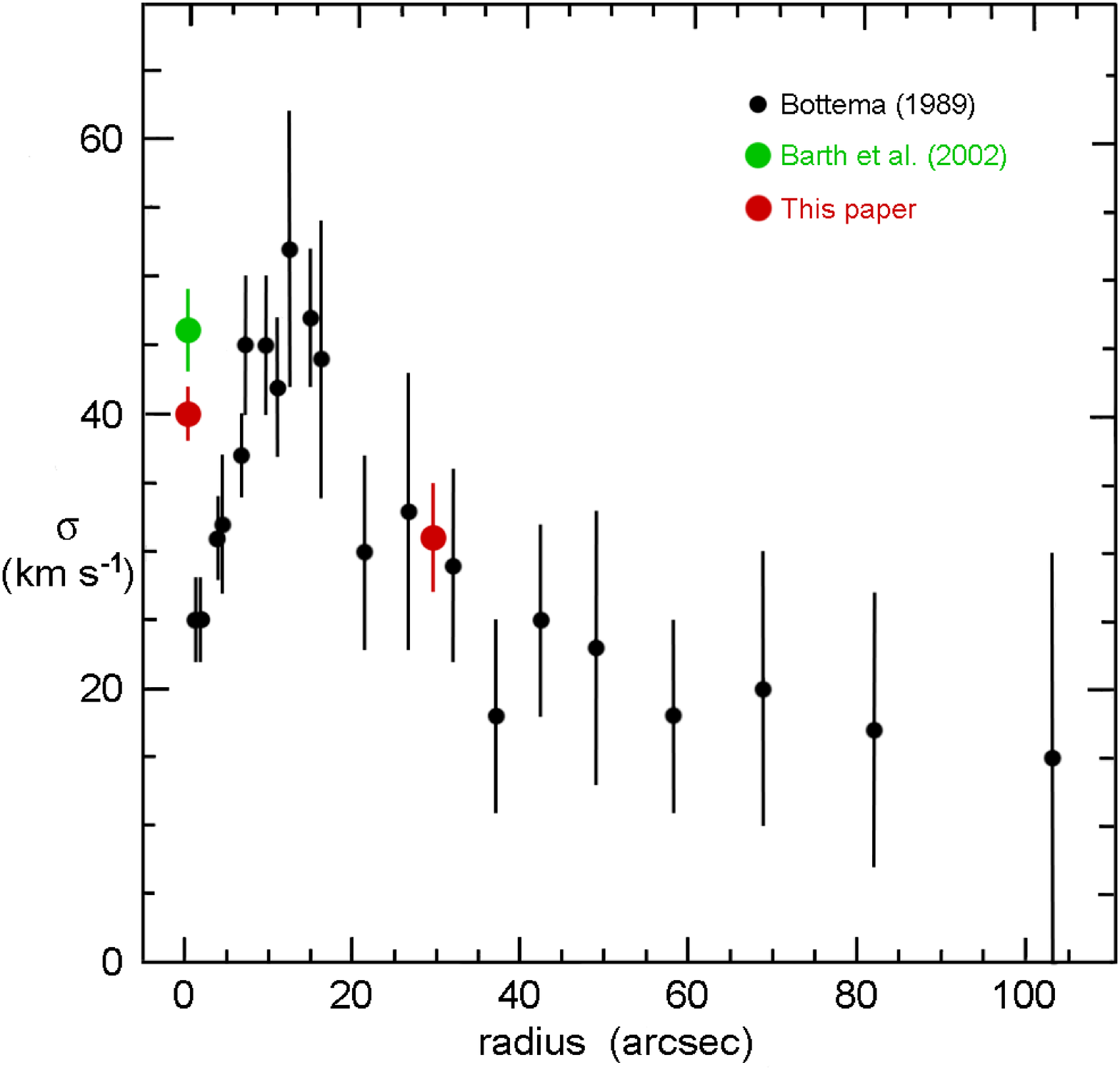}

\figcaption[]
{Velocity dispersion versus major-axis radius in NGC 6503 as measured by Bottema (1989; 
see also Bottema \& Gerritsen 1997).  Our value of the disk dispersion at a true radius of
30$^{\prime\prime}$ almost along the minor axis agrees well with Bottema's measurements along the
major axis.  But we do not confirm the central $\sigma$ drop.  Rather, our measurement of the 
central velocity dispersion is consistent with that of Barth \etal (2002; see also Ho \etal 2009).
\lineskip=-4pt \lineskiplimit=-4pt
}

\eject

      Some aspects of the observations favor the larger central $\sigma$.  The strongest
argument is that the two results based on the best instrumental $\sigma_{\rm instr}$ =
25 km s$^{-1}$ (Barth \etal 2002) and $\sigma_{\rm instr}$ = 8 km s$^{-1}$ (this paper) agree 
with each other in disagreeing with Bottema's result based on $\sigma_{\rm instr}$ = 35 km s$^{-1}$.  
A possible danger in Bottema's measurement is (1) that his slit width controlled his 
instrumental resolution and (2) that his slit may have been wider than the image of the nucleus.  
If the nucleus underfills the slit, then it is easy to underestimate its $\sigma$.  Another possible 
problem is the wavelength region used.  We and Barth observed at $\sim 8500$ \AA, where an 
admixture of young stars has almost no effect on $\sigma$ measurements.  Bottema observed at 5020 \AA.  
Cid Fernandes \etal (2005) show that the center is dominated by intermediate-age stars.  Their 
blue spectrum may not be as well matched by standard stars or -- for broad absorption lines -- as 
suitable for $\sigma$ measurements as are near-infrared spectra.  Finally, Bottema measured $\sigma$ 
via cross-correlation; this is less robust than FCQ (this paper) or than fitting broadened star 
spectra to the galaxy spectrum in pixel space (Barth \etal 2002).  Of course, these caveats are 
not conclusive.

      Some aspects of the observations probably did not cause the disagreement.
Bottema's exposure time was a heroic 35,400~s (9 h 50 m) taken in 1200 s chunks with the 2.5 m 
Isaac Newton telescope at La Palma.  Barth's spectrum was a 3600 s exposure taken with the Palomar 
5 m Hale telescope, and ours is a 600 s exposure taken with the 9 m Hobby-Eberly telescope.  
Barth got high $S/N$, and we got $S/N = 91$ per resolution element.  However, Bottema's
spectrum had good enough $S/N$ to allow him to measure small velocity dispersions at large
radii.  So it is unlikely that the central $S/N$ was a problem.  Also, our measurement of the disk 
dispersion at a major-axis radius of 30$^{\prime\prime}$ agrees well with Bottema's results
(Figure 2).  There is no reason to believe that either set of observations is 
unable to measure small dispersions.  Note that we measure mostly the radial velocity dispersion 
near the minor axis of the disk, whereas Bottema measured mostly the azimuthal dispersion; we do 
not expect them to be exactly equal.  Finally, spatial resolution is not the issue: Bottema had 
1\farcs0 to 1\farcs7 seeing; Barth had 1\farcs5 to 2\farcs0 seeing, and we had $1\farcs7$ FWHM 
seeing.  Moreover, our aperture was 3$^{\prime\prime}$ in diameter, and Bottema (1989) 
and Barth \etal (2002) binned their spectra in spatial pixels of 2\farcs64 and 3\farcs74, 
respectively.  So all three observations have poor spatial resolution.  It is remarkable 
that Bottema saw a dispersion drop over a radius of $\pm 2$ pixels of 3\farcs74 each.

      Alternatively, could all results in Figure 2 be correct?  This could happen as a consequence
of the fact that we and Barth observed near $\lambda = 8500$ \AA~whereas 
Bottema observed near 5020 \AA.  If the nucleus is colder than the pseudobulge and if
it is brighter in the blue than in the infrared, then it could dominate Bottema's result
but not ours.  No blue-band HST image is available to check whether the $I$-band photometry 
derived in \S\thinspace3.4 (Figure~16) is relevant for understanding Bottema's data.  However, 
the brightness contrast of the nucleus plus pseudobulge in Bottema's spectrum 
(his Figure 2) looks higher than the contrast in our spectrum.  Also, recall that Cid Fernandes
\etal (2005) found that the center is dominated by intermediate-age stars.  It is possible that 
Bottema measured a different stellar population than we did or than Barth did.

      We therefore do not know whether NGC 6503 has a central drop in $\sigma$.  In \S\thinspace3.4,
we use both $\sigma$ values to derive nuclear mass-to-light ratios.  The one based on $\sigma = 25 \pm 3$ 
km s$^{-1}$ is more plausible.  But the conservative choice is to adopt our measurement,
$\sigma = 40 \pm 2$~km~s$^{-1}$.  The Wolf \etal (2010) mass estimator 
used in \S\thinspace3.4 is valid for any set of test particles -- even ones that contribute no 
significant mass -- provided that $\sigma$ and the brightness distribution are measured for the 
same stars.  This means that we {\it must} use our $\sigma$ measured in $I$ band to match the HST 
surface photometry and to derive the nuclear mass $M_{\rm nuc}$.  We also use it to get our $M_\bullet$ limit.

\subsubsection{NGC 6946}

      NGC 6946 has almost the same redshift and therefore almost the same data reduction
problems as NGC 6503 (see the first three paragraphs of the previous section).  However, we
have much higher-$S/N$ spectra of NGC 6946, because we have four, 900 s exposures, obtained,
as always, on different nights.  The nuclear contrast is better than for NGC 6503 also: the 
nuclear spectra are 9 -- 11 times brighter than the ``sky'' spectra.
Since the latter are taken at 10$^{\prime\prime}$ distance from the nuclear aperture, they
are, as usual, well within the galaxy.  They were positioned in the transition region between 
what will turn out to be a tiny pseudobulge and the galaxy's exponential disk.  However, galaxy 
absorption lines are negligible in the sky spectra, and sky subtraction was straight-forward.  

      For this galaxy, the best sky subtraction was obtained by scaling the two spectra 
given by the sky apertures to have the same average emission-line intensities as the
nuclear spectrum using a single scaling factor for each aperture (the spectra were taken 
over a period of only six nights).  The scale factors were determined for the two apertures
by measuring the strengths of 239 and 297 lines in the four NGC 6946 spectra and in two
M{\thinspace}33 spectra obtained during the same nights.  (The number of lines used is not 
the same for the two sky apertures because different pixellation of almost-unresolved lines 
causes different problems -- for example, with blends -- for different lines.)  The scale 
factors are determined to $<$ 1\thinspace\%.  The sky-subtracted spectra are very clean.

      However, the velocity dispersion in NGC 6946 is slightly larger than that in
NGC 6503, and the redshift is slightly different, too.  The wings of the Ca triplet
lines reach closer to the ends of the spectral orders, so continuum-fitting was more
of a problem.  The best single spectrum yielded an FCQ velocity dispersion of
$\sigma = 55.7 \pm 1.1$ km s$^{-1}$.  All four spectra summed but analyzed only
using the safest (8494 \AA) line gave $\sigma = 56.4 \pm 0.9$ km s$^{-1}$.  The
signal-to-noise ratios in the best spectrum and in the sum of four spectra were
292 and 416 per resolution element, respectively.  The errors in $\sigma$ are
completely dominated by problems with the continuum removal.  They may be
underestimated by FCQ, which bases its error estimates on $S/N$ and on the 
quality of the star-galaxy spectral match.  

      We adopt $\sigma = 56 \pm 2$ km s$^{-1}$.  We therefore confirm the result in
Ho \etal (2009),  $\sigma = 55.8 \pm 9.4$ km s$^{-1}$.

\subsubsection{NGC 5457 = M{\thinspace}101}

      NGC 5457 is the most difficult galaxy in our sample: at a heliocentric radial 
velocity of 240 km s$^{-1}$ (NED,~HyperLeda), the redshifted Ca triplet lines are almost 
exactly centered on sky emission lines.  Any oversubtraction or undersubtraction of 
the sky spectrum would result, respectively, in an underestimate or an overestimate of $\sigma$.  
Therefore, for each of our three, 600 s exposure spectra, we measured the sky spectra 
scaling factors using 45 -- 55 emission lines.  These factors produced relatively clean 
sky-subtracted absorption-line profiles.  Each spectrum was reduced individually through 
both the FQ and FCQ programs.  This provides a consistency check for the three separate 
sky subtractions.

\vskip 0.1pt\eject

\def\cl{\centerline}

\begin{figure*}[b]

\vfill

\includegraphics{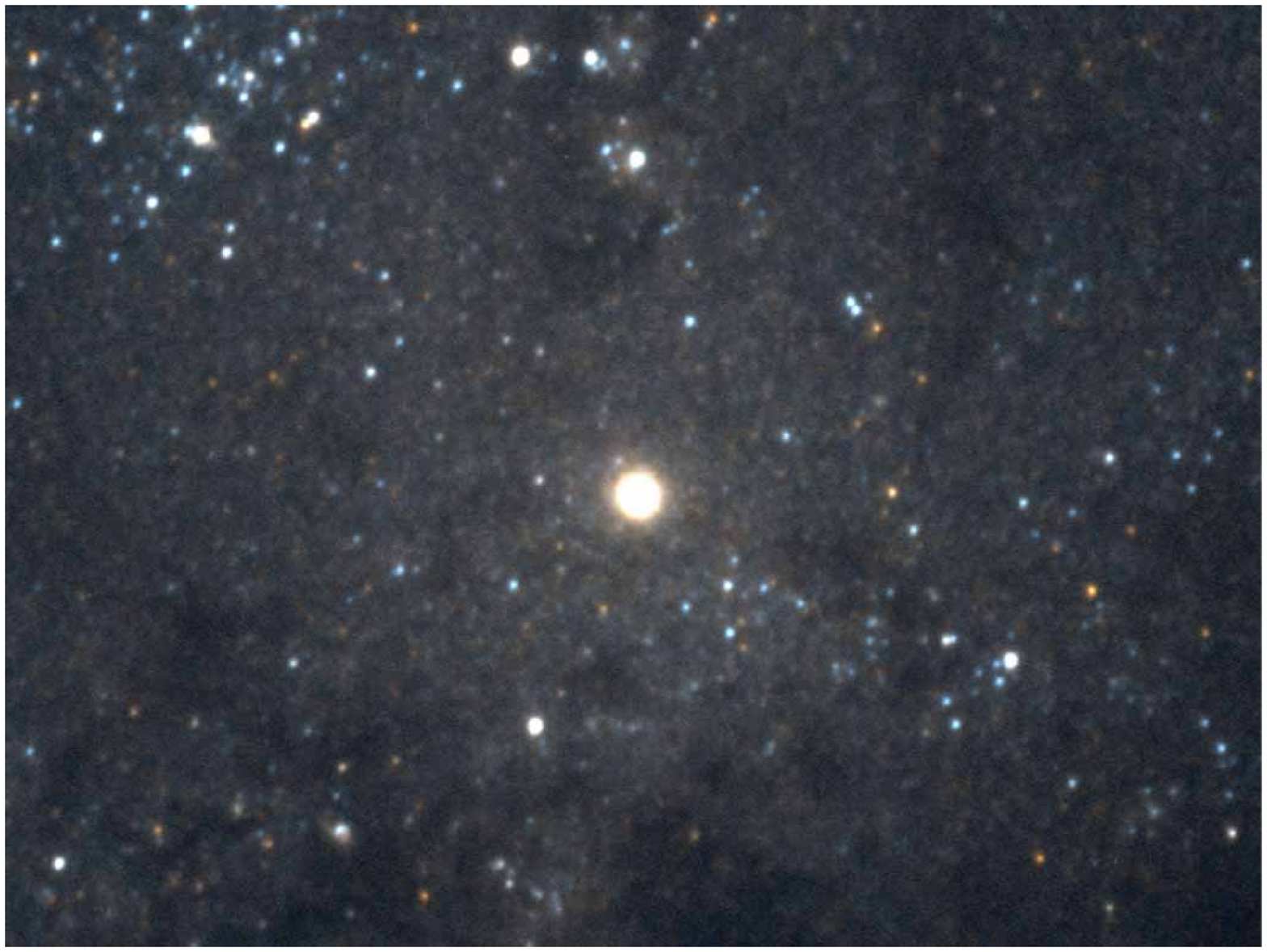}

\figcaption[]
{Central 69$^{\prime\prime} \times 110^{\prime\prime}$ of M{\ts}33, in a $B$- and $R$-band 
color version of Figure 1 in Kormendy \& McClure (1993).  The nucleus is a dense, central 
star cluster that is very distinct from the galaxy's disk.  There is certainly no classical
bulge in M{\ts}33, and there is arguably no pseudobulge (Kormendy \& Kennicutt 2004).  So
M{\ts}33 is a moderate-sized (rotation velocity $\sim 135 \pm 10$ km s$^{-1}$:
Corbelli \& Schneider 1997;
Corbelli \& Salucci 2000;
Corbelli 2003)
example of a pure-disk galaxy.
\lineskip=-4pt \lineskiplimit=-4pt
}
\end{figure*}

      However, sky subtraction proves not to be the biggest problem with the NGC 5457 spectral
reductions.  Instead, continuum fitting is especially difficult because the galaxy's redshift 
puts the two red triplet lines too close to the ends of orders.  In fact, both lines have 
wavelengths that appear at the blue end of one order and the red end of the adjacent order.
In combining and averaging continuum-divided orders, we kept these lines only in the order in
which they were farther from the end of the wavelength range covered by that order.  Nevertheless,
NGC 5457 -- even more than our other galaxies -- is best measured by the $\lambda = 8498$ \AA~line,
which is fortuitously located in the middle of its spectral order.

      Another problem -- not well known but correctly emphasized by Barth \etal (2002), by Walcher 
\etal (2005), and by Ho \etal (2009) --
is that the Ca triplet lines are intrinsically broad.  Averaged over our standard stars, the
intrinsic width of the $\lambda = 8498$ \AA~line is $\sigma_{\rm intr} \simeq 28$ km s$^{-1}$; 
that of $\lambda = 8542$ \AA~is $\sigma_{\rm intr} \simeq 50$ km s$^{-1}$; and
that of $\lambda = 8662$ \AA~is $\sigma_{\rm intr} \simeq 35$ km s$^{-1}$.
Once $\sigma_{\rm instr} < \sigma_{\rm intr}$, the intrinsic widths of the absorption
lines and not the instrumental resolution of the spectrograph largely control the smallest
dispersions that we can measure.  High $S/N$ overcomes most of the problem; this is why 
we had no trouble with M{\ts}33.  Our $S/N$ for NGC 5457 is good; it ranges from 160 to 165 
per resolution element for the three spectra.  Nevertheless, the intrinsic narrowness of the 
$\lambda = 8498$ \AA~line is another reason why results from this line alone are more reliable 
than those from other lines or from the whole spectrum.  This remark applies to some extent to
all of our galaxies with small velocity dispersions but not (for example) to NGC 3338.

      With this background, our results are as follows:

      For the complete wavelength range, FCQ gives 
$\sigma = 36.1 \pm 1.3$ km s$^{-1}$ for the best spectrum and
$\sigma = 36.6 \pm 1.2$ km s$^{-1}$ and
$\sigma = 34.2 \pm 1.2$ km s$^{-1}$ for the other two spectra.  The mean is
$\sigma = 35.6 \pm 1.4$ km s$^{-1}$.  At such small $\sigma$, comparison to FQ is important.  
                                      It gives a mean value of
$\sigma = 36 \pm 2$ km s$^{-1}$, in good agreement with FCQ.  However:

\vskip 4.2truein

      In marked contrast, for the $\lambda = 8498$ \AA~line alone, FCQ gives 
$\sigma = 27.3 \pm 2.0$ km s$^{-1}$ for the best spectrum and
$\sigma = 29.8 \pm 1.3$ km s$^{-1}$ and
$\sigma = 29.8 \pm 1.6$ km s$^{-1}$ for the other two spectra.  The mean is
$\sigma = 29.0 \pm 1.9$ km s$^{-1}$.  Omitting only the $\lambda = 8542$ \AA~line, FQ gives 
$\sigma = 25.2 \pm 3.5$ km s$^{-1}$.  For the best spectrum, FQ gives 
$\sigma = 23.3 \pm 3.1$ km s$^{-1}$.

      Recognizing that the error estimates given by FQ and FCQ do not take into account any
problems with continuum fits, we conservatively adopt 
$\sigma = 27 \pm 4$ km s$^{-1}$.  This agrees  with 
$\sigma = 23.6 \pm 8.7$ km s$^{-1}$ obtained by Ho \etal (2009).  
\vskip -15pt

\null

\subsubsection{Adopted Velocity Dispersions}

      Table 1 lists our $\sigma$ measurements and the masses derived from them in \S\thinspace3.  
For M{\thinspace}33, $M_\bullet \lesssim 1500$ $M_\odot$ based on HST spectroscopy was derived 
by Gebhardt \etal (2001). NGC 3338 and NGC 3810 are too far away to yield useful $M_\bullet$ limits.

      Table 1 provides an independent test of the $\sigma$ measurements in
Ho \etal (2009).  Note that our smallest velocity dispersions are smaller than
$\sigma_{\rm intr}$ even for the $\lambda = 8498$ \AA~line.  Therefore we emphasize: 
{\it Ho \etal (2009) are not much less able to measure small velocity 
dispersions with $\sigma_{\rm instr} = 42$ km s$^{-1}$ than we are with
$\sigma_{\rm instr} = 8$ km s$^{-1}$.}  Moreover, the excellent agreement of our 
measurement and Ho's of $\sigma$ in M{\thinspace}33 implies that systematic errors in 
Ho \etal (2009) are small even at the smallest $\sigma$.  Ho \etal (2009) actually 
have important advantages over our measurements: (1) They used a long-slit spectrograph, 
so they can more accurately subtract sky and galaxy light from near the nucleus.
(2) Their spectrograph is not an echelle, so they have no problems with continuum fits 
and do not need to combine spectral orders.  Finally, (3) their 2$^{\prime\prime}$ 
slit is narrower than our 3$^{\prime\prime}$-diameter aperture, and their seeing
at Palomar Observatory likely was better than ours at the HET for most observations.  
Our results correct one dispersion value in Ho \etal (2009).  And we generally have 
smaller estimated errors.  But one of our main contributions is to provide independent,
high-resolution verification of the large $\sigma$ database in Ho \etal (2009) .

\vfill

\includegraphics{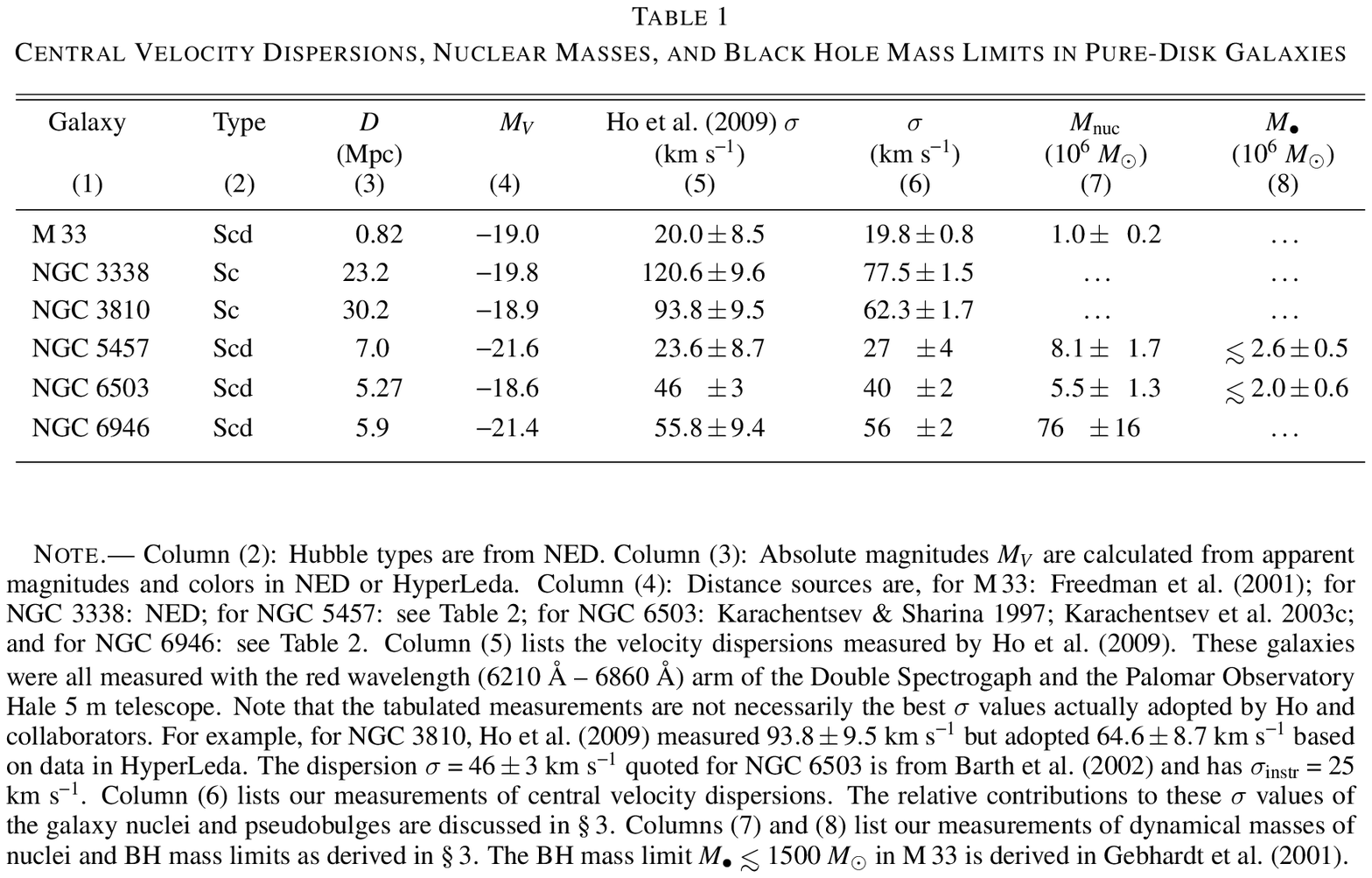}

\cl{\null}

\cl{\null}

\vskip 10.truein

\section{Properties of Nuclei and Pseudobulges}

      In this section, we measure surface brightness profiles of NGC 5457, NGC 6946, and NGC 6503.
To do this, we combine archival HST images with ground-based images.  The photometry allows us to 
identify and measure the properties of the central stellar components.  We show that all three 
galaxies have nuclear star clusters embedded in tiny pseudobulges.   We combine the photometry 
with the $\sigma$ measurements from the previous section to measure the masses of the nuclei.  
And we derive BH mass limits for NGC 5457 and NGC 6503.  First, as an illustration, we apply this
machinery to M{\ts}33.

\subsection{M{\ts}33}

      The nucleus of M{\ts}33 is illustrated in Figure 3.  Because it is both very compact
and very cold ($\sigma = 20 \pm 1$ km s$^{-1}$), strong upper limits on $M_\bullet$ have been 
derived.  Kormendy \& McClure (1993) found $M_\bullet \lesssim 5 \times 10^4$ $M_\odot$ from 
ground-based photometry and spectroscopy; Lauer \etal (1998) improved this to $M_\bullet 
\lesssim  2 \times 10^4$ $M_\odot$ by adding HST photometry; Merritt \etal (2001) got 
$M_\bullet \lesssim 3000$ $M_\odot$ using spatially resolved HST STIS spectroscopy and
improved dynamical modeling, and Gebhardt \etal (2001), also using STIS spectroscopy
and three-integral dynamical models, derived the strongest $M_\bullet$ upper limit in any
galaxy to date: $M_\bullet \lesssim 1500$ $M_\odot$.

       The stellar mass of the nucleus was not measured in any of the above papers.  We do so here.  
We begin by decomposing~the

\vskip 210pt

\cl{\null}

\cl{\null}

\cl{\null}

\cl{\null}

\cl{\null}

\cl{\null}

\cl{\null}

\cl{\null}

\cl{\null}

\cl{\null}

\cl{\null}

\cl{\null}

\cl{\null}

\cl{\null}

\cl{\null}

\vskip -5.5pt

\noindent HST plus ground-based profile of M{\ts}33 (Gebhardt \etal 2001) into a 
S\'ersic function plus an exponential.  The fit RMS is 0.06 mag arcsec$^{-2}$ and 
$r_e = 0\farcs36 = 1.4$ pc for the nucleus.  Its total magnitude is $V_{\rm nuc} = 14.05 \pm 0.07$ or
\hbox{$I_{\rm nuc} = 13.05 \pm 0.07$} from Kormendy \& McClure (1993) and from the above decomposition using 
colors from Lauer \etal (1998).  The Merritt \etal (2001) dynamical models give $M/L_V = 0.35$ 
and $M_{\rm nuc} = 0.54 \times 10^6$~$M_\odot$.  The Gebhardt \etal (2001) dynamical models give 
$M/L_I = 0.68$ and $M_{\rm nuc} = 1.30 \times 10^6$ $M_\odot$.  In the rest of this section, we 
measure masses using the Wolf \etal (2010) estimator of the nuclear half-mass, 
$M_{1/2} = 4 \sigma^2 r_e / G$ (see the next section for a discussion).  For M{\ts}33, it implies that
$M_{\rm nuc} = 2 M_{1/2} = 1.04 \times 10^6$ $M_\odot$.  Note that this compares well with the 
mean of the results from the Merritt and Gebhardt dynamical models.  We adopt the mean of all 
three determinations, \hbox{$M_{\rm nuc} = (1.0 \pm 0.2) \times 10^6$ $M_\odot$} (Table 1).

      In \S\S3.2 and 3.4, we base $M_\bullet$ limits on the minimum possible mass for a spherical stellar 
system plus BH.  Merritt (1987) shows that this limit is achieved if all of the mass is in the central point; 
then \hbox{$M_\bullet \lesssim 3 \sigma^2 <$\null$1/r$\null$>^{-1}{\kern -3pt}/G$.}
The harmonic mean radius of the M{\ts}33 nucleus is \hbox{$<$\null$1/r$\null$>^{-1} = 0\farcs19 = 0.76$~pc} 
and $M_\bullet \lesssim 2.1 \times 10^5$ $M_\odot$.  This is not competitive with HST-based limits.  But 
even our modest limits based on such virial theorem arguments can be useful for BH demographic studies. 

\vskip 280pt

\begin{figure*}[b]

\vskip 3.8truein

\includegraphics{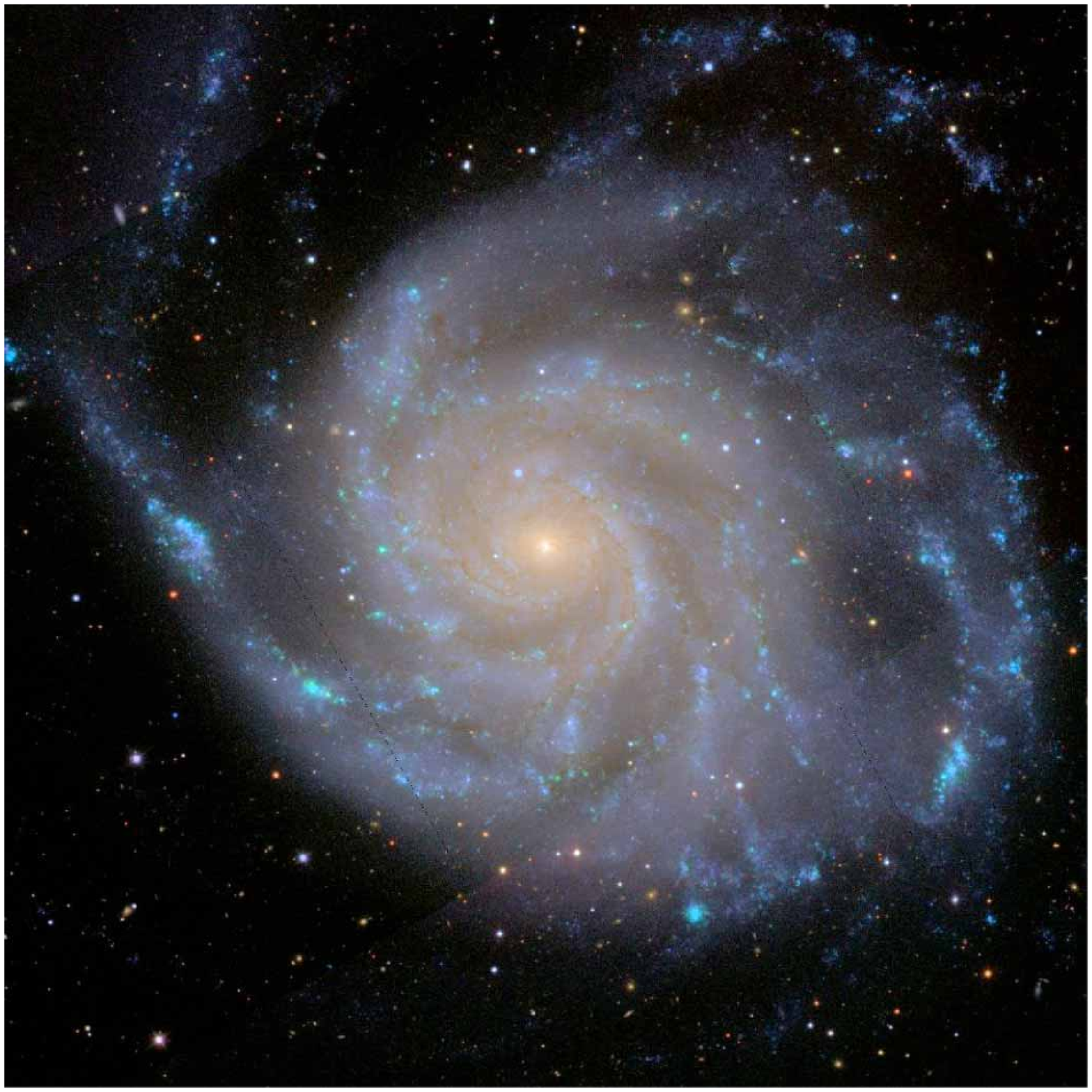}

\figcaption[]
{SDSS color image of NGC 5457 ({\tt http://www.wikisky.org}).  This image emphasizes how 
much this giant galaxy ($V_{\rm circ} \simeq 210 \pm 15$ km s$^{-1}$: Table 2) is dominated 
by its disk.  The tiny, bright center is the pseudobulge; it makes up 2.7\ts\% of the $K$-band 
light of the galaxy (see text).  The nucleus whose properties we use to constrain $M_\bullet$ 
makes up only 0.03\ts\% of the $K$-band light of the galaxy and is completely invisible here. 
It is illustrated in Figure 5.
\lineskip=-4pt \lineskiplimit=-4pt
}
\end{figure*}

\vfill\eject
 
\subsection{NGC 5457 = M{\ts}101}

      Giant pure-disk galaxies present the biggest challenge to our picture of galaxy formation,
because they require the most hierarchical halo growth without converting any pre-existing
stellar disk into a classical bulge.  They also provide important constraints on BH correlations 
with host galaxies.  This paper emphasizes such galaxies.  However, the biggest galaxies are the 
rarest galaxies.  Few are close enough for $M_\bullet$ measurements.  Three giant, unbarred Scd 
galaxies stand out as being potentially useful.  IC 342 has a published $M_\bullet$ limit (B\"oker 
\etal 1999).  NGC 6946 is the subject of \S\ts3.3.  And NGC 5457 -- the best galaxy in many ways -- 
is discussed here.

      Figure 4 shows that the galaxy is completely disk-dominated.  The reddish, 
high-surface-brightness center would traditionally be identified as a tiny bulge; this 
defines the Scd Hubble type.  We will find that it is a pseudobulge: it has the properties 
of bulge-like central components that were manufactured by star \phantom{00000}

\vskip 4.0 truein

\noindent formation following secular inward treansport of gas (see Kormendy 1982, 1993; 
Kormendy \& Kennicutt 2004 for reviews).  The plausible engine for secular evolution is, in this case,
spiral structure that lacks an inner Lindblad resonance.  We will find that the pseudobulge
makes up 2.7\ts\% of the light of the galaxy.  At its center, HST images reveal a distinct 
nuclear star cluster (a ``nucleus'') that makes up only 0.03\ts\% of the light of the galaxy. 
It is too small to be visible in Figure 4, but it is illustrated in Figure 5.  Its properties 
provide our $M_\bullet$ limit.

      To understand our $\sigma$ measurements and to estimate $M_\bullet$, we need surface
photometry of all components in the galaxy.  That is, we need a composite brightness profile
that measures the nucleus at the highest possible spatial resolution but that also reaches
large radii.  It would be best (1) if this profile were observed in approximately the same
wavelength range as the spectroscopy and (2) if it were insensitive to the young stars and
dust that are clearly present (Figure 5).

\vfill\eject

\cl{\null}

\vskip 3.38truein

\includegraphics{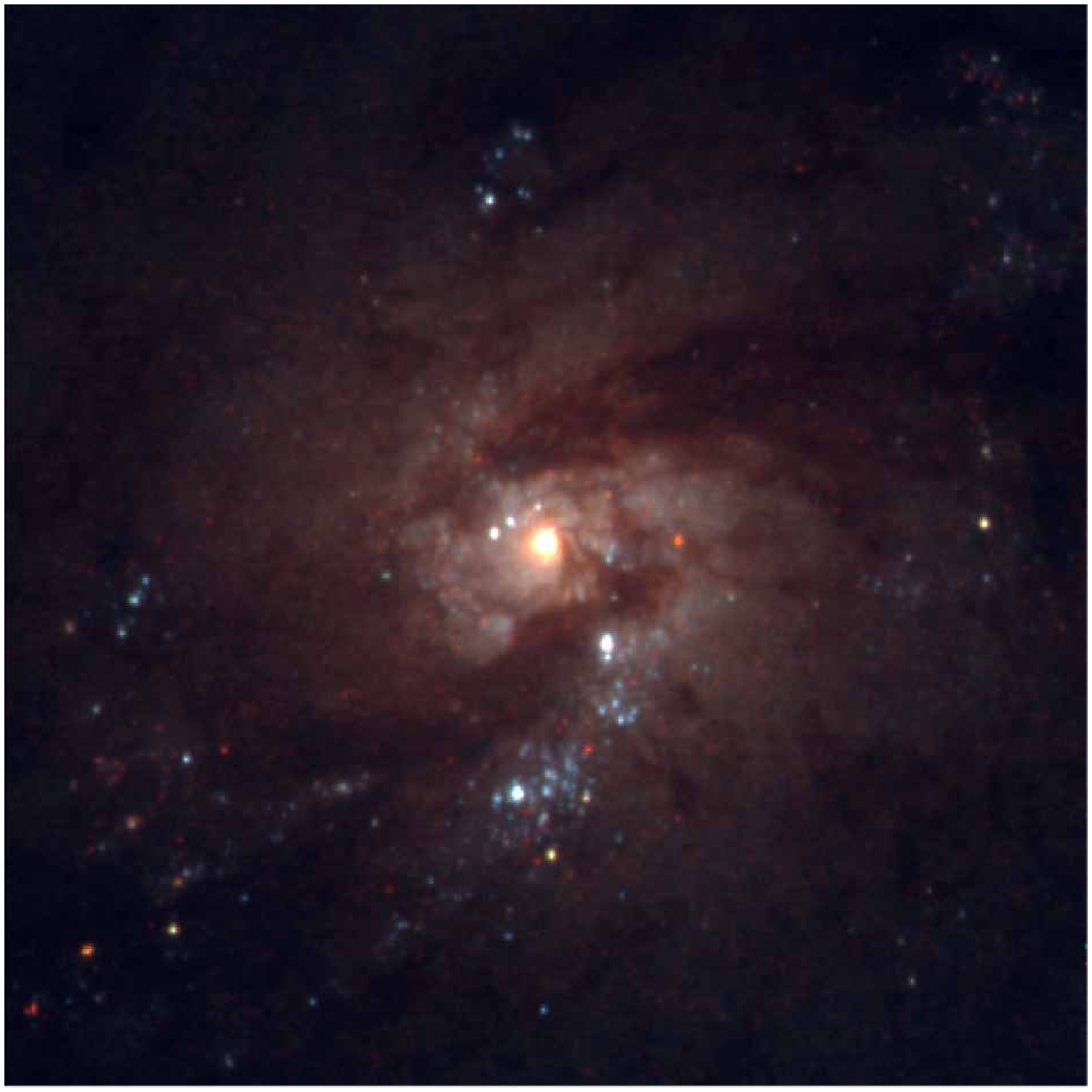}

\figcaption[]
{Color image of the central $20\farcs5 \times 20\farcs5$ of NGC 5457 made from 
$B$-, $V$-, and $I$-band, HST ACS images.  The nucleus is overexposed at the center.
As in M{\ts}33, the nucleus is clearly distinct from the lower-surface-brightness
center of the star-forming pseudobulge (see also Figure 6).  Spiral dust lanes are 
canonically interpreted as regions where gas is being channeled toward the center 
(e.{\ts}g., Athanassoula 1992).
\lineskip=-4pt \lineskiplimit=-4pt
}

\vskip 17pt

      We obtained spectroscopy at the Ca triplet ($\lambda \simeq 8550$ \AA), so
$I$-band photometry sees approximately the same stars.  The HST archives contain two 
$I$-band ACS images that are ideally suited to our purposes.  We use these for the 
central profile.

      However, $K$-band images would more securely provide a brightness distribution
that is proportional to the stellar mass distribution.  Therefore, we constructed a
$K$-band composite profile by grafting a central profile measured using an HST archive 
NICMOS F190N image (brown crosses in Figure 6) to an outer $K$ profile from the
2MASS\footnote{The 2MASS survey uses a $K_s$ bandpass whose effective wavelength is
$\sim 2.16$ $\mu$m (Carpenter 2001; Bessell 2005).  Following the above papers,
we assume that $K_s = K - 0.044$.  Then the $K_s$-band absolute magnitude of the Sun is
3.29 (Cox 2000).  Except in this footnote, we call the 2MASS $K_s$ band ``$K$'' for convenience.}
Large Galaxy Atlas (Jarrett \etal 2003) (brown filled circles in Figure 6.)  The problem
is that the NICMOS PSF and the NIC3 scale of 0\farcs2 pixel$^{-1}$ substantially smooth
the (as it turns out) tiny nucleus.  Therefore we used the ACS $I$-band profile interior
to $2\farcs5$, and we verify that $I$ band is an accurate surrogate for $K$ band in Figure 6.

      Even in HST $I$ band, the PSF causes significant smoothing.  We therefore 
applied 40 iterations of Lucy-Richardson deconvolution (Lucy 1974, Richardson 1972),
as in Lauer \etal (2005).  We used a {\tt VISTA} program that was written and kindly
made available by T.~R.~Lauer; it was thoroughly tested in Lauer \etal (1992, 1995, 
1998).  The composite profile constructed from the deconvolved $I$-band profile at radii 
$r \leq 11\farcs0$ and from the 2MASS $K$ profile at $r \geq 2\farcs5$ is illustrated
in Figure 6.

      The next step was to decompose this profile outside the nucleus 
at $0\farcs65 \leq r \leq 370^{\prime\prime}$ into an exponential disk and
a S\'ersic (1968) $\log{I(r)} \propto r^{1/n}$ function (pseudo)bulge.  The
S\'ersic and exponential fits are shown by black dashed curves in Fig.~6;
their sum is the black solid curve.  It fits the observed profile to an RMS of
0.069 $K$ mag arcsec$^{-2}$ in the fit range.  Since we need to constrain
the (pseudo)bulge properties accurately, we did not worsen the disk
fit by including the outermost three points.

\cl{\null}

\vskip 3.22 truein

\includegraphics{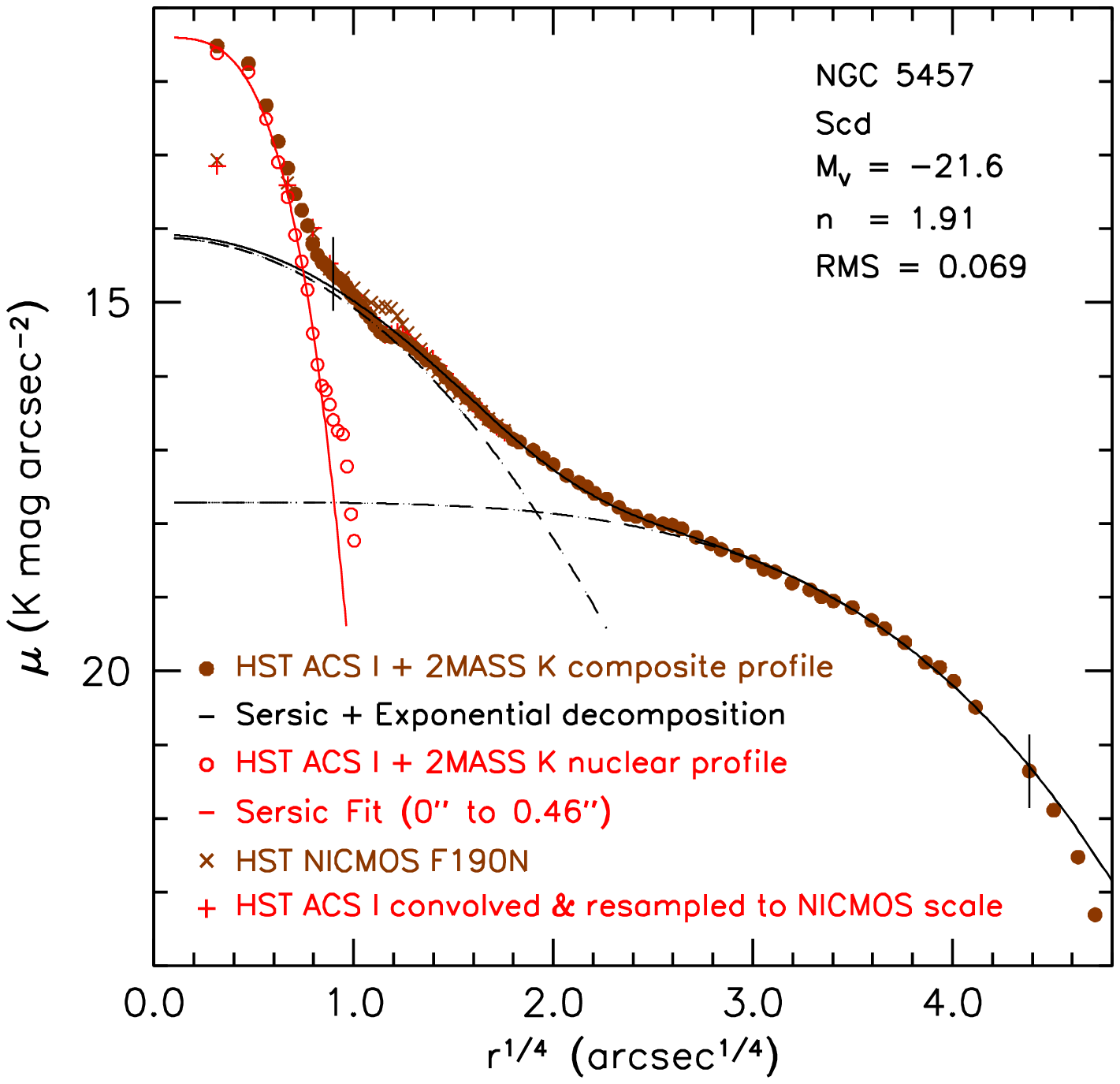}

\figcaption[]
{The brown points show the major-axis, $K$-band brightness profile of NGC 5457.  They are a
composite of the 2MASS Large Galaxy Atlas profile (Jarrett \etal 2003) at large radii and
a deconvolved $I$-band HST ACS profile shifted in $\mu$ to agree with the outer profile 
where they overlap (2\farcs5 -- 11\farcs0).  Also shown is an HST NICMOS F190N
profile (crosses) similarly shifted to the 2MASS outer profile.  The black lines show a 
decomposition of the profile outside the nuclus into an exponential disk and a S\'ersic 
function (pseudo)bulge (dashed black lines).  Their sum (solid black line) fits the observed 
profile in the fit range (vertical dashes) with an RMS of 0.069 $K$ mag arcsec$^{-2}$.  The
 S\'ersic index of the inner component is $n = 1.91$ (key).  Subtracting the fit from the
observed brightness profile provides the brightness profile of the nucleus (open red circles). 
A S\'ersic function fitted to the well defined inner part of the nucleus gives the red curve 
($n = 1.1$).   The nuclear profile is determined entirely from the deconvolved ACS $I$-band 
image.  To check that this accurately represents the inner $K$-band light, we convolved the 
deconvolved ACS $I$ image with the NICMOS PSF and resampled the resulting image at NIC3 scale.  
This gives the profile of the nucleus that is shown by the red plus signs.  It agrees 
well with the profile measured in the NICMOS image.  This shows that the $I$-band image 
is a good high-spatial-resolution surrogate for the $K$-band light.
\lineskip=-4pt \lineskiplimit=-4pt
}

\vskip 10pt
 
      The S\'ersic index of the (pseudo)bulge is 1.91.  Many authors have found that
classical (we believe:~merger-built) bulges almost all have $n \geq 2$
and that observing $n < 2$ correlates with other indicators that the
``bulge'' was built  out of the disk by secular evolution 
(Courteau \etal 1996;
Carollo \etal 1997, 1998, 2001, 2007;  
Carollo 1999;
MacArthur \etal 2003; 
Balcells \etal 2003;  
Fathi \& Peletier 2003;  
Kormendy \& Kennicutt 2004;
Kormendy \& Cornell 2004;
Scarlata \etal 2004;
Peletier 2008;
Fisher \& Drory 2008, 2010;
Gadotti 2009;         
Ganda \etal 2009;     
Weinzirl \etal 2009;
Mosenkov \etal 2010). 
We conclude that NGC 5457 contains a pseudobulge.  Further evidence for a pseudobulge
is provided by the fact that the parameters do not fit the fundamental plane correlations
for classical bulges and ellipticals (Kormendy \etal 2009; Kormendy 2009).  Finally, star 
formation and spiral structure (Fig.~5) are additional pseudobulge indicators (Kormendy 
\& Kennicutt 2004).  

      The magnitude of NGC 5457 obtained by integrating the profile in Fig.~6 to its 
outermost point is $K = 5.530$.  This agrees well with the total magnitude, $K = 5.512$,
given by the 2MASS Large Galaxy Atlas.  The total magnitude of the pseudobulge given by our
decomposition is 9.42.  So the pseudobulge-to-total luminosity ratio is $PB/T = 0.027$ 
(Table 2).

      To use $\sigma$ to derive an $M_\bullet$ limit, we need the properties of the nucleus.  
This is much smaller and denser than the already tiny pseudobulge (Figures 5 and 6).  We 
derive the brightness profile of the nucleus by subtracting the pseudobulge-plus-disk model 
(black curve in Figure 6) from the observed profile.  The result is the profile shown by 
the red open circles in Figure 6.

      The total magnitude of the nucleus obtained by integrating its profile and taking
into account its average axial ratio, $b/a = 0.8$, is $K = 14.37$.  Because this result is 
very sensitive to small wiggles in the observed profile caused by azimuthally averaging star
formation, dust absorption, and noise, we also fitted a S\'ersic function to the well
defined, inner parts of the profile (red solid curve in Figure 6). This gives
a total nuclear magnitude of $K = 14.54$.  The corresponding ratios of nuclear to total
light are $N/T = 0.00029$ and 0.00025, respectively.  So the nucleus contains
$0.027\pm0.002$\ts\% of the light of the galaxy.  This is approximately one-quarter of the
typical ratio $M_\bullet/M$ of supermassive BHs to the mass of their host elliptical 
galaxies and near the bottom end of the range of observed $M_\bullet/M$ values (e.{\ts}g.,
Merritt \& Ferrarese 2001;
Laor 2001;
McLure \& Dunlop 2002;
Kormendy \& Bender 2009, in which the correlation between $M_\bullet/M$ and core missing
light -- their Fig.~2 -- adds additional support for small $M_\bullet/M$ values).

      We need to address one more issue before deriving a nuclear mass and $M_\bullet$ limit.
This is the appropriate value of $\sigma$ to use.~Ho \etal (2009) find a nuclear dispersion 
$\sigma = 23.6 \pm 8.7$ km s$^{-1}$.  We observe $\sigma = 27 \pm 4$ km s$^{-1}$ in a 
3$^{\prime\prime}$-diameter aperture centered on the nucleus.  These values are consistent, 
but we briefly explore the difference.  
Integrating the total composite profile of the galaxy (brown filled circles in Figure 6) 
inside $r = 1\farcs5$ gives a $K$-band magnitude of 12.81.  Comparing this to the above
total magnitude of the nucleus implies that the fraction of the light seen by our spectral 
aperture that comes from the nucleus is 0.24 from the integration of the nuclear profile or 
0.20 from the S\'ersic fit.  
Moreover, the seeing FWHM as measured from the setup exposures was about $2^{\prime\prime}$.
This blurs more nuclear light out of our aperture.  We conclude that we measured the
central velocity dispersion of the pseudobulge.  Ho \etal (2009) had better seeing at the 
Palomar 5 m telescope and better sky$+$pseudobulge subtraction via their long-slit spectra.  
Their central $\sigma$ may be a better measurement of the nucleus.  In particular, it
may be a hint that the velocity dispersion of the nucleus is smaller than that of the pseudobulge.  
This would be consistent with other observations of $\sigma$ drops in nuclei and in pseudobulges 
(see references in \S\thinspace2.3.3).  It favors the conclusion that BHs are small in bulgeless 
galaxies.  However, given measurement errors, we adopt the weighted
mean of the two measurements, $\sigma = 26.4 \pm 3.6$ km s$^{-1}$, for the nucleus.

      First, we estimate the mass $M_{\rm nuc}$ of the nuclear star cluster.
Wolf \etal (2010) present a new mass estimator,
$$
M_{1/2} = 3 \sigma^2 r_{1/2} / G \simeq 4 \sigma^2 r_e / G, \eqno{(1)}
$$
where $M_{1/2}$ is the mass contained within $r_{1/2}$, the radius of the sphere that 
contains half of the light of the unprojected light distribution.  Also, $\sigma$ is the 
line-of-sight projected velocity dispersion, $r_e$ is the half-light radius of the projected 
light distribution, and $G$ is the gravitational constant.  This estimator has two 
virtues for our case: (1) It uses self-consistent properties $r_{1/2}$, $r_e$, and $\sigma$
of any tracer population -- in this case, the stars that contribute most of the light -- 
even when these have a radial distribution that is different from the unknown radial 
distribution of mass.  That is, it does not require the assumption that mass follows
light.  (2) Wolf \etal (2010) show that $r_{1/2}$ is a ``sweet spot'' radius whose choice
ensures that $M_{1/2}$ is minimally sensitive to unknowns like the 
velocity anisotropy of the tracer particles.  We then assume that $M_{\rm nuc} = 2 M_{1/2}$.

\lineskip=-4pt \lineskiplimit=-4pt

      The nucleus of NGC 5457 has $r_e \simeq 0\farcs223$ from an integration of the observed 
PSF-deconvolved brightness profile.  Multiplying by $\sqrt{0.8}$, the mean $r_e = 0\farcs200 = 
6.8$ pc.  Then equation (1) gives 
$M_{1/2} = 4.4 \times 10^6$ $M_\odot$.  
The integral of the nuclear profile also gives 
$K_{\rm nuc} = 14.37$, 
$M_{K,\rm nuc} = -14.86$, and
hence a {\it total} nuclear luminosity of 
$L_{K,\rm nuc} = 18.1 \times 10^6$ $L_{K\odot}$.
Half of this is $L_{K,1/2} = 9.0 \times 10^6$ $L_{K\odot}$ 
and so the nucleus has a global mass-to-light ratio of 
$M_{1/2}/L_{K,1/2} = 0.49$.

Similarly, the S\'ersic fit to the nuclear profile gives $r_e \simeq 0\farcs191$
corresponding to a mean $r_e$ of 5.8 pc.  Equation (1) gives
$M_{1/2} = 3.8 \times 10^6$ $M_\odot$ and an integral of the S\'ersic fit gives
$K_{\rm nuc} = 14.54$, 
$M_{K,\rm nuc} = -14.69$, and
hence a {\it total} nuclear luminosity of 
$L_{K,\rm nuc} = 15.5 \times 10^6$ $L_{K\odot}$.
Half of this is $L_{K,1/2} = 7.75 \times 10^6$ $L_{K\odot}$ 
and so the nucleus has a global mass-to-light ratio of 
$M_{1/2}/L_{K,1/2} = 0.49$.

      In the above, we assumed that the distance to NGC 5457 is $7.0$ Mpc (see Table 2).
Also, the absolute magnitude of the Sun is $M_{Ks\odot} = M_{K\odot} - 0.044 = 3.286$
(Cox 2000; footnote 7 here).

      A check on the above $M/L_K$ ratio is provided by estimating the mass $M(r_c)$ and light 
$L(r_c)$ inside the core radius $r_c$. An approximate $M(r_c)$ is provided by King (1966) core fitting,
     $(M/L)_0 \simeq 9 \sigma^2 / 2 \pi G \Sigma_0 r_c$, 
where $\Sigma_0$ is the central surface brightness and $r_c$ is the radius at which the surface 
brightness has fallen by a factor of 2 from the central value.  From the S\'ersic fit to the 
nucleus, we derive an upper limit on the core radius, $r_c \lesssim 0\farcs064$ and a lower 
limit on the central surface brightness, $\Sigma_0$ corresponding to 11.41 $K$ mag arcsec$^{-2}$.
The product $\Sigma_0 r_c$ is much less sensitive to resolution than either value is individually
(Kormendy \& McClure 1993).  This gives $(M/L_{K})_0 = 0.45$.  Note that this 
is an estimate of the central volume (not projected) $M/L$ ratio.   The core 
mass-to-light ratio of the nucleus is remarkably similar to the global value.  This strengthens 
the justification that our measurements of $M/L$ ratios and masses are realistic.
The uncertainty is that we had to assume that $\sigma$ is independent of radius.  This has been 
verified in M{\ts}33 (Kormendy \& McClure 1993; Gebhardt \etal 2001; contrast Merritt \etal 2001) 
but not in our present galaxies. 

      The above mass-to-light ratios are intermediate between values of $M/L_K \sim 1$ 
that are normal for old stellar populations and the smallest values $M/L_K \simeq 0.05$ observed
for the youngest stellar populations (B\"oker \etal 1999).  We need this $M/L_K$
in order to understand the stars.  Assuming below that $M/L_K = 0$ therefore gives a strong 
upper limit on $M_\bullet$.

      A limit on $M_\bullet$ can be derived by making dynamical
models of the light distribution and the luminosity-weighted total $\sigma$ with $M/L$ and $M_\bullet$
as free parameters.  Merritt (1987) shows that the total mass is minimized by putting all of the mass
into a point at the center.  Independent of velocity anisotropy, this minimum is
$M_{\rm min} = <{\null}V^2\null>/G <\null1/r\null>$, where $<\null V^2\null>$ is the mean-square stellar 
velocity and $<\null1/r\null>$ is the harmonic mean radius of the cluster.  We assume isotropy and 
adopt $M_\bullet \lesssim 3 \sigma^2 / G <\null1/r\null>$.  Barth \etal (2009) arrive at the same conclusion
by using Jeans models to
explore the tradeoff between $M/L$ and $M_\bullet$ for the nucleus of the Sd galaxy NGC 3621; for this
example, the range of masses obtained for plausible anisotropies is small.  For the nucleus of
NGC 5457, we measure $<\null1/r\null>^{-1} = 0\farcs18 \pm 0\farcs01$.  Correcting  for flattening,
$<\null1/r\null>^{-1}  = 5.4 \pm 0.2$ pc.  Therefore we conclude that $M_\bullet \lesssim (2.6 \pm 0.5)
\times {\kern -2pt}10^6$ $M_\odot$.  In comparison, the mass of the nucleus is 
$M_{\rm nuc} = (8.1 \pm 1.7) \times {\kern -2pt}10^6$ $M_\odot$\ts(Table\ts1).

\begin{figure*}[b]

\vskip 3.8truein

\includegraphics{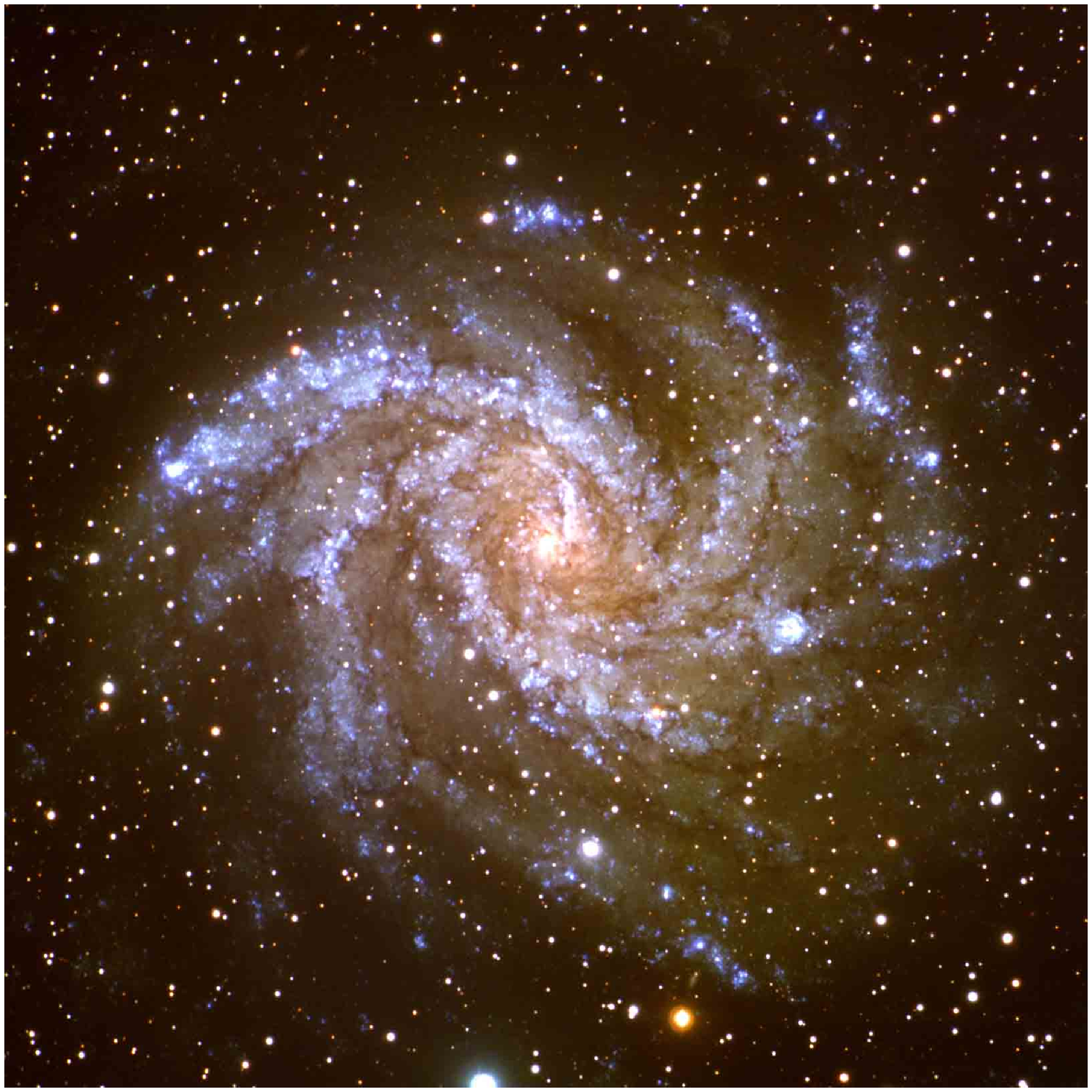}

\figcaption[]
{Color image of NGC 6946 taken with the Large Binocular Telescope 
({\tt http://medusa.as.arizona.edu/lbto/astronomical.htm}).   This galaxy is very similar to
NGC 5457: it is a giant galaxy ($V_{\rm circ} \simeq 210 \pm 10$ km s$^{-1}$: Table 2), but it is 
completely dominated by its disk (Hubble type Scd).  As in NGC 5457, the tiny, bright center visible
in this image proves to be a pseudobulge that makes up 2.4\ts\% of the $I$-band light of the galaxy 
(see text).  The nucleus whose dispersion we measure makes up only 0.12\ts\% 
of the $I$-band light of the galaxy.  It is completely invisible here but is illustrated in Figure 8.
\lineskip=-4pt \lineskiplimit=-4pt
}
\end{figure*}

\vfill\eject

\subsection{NGC 6946}

     Globally, NGC 6946 is very similar to NGC 5457.  It has the same Scd Hubble 
type.  It has almost the same luminosity ($M_V \simeq -21.4$ versus $-21.6$ for NGC 5457),
inclination-corrected maximum rotation velocity ($V_{\rm max} = 210 \pm 10$ km s$^{-1}$ versus
$210 \pm 15$ km s$^{-1}$ for NGC 5457), and distance (5.9 Mpc versus 7.0 Mpc for NGC 5457; see
Tables 1 and 2).  It is less well known than NGC 5457 because it is heavily obscured by our Galactic disk.
We adopt absorptions $A_V = 1.133$, $A_I = 0.663$, and $A_K = 0.125$ (NED, following Schlegel \etal 1998).

     Figure 7 illustrates the similarity to NGC 5457.~We tried to match the color scheme of 
Figure 4 but did not fully~succeed: the bandpasses are different, and the correction for
foreground reddening is not perfect.  In fact, the galaxies have similar dereddened total colors: 
$(B - V)_{T0} \simeq 0.46$ for NGC 6946 and 0.44 for NGC 5457.
Both disks are dominated by ongoing star formation. A difference is that NGC 6946 has a compact 
central concentration of molecular gas and a nuclear starburst; we will detect this gas dynamically.
We will not find a secure $M_\bullet$ limit.

\vskip 4.0truein

     Like NGC{\ts}5457, NGC{\ts}6946 has no hint of a classical bulge.  In photometry
discussed below, the overexposed red center shown in Figure 7 proves to be a pseudobulge.  
As in NGC{\ts}5457, it is easy to identify an engine for secular evolution: the spiral
structure and associated dust lanes reach the nucleus, so there is no effective inner 
Lindblad resonance (see Kormendy \& Norman 1979) that acts as a barrier to inflowing gas.  
However, we expect that secular evolution is slow in a barless Scd galaxy (Kormendy \& 
Kennicutt 2004; Kormendy \& Cornell 2004).  So, as in  NGC 5457, it is no surprise 
that the pseudobulge of NGC 6946 is tiny.  It adds up to 2.4\ts\% of the $I$-band 
light of the galaxy.  

      At the center of NGC 6946 is an even tinier nucleus (Fig.~8) that is seen in the $V$-band 
decomposition of Fisher \& Drory (2008) but that is still more obvious in $I$ band.  Large color 
gradients in NGC 6946 imply (in contrast to NGC 5457) that the nucleus is dominated by young stars.  
To measure its mass, it is important that we measure its brightness profile at the same wavelength 
that we used in our spectroscopy to measure $\sigma$.  We~therefore work in $I$ band. 

\vfill\eject

\cl{\null}

\vskip 3.38truein

\includegraphics{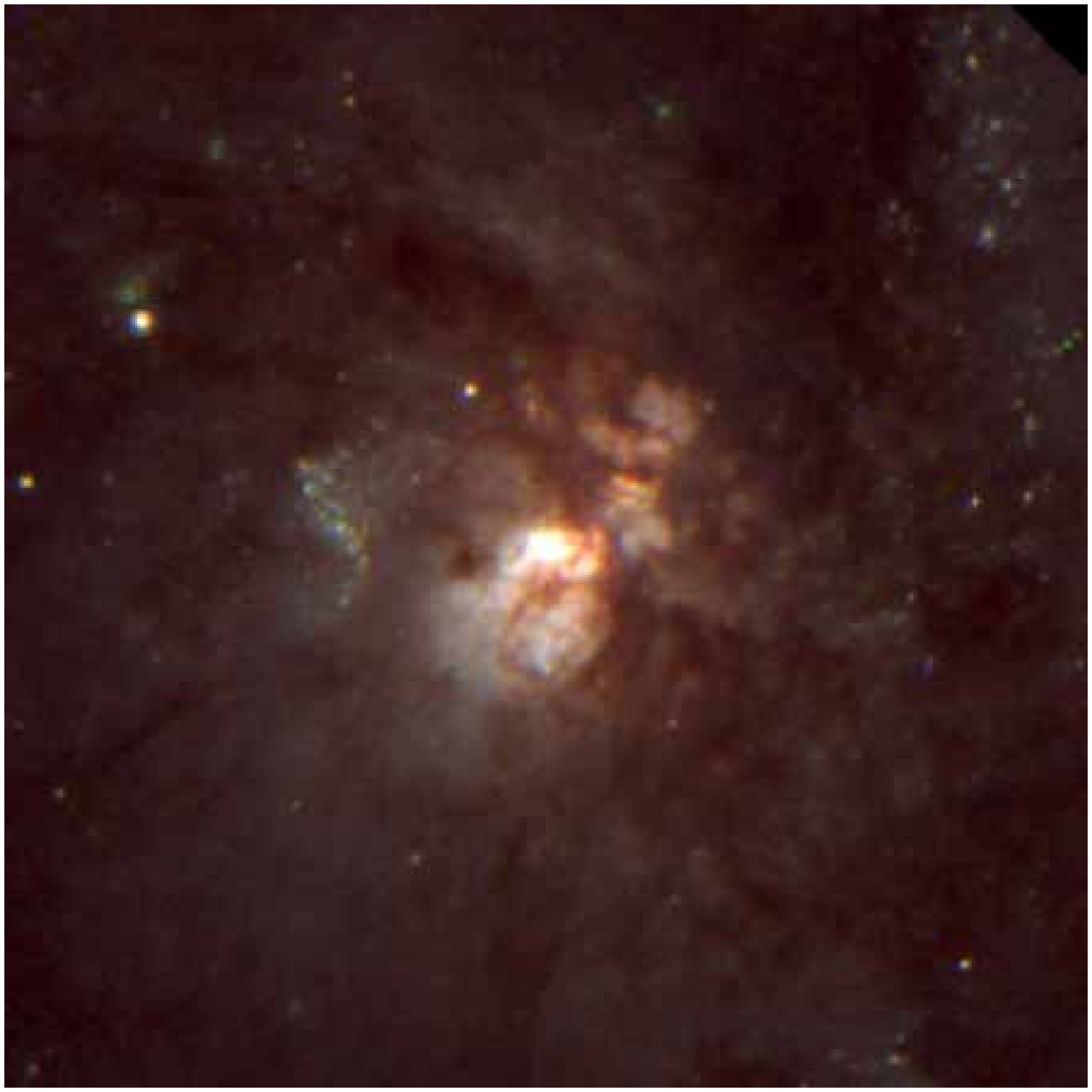}

\figcaption[]
{Color image of the central $20\farcs5 \times 20\farcs5$ of NGC 6946 made from 
F547M-, F606W-, and F814W-band, HST WFPC2 images.  North is up and east is at left.
The nucleus is overexposed at the center.  As in M{\ts}33 and NGC 5457, the nucleus is 
clearly distinct from the lower-surface-brightness center of the star-forming pseudobulge 
(see also Figures 9 -- 11).  The pseudobulge is irregular due to patchy star formation and 
differently patchy absorption.  Its SE\ts--{\ts}NW elongation causes the $\epsilon$ maximum at 
$r^{1/4} \simeq 1.3$ in Figure 9; this was called a ``nuclear bar'' by Elmegreen 
\etal (1998).  The galaxy looks less patchy in $H$ and $K$ bands, but it continues to be brightest
at the same nucleus.
\lineskip=-4pt \lineskiplimit=-4pt
}

\vskip 17pt

\subsubsection{$I$-Band Photometry and Nuclear Mass Estimates}

      Figures 9\ts--\ts11 show the brightness profile of NGC 6946.  The individual measurements
are shown in Fig.~9; their average~is in Fig.~10 and 11.  Figure 9 also shows 
ellipticity profiles $\epsilon(r)$.
 
      At  $r \geq 6\farcs0$, we used an $I$-band profile from the McDonald 0.8 m 
telescope measured and kindly provided by Fisher \& Drory (2008).  At $r \leq 23\farcs2$,
we measured the profile in an HST ACS F814W image, and at $r \leq 15\farcs9$, we measured
it in the WFPC2 F814W image used in Figure~8.  Where they overlap, the HST profiles agree 
almost perfectly (RMS difference $=$ 0.030 $I$ mag arcsec$^{-2}$ for 49 overlapping points
omitting one deviation of 0.138 mag arcsec$^{-2}$).  We also measured the ACS
profile after 40 iterations of Lucy-Richardson deconvolution.   However, HST
easily resolves the central flat profile, so deconvolution makes no 
significant difference.  We adopt the undeconvolved profile.  

      We used the $I$-band VEGAmag zeropoint 25.53561 mag 
({\footnotesize\tt http://www.stsci.edu/hst/acs/analysis/zeropoints})
for ACS observations taken before 2006 July 4.  To estimate the total magnitude of the galaxy, 
we extended the observed profile using the exponential fit in Figures 9 -- 11.
Integrating this extended profile together with the ellipticity profile gives a total 
apparent magnitude of $I_{\rm tot} = 7.43$.  This compares very well with $I_{\rm tot} = 7.46$
found by Makarova (1999).  Also, our exponential disk fit in Figures 9 -- 11 has an 
apparent central surface brightness of 19.44 $I$ mag arcsec$^{-2}$ and a scale length of 
115\farcs9. Makarova got 19.41 $I$ mag arcsec$^{-2}$ and 113\farcs1, respectively.  
Springob \etal (2007) got $I_{\rm tot} = 7.33 \pm 0.04$ extrapolated to 8 disk scale lengths.
However, we and Springob extrapolate the surface brightness profile to
23.5 $I$ mag arcsec$^{-2}$ at $r = 425\farcs6$ and $423\farcs2$, respectively.  So the
agreement in zeropoints, parameters, and total magnitudes is good.  

\cl{\null}
\vskip 4.71truein

\includegraphics{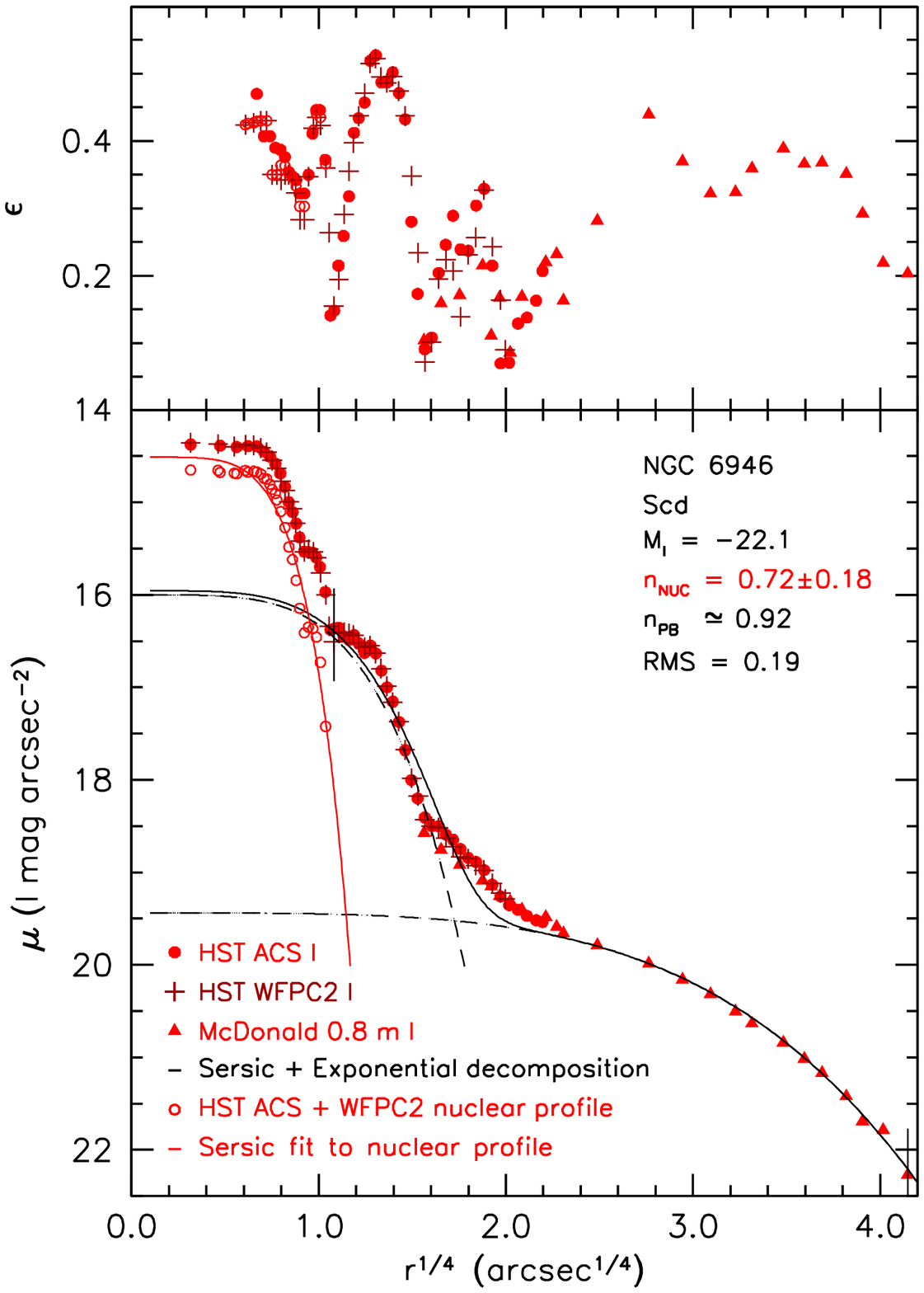}

\figcaption[]
{Major-axis, $I$-band brightness profile of NGC 6946.~Black lines show a decomposition 
in the fit range (vertical dashes) into an exponential disk and a S\'ersic pseudobulge 
(dashed black lines).  Their sum is the solid black line.  The S\'ersic index of the 
pseudobulge is $n_{\rm PB} = 0.92$~(key).  Subtracting the fit from the observed profile 
gives the profile of the nucleus (open red circles).  A S\'ersic fit to the nuclear
profile (red curve) has $n_{\rm nuc} = 0.72 \pm 0.18$.  
\lineskip=-4pt \lineskiplimit=-4pt
}

\vskip 14pt

      However, it is unrealistic to think that we know the total magnitude to better
than $\sim 0.1$ mag.  Reasons include the irregularities introduced by patchy star formation
and dust absorption, spiral structure, and the overall disk asymmetry.  The ellipticity
measurements are uncertain at large radii.  The brightness and
the ellipticity profiles must be extrapolated to get the total magnitude; we do not
know whether the disk has an outer cutoff.  Even the uncertainties of foreground star 
removal are not negligible.  We adopt $I_{\rm tot} = 7.43$ from our photometry.  
With $A_I = 0.663$ and an adopted distance of 5.9 Mpc (Table 2), the total absolute 
magnitude of NGC 6946 is $M_I = -22.1$ (keys to Figures 9 -- 11).

      To derive $M_{\rm nuc}$, we need the $I$-band total luminosity and the effective radius of the nucleus.  
Also, we need to know that our spectroscopy measured its velocity dispersion.  And we need a reliable 
classification and total luminosity of the (pseudo)bulge.  All of these require decomposition of the 
observed brightness distribution into nuclear, bulge, and disk contributions, with due regard to 
uncertainties introduced by the patchy light distribution in Figure 8.

      Figures 9 -- 11 show three decompositions.  The disk fit is identical in all three.  
The overall fit to the (pseudo)bulge is best in Figure 9:~we fit all of the profile outside~the~nucleus.  
Figures 10 and 11 provide error bars on the (pseudo)bulge parameters. \phantom{00000}
\vskip 10pt\centerline{\null}

\vskip 3.18 truein

\includegraphics{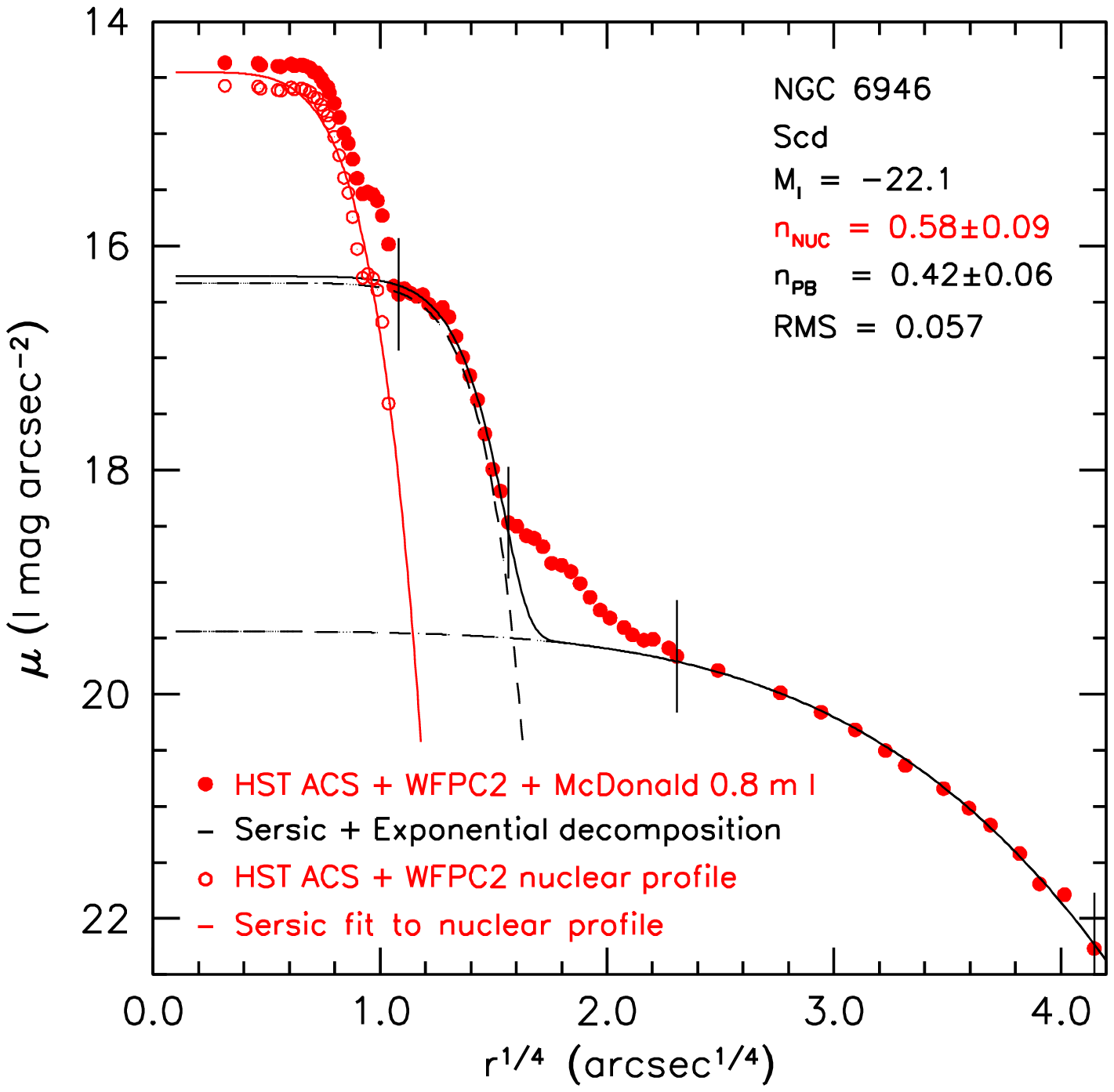}

 \figcaption[]
{Composite brightness profiles of NGC 6946.  Black lines show a S\'ersic -- exponential 
decomposition of the extra-nuclear profile in two radial ranges (vertical dashes) that
omit points between $r^{1/4} = 1.60$ and 2.27.  This gives
(solid black line) a better fit to the central pseudobulge profile and a
more accurate extrapolation into the nucleus.  The nuclear profile and
S\'ersic fit were then calculated as in Figure 9.
\lineskip=-4pt \lineskiplimit=-4pt
}
      
\vskip 10pt

\noindent The good fit to the central, almost-constant-surface-brightness part of the (pseudo)bulge 
in Figure 10 provides the best extrapolation into the nucleus and therefore the best brightness 
profile of the nucleus.  The decompositions in Figures 9 and 11 are used to provide error bars on 
the nuclear parameters.  The profile of the nucleus is so steep that these errors are small.

      Figure 9 shows a decomposition of all of the profile outside the nucleus.  
Between the vertical tics ($1\farcs3 \leq r \leq 296^{\prime\prime}$), an outer 
exponential profile $+$ an inner S\'ersic (1968) \hbox{$\log{I(r)} \propto r^{1/n}$}
function fit the data with an RMS of 0.19 $I$ mag arcsec$^{-2}$.  The RMS is 
dominated by the poor (pseudo)bulge fit; the fit to the disk is good to a few percent.
The measurements of the (pseudo)bulge are accurate -- the ACS and WFPC2 profiles
agree almost perfectly -- but given the asymmetric and patchy star formation and dust,
the idea that the brightness distribution in Figure 8 can be described by $\mu(r)$, 
$\epsilon(r)$, and a position angle profile is more approximate than usual.  

      Happily, the fit in Figure 9 is easily adequate for our needs.  
We {\it do not\/} use it to measure the (pseudo)bulge magnitude.  We use it only
to help classify this component and to estimate how much light it adds
to the nucleus.  First, the classification: Its S\'ersic index 
is $n_{\rm PB} \simeq 0.92$.  The range in Figures 10 and 11 is $n_{\rm PB} = 0.42 \pm 0.06$ 
to $n_{\rm PB} = 0.92 \pm 0.37$.  Robustly,  $n_{\rm PB} \ll 2$.  Many papers (cited in \S\ts3.2)
have shown that this implies that the component is a pseudobulge.

      Note:~There is no sign of a classical bulge in NGC~6946: i.{\ts}e., one 
that has  $n \gtrsim 2$ and that satisfies the fundamental plane correlations for
elliptical galaxies (see Carollo 1999, Kormendy \& Fisher 2008, Fisher \& Drory 2008,
and Gadotti 2009 for bulge-pseudobulge comparisons).

      Next, we need to derive the brightness profile and the structural parameters of
the nucleus.  Extrapolating the sum of the exponential and S\'ersic-function
fits in Figures 9\ts--\ts11 (solid black curves) to smaller radii provides three 
estimates of the amount of pseudobulge light that underlies the nucleus. 
\phantom{000000000000}

\vskip 3.18 truein

\includegraphics{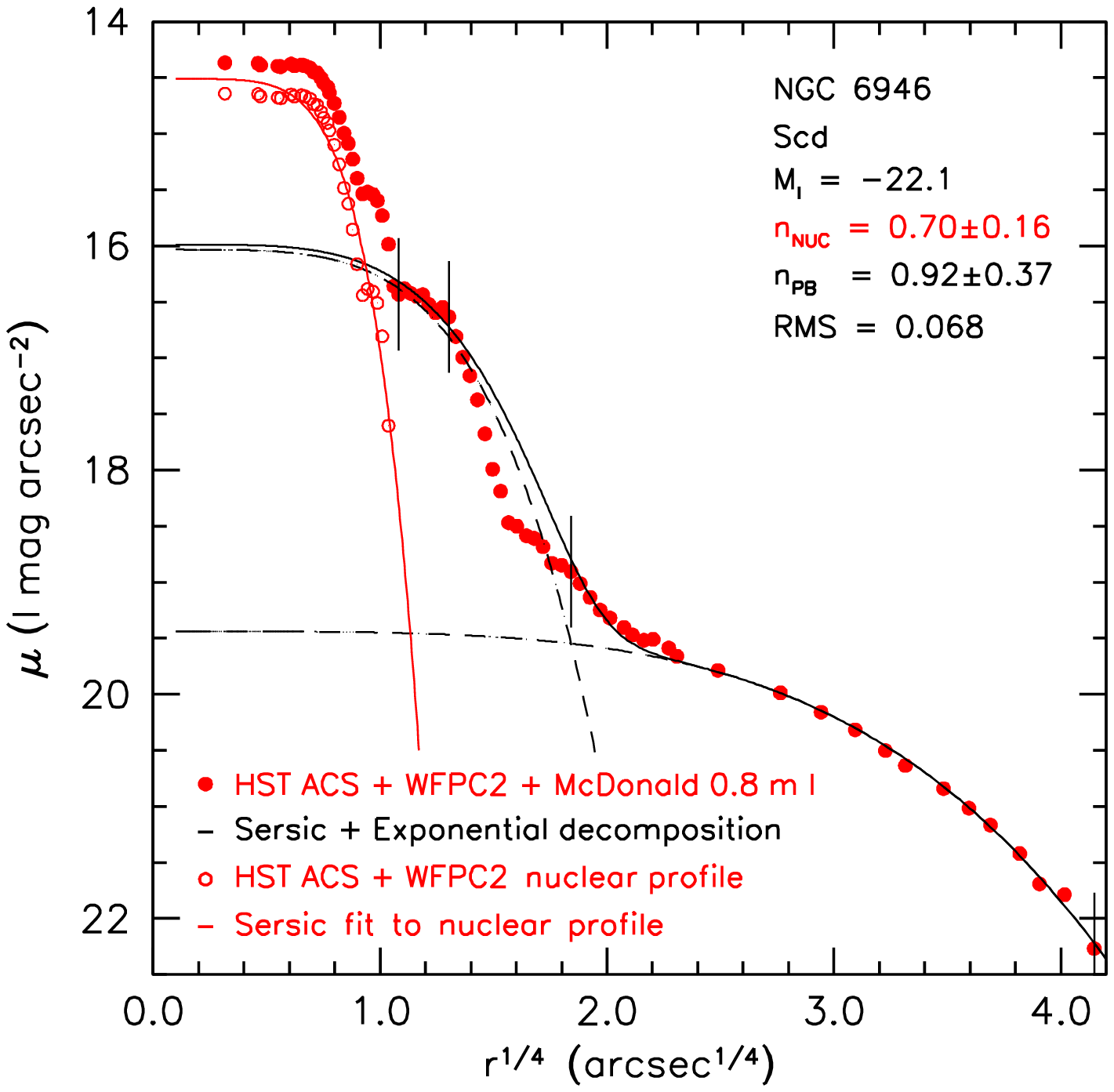}

 \figcaption[]
{Composite brightness profile of NGC 6946.  Black lines show a S\'ersic -- exponential 
decomposition of the extra-nuclear profile in two radial ranges (vertical dashes) 
that omit points between $r^{1/4} = 1.33$ and 1.80.  This gives the largest
(pseudo)bulge $n_{\rm PB} = 0.92 \pm 0.37$ that is consistent with the data.  The 
extrapolation into the nucleus is fortuitously almost identical to that in Figure 9.
The nuclear profile and S\'ersic fit were calculated as in Figure~9.
\lineskip=-4pt \lineskiplimit=-4pt
}
      
\vskip 10pt

\noindent Subtracting these from the observed profile gives the nuclear profile.  It is 
shown by open circles in Fig.~9\ts--\ts11.  Fortunately, the pseudobulge contributes
little light underlying the nucleus in $I$ band.  So uncertainties in 
the above extrapolation are small.  S\'ersic fits to the nuclear profiles are
shown by the red curves in Figures 9\ts--\ts11.  The most accurate inner pseudobulge
fit and therefore plausibly the best extrapolation is the one in Figure 10.  The 
resulting S\'ersic index of the nucleus is $n_{\rm nuc} = 0.58 \pm 0.09$.  The
nuclear profile at $r \gtrsim 0\farcs5$ falls off almost as steeply as a Gaussian 
($n = 0.5$).  The other two decompositions provide error bars.  We conclude that 
$n_{\rm nuc} = 0.6_{-0.1}^{+0.2}$.  One consequence is that the effective and 
harmonic mean radii are well constrained.  This improves our mass estimates.

      The S\'ersic fits in Fig.~9 -- 11 give major-axis effective radii of the nucleus, 
$r_e =  0\farcs63 \pm 0\farcs11$, 
       $0\farcs62 \pm 0\farcs10$, and 
       $0\farcs60 \pm 0\farcs10$,
respectively.  Alternatively, integrating the observed profiles implies that 
$r_e =  0\farcs52 \pm 0\farcs02$, 
       $0\farcs52 \pm 0\farcs02$, and
       $0\farcs51 \pm 0\farcs02$.
We adopt $r_e = 0\farcs57 \pm 0\farcs11$ along the major axis.  Since the ellipticity of the
nucleus is $\epsilon \simeq 0.35 \pm 0.05$ (Figure 9), the mean effective radius that
is relevant for Virial theorem arguments is $<$\null$r_e$\null$>$ = $r_e \sqrt{1 - \epsilon}$ =
$0\farcs46 \pm 0.09$ $\simeq$ $13.2 \pm 2.6$ pc.

      Averaging all three decompositions, the total magnitude of the nucleus corrected for its
flattening is $I_{\rm nuc} = 14.70^{+0.06}_{-0.10}$.  The~corresponding absolute magnitude is
$M_{I,\ts\rm nuc} = -14.8$.  The nucleus-to-total luminosity ratio is 
$(N/T)_I = 0.00124^{+0.00013}_{-0.00016}$.

      Getting the pseudobulge-to-total luminosity ratio is trickier, because none of the 
three decompositions is adequate to provide the pseudobulge magnitude from the S\'ersic fit.  
Instead, we exploit the excellent fit of the exponential to the disk~profile. The disk
dominates at $r^{1/4} > 2.4$.  We therefore measure the pseudobulge $+$ nucleus contribution 
by integrating the observed brightness and ellipticity profiles out to the above radius and 
subtracting the exponential disk fit integrated to the same radius with 
$\epsilon_{\rm disk} = 0.35$ (Figure 9).  We then subtract the nucleus.  This gives the
pseudobulge apparent and absolute magnitudes, $I_{\rm PB} = 11.47$ and $M_{I,\ts\rm PB} = -18.0$.  
The pseudobulge-to-total luminosity ratio is $(PB/T)_I = 0.024$.  Large color gradients in 
NGC 6946 imply that $N/T$ and $PB/T$ are different in other bandpasses.  But $(PB/T)_I$ in 
NGC 6946 is similar to $(PB/T)_K$ in NGC 5457.  Both pseudobulges are tiny.

      Before estimating masses, we need to check whether our
spectra adequately measure $\sigma$ in the nucleus of NGC 6946.  Comparing $I_{\rm nuc} =
14.70$ with the integral of the total brightness and ellipticity profiles out to the radius
$r = 1\farcs5$ of our input fiber implies that 52\ts\% of the light in our spectra came from the
nucleus.  The real contribution is slightly smaller because of seeing.  However, recall
that we measured $\sigma = 56 \pm 2$ km s$^{-1}$, whereas Ho \etal (2009) got
$\sigma = 55.8 \pm 9.4$ km s$^{-1}$.  We expect that the Ho \etal (2009) spectra have better
spatial resolution than our own.  One possible concern is that nuclear velocity dispersions may
be slightly smaller than pseudobulge dispersions in NGC 5457 (\S\ts2.3.5) and NGC 6503~(\S\ts2.3.3). 
But our excellent fits of broadened standard star spectra to the line-of-sight velocity 
distributions exclude roughly equal contributions to our NGC 6946 spectra from two components 
that have very different velocity dispersions.  We therefore feel safe in adopting 
$\sigma = 56 \pm 2$ km s$^{-1}$ for the nucleus.

      We now derive the dynamical mass of the nucleus using the Wolf \etal (2010) estimator.
As in NGC 5457, the nucleus has a steep enough profile -- steeper than $I \propto r^{-2}$ deprojected~--
so that we can treat it as an independent dynamical system (Tremaine \& Ostriker 1982).  Then 
$<$\null$r_e$\null$>$ = $13.2$ pc and $\sigma = 56$ km s$^{-1}$ imply that the nuclear half-mass is 
$M_{1/2} = (38 \pm 8) \times 10^6$ $M_\odot$.  From $M_{I,\rm nuc} = -14.82$, half of the $I$-band 
luminosity is $L_{1/2} = (19.1^{+1.7}_{-1.0}) \times 10^6$ $L_{\ts I\odot}$.  So the mass-to-light 
ratio inside a sphere that contains half of the mass of the nucleus is $(M/L)_{\ts I} = 2.0^{+0.5}_{-0.4}$.

      A check on this result is provided by the core $M/L$ ratio.  It is better determined in
NGC 6946 than in NGC 5457 because a flat profile is well resolved at the center.  Its physical origin 
is unlikely to be the same as those of the cores in globular clusters or in elliptical galaxies.  
However, estimates of how much gravity is required to bind the near-central stars do not 
depend on this physics.  Also, $r_c \simeq 0\farcs48 \simeq 14$ pc 
is small; if the three-dimensional velocity dispersion is $\sqrt{3} \sigma$, then a typical
star travels a distance of $r_c$ in $\sim$\ts140,000 yr.  This is much less than the 
lifetimes of even the most massive stars.  It seems safe to assume that the core and, 
indeed, all of the nucleus is well mixed and in dynamical equilibrium.  The central 
surface brightness is $14.37 \pm 0.05$ $I$ mag arcsec$^{-2}$, and the mean core radius is
$<$\null$r_c$\null$>$ = $r_c \sqrt{1 - \epsilon} = 0\farcs39 = 11.2$ pc.  Then the core mass-to-light 
ratio is $(M/L)_{I,0} = 9 \sigma^2 / 2 \pi G \Sigma_0 r_c$ = 
$(1.9 \pm 0.1)$ $(M/L)_{\ts I\ts\odot}$.  This is in excellent agreement with the global 
mass-to-light ratio $(M/L)_{\ts I} = 2.0^{+0.5}_{-0.4}$ estimated above.  Note that $r_c \approx r_e$,
so this is a check on our machinery rather than a check on whether $M/L_I$ depends on radius.
The core mass is $M(r_c) = 1.074 \Sigma_0 r_c^2 (M/L)_{\ts I,0} = (12.3 \pm 0.6) \times 10^6$ $M_\odot$.  
All estimated errors here are internal; they do not include distance, magnitude zeropoint, or model 
assumption errors. 

      The question is:~What objects dominate $M_{1/2}$ and $M(r_c)$?  Possibilities include stars 
(which can be obscured by dust), gas, and a central BH.  For stars, $M/L_{I,0} = 2.0 \pm 0.1$ is normal 
for an old stellar population in a globular cluster (Wolf \etal 2010, see Fig.~4) or a small early-type 
galaxy (Cappellari \etal 2006, see Fig~8).  Star formation is in progress in NGC 6946, but our Figure 8 
also shows patchy absorption.  We need to look at the situation in more detail to see how consistent 
our results are with the sum of a central concentration of gas plus a partly absorbed, mixed-age stellar 
population.  

\subsubsection{$H$- and $K$-Band Photometry and Mass Estimates}

      We therefore measured $K$-, $H$-, and $V$-band profiles (Fig.~12).  The $K$-band 
profile was measured using an HST NICMOS NIC3 F190N image zeropointed to the 2MASS Large Galaxy Atlas 
$K_s$ profile (Jarrett \etal 2003).~We abbreviate $K_s$ as~``$K$''.  The $H$-band profile was 
measured using a NICMOS NIC2 F160W image zeropointed to the 2MASS $H$-band profile.  
Unlike the $I$- and $V$-band profiles, the HST $H$ and $K$ profiles required 40 iterations 
of Lucy-Richardson deconvolution using PSF stars in the images.  Finally, we remeasured the 
$V$-band profile using an HST WFPC2 PC F547M image.  The zeropoint is from Dolphin (2009).  
In Figure 12, the outer $V$-band profile is from Fisher \& Drory (2008) shifted to our zeropoint.

      The contrast of the nucleus above the pseudobulge is smaller at $H$ and $K$ than at
$I$, so we do not try profile decomposition.  Instead, we estimate core masses and $M/L$ ratios
using the total profiles in Figure 12.  In $H$ band, $<$\null$r_c$\null$>$ = 0\farcs60 = 17~pc;
$M/L_H = 0.37$; $M(r_c) = 19 \times 10^6$ $M_\odot$.  In $K$ band, $<$\null$r_c$\null$>$ = 0\farcs69 = 20 pc; 
$M/L_K = 0.15$; $M(r_c) = 22 \times 10^6$ $M_\odot$. 
These infrared mass-to-light ratios are smaller than those of unobscured, old stellar populations. 
This suggests that the nucleus contains young stars that are partly obscured at optical wavelengths.

\vfill

\includegraphics{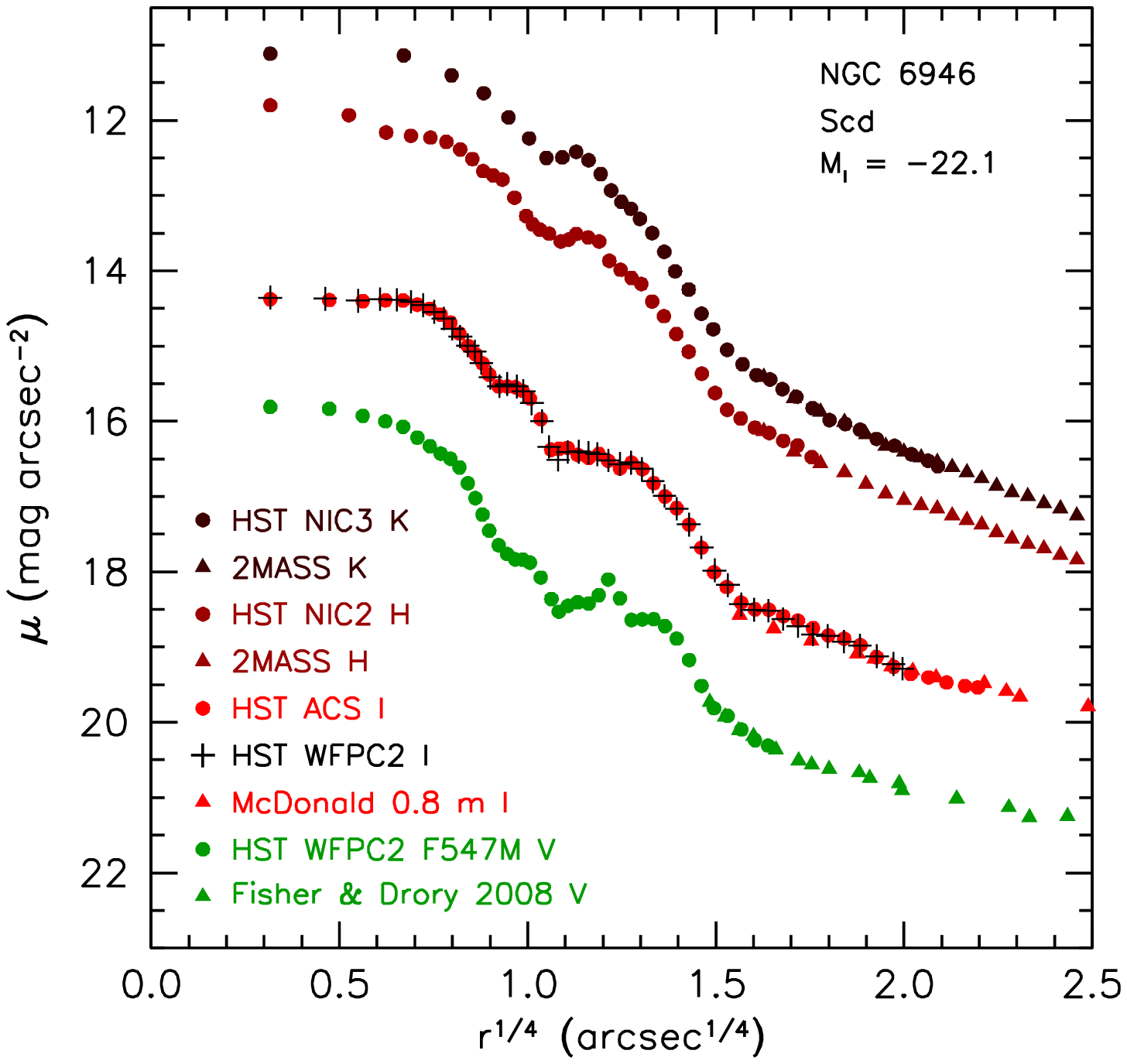}

 \figcaption[]
{Composite $K$-, $H$-, $I$-, and $V$-band, major-axis brightness profiles of NGC 6946. 
All individual profiles that are used in this paper are shown.
\lineskip=-4pt \lineskiplimit=-4pt
}

\vskip 5pt

\subsubsection{Stellar Population Models and the\\Stellar Mass of the NGC 6946 Nucleus}

      We therefore compare our results with models of stellar populations that include starbursts and 
internal absorption.  The modeling machinery from Drory \etal (2004a, b) was used to fit the central 
surface brightnesses, 
$\mu_V = 15.81 \pm 0.05$ $V$ mag arcsec$^{-2}$,
$\mu_I = 14.37 \pm 0.05$ $I$ mag arcsec$^{-2}$,
$\mu_H = 11.80 \pm 0.08$ $H$ mag arcsec$^{-2}$, and
$\mu_K = 11.11 \pm 0.10$ $K$ mag arcsec$^{-2}$.
Relative errors are estimated from plausible zeropoint and photometry errors.  The foreground
extinctions were assumed to be $A_V = 1.133$, \hbox{$A_I = 0.663$,} $A_H = 0.197$, and $A_K = 0.125$ 
(Schlegel et al. 1998).  An example of such a model is shown in Figure 13.  

     Each model consists of the sum of a starburst with constant star formation rate for the past 50~Myr 
and a stellar population of intermediate to old age.  Its spectrum is synthesized using the Charlot \& 
Bruzual (2010) stellar population synthesis library that incorporates an improved treatment of thermally-pulsating
      
\clearpage

\begin{figure*}[ht]

\vskip 8.45 truein

\includegraphics{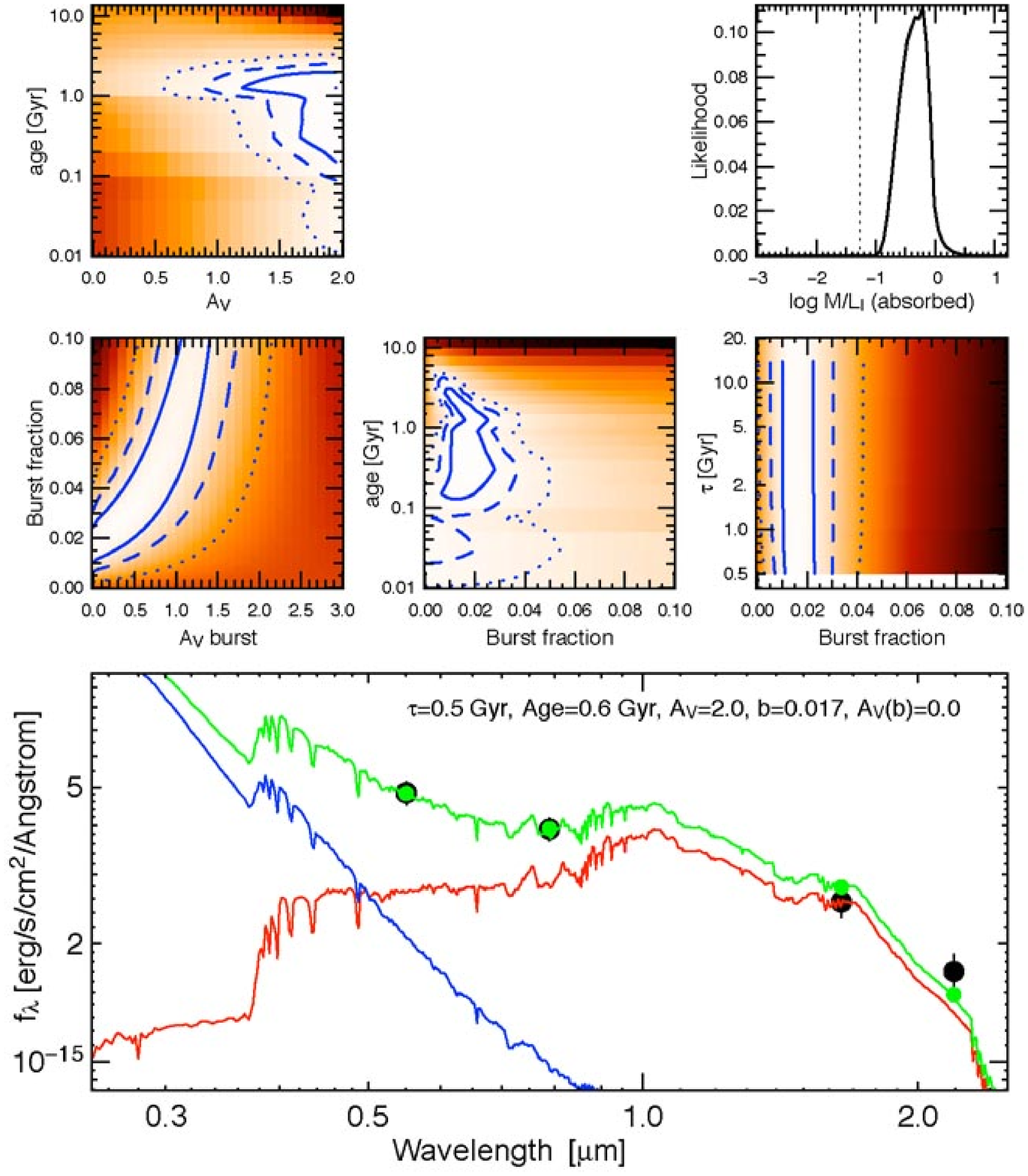}

 \figcaption[]
{Example of a stellar population model fitted to the central surface brightnesses of NGC 6946 in
$V$, $I$, $H$, and $K$ bands after correction for foreground Galactic extinction.  In the bottom panel, 
the green points are the model fits to the observations (black points).  They are synthetic 
surface brightnesses calculated from the sum (green spectrum) of a starburst (blue spectrum) that 
has had a constant star formation rate for the past 50 Myr and an intermediate-age stellar population 
(red spectrum).   The intermediate-age stellar population has an age of 0.4 Gyr, and its star 
formation rate decays with an exponential $e$-folding time of $\tau = 0.5$ Gyr.  It is extincted
by $A_V = 2.0$ mag.  The mass fraction in the starburst is $b = 0.017$, and the burst is not extincted:
$A_V(b) = 0$.  The upper panels show $\chi^2$ values (orange shading: darker means less likely) and 
$\chi^2$ contours that illustrate the coupling between the various parameters.  Blue solid, dashed, 
and dotted contours are 1-$\sigma$, 2-$\sigma$ and 3-$\sigma$, respectively.~The upper-right panel 
shows the likelihood distribution of total $I$-band mass-to-light ratio including internal extinction; 
i.{\ts}e., as we observe~it.  The dashed line shows the most likely unextincted mass-to-light ratio.
Some parameters are strongly coupled  (for example -- as expected -- $b$ and $A_V$).  Some parameters
are poorly constrained (e.{\ts}g., the age of the intermediate-age population).    But the extincted 
mass-to-light ratio has a median value of $M/L_I = 0.42$ and a most likely value of $M/L_I = 0.61$.  
It is robustly less than 1.  All these values are much smaller than the $M/L_I = 2.0 \pm 0.1$ 
that we observe.  The population modeling machinery is from Drory \etal (2004b, see also 2004a).
The fit shown has $\chi^2 = 0.91$ per degree of freedom.
\lineskip=-4pt \lineskiplimit=-4pt
}
\end{figure*}

\vfill\clearpage

\noindent asymptotic giant branch (TP-AGB) stars.  They dominate intermediate-age 
\hbox{(0.5\ts--\ts2} Gyr old) populations from the $I$ band through the near-infrared (Maraston 2005; 
Maraston et al.\ 2006).  They can lead to changes of up to a factor of 2 -- 3 in the stellar mass estimates 
for starbursts such as the one in NGC 6946.  In Figure 13, the burst fraction $b$, the internal absorptions $A_V(b)$ 
and $A_V$ of the burst and of the older population, the latter's age, and its star formation $e$-folding 
time $\tau$ are free parameters.  We want to know the total mass-to-light ratio of the extincted
model.  The upper-right panel of Figure 13 shows the likelihood distribution of the extincted $M/L_I$
marginalized over all other parameters.  The most likely {\it unextincted} $M/L_I = 0.05$ (dashed line).  
But the extincted ratio has a median value of $M/L_I = 0.42$ and a most likely value of $M/L_I = 0.61$.  
It is robustly less than the observed value $M/L_I = 2.0 \pm 0.1$. 

      Stellar populations have much more freedom than Figure 13 explores.  With only four points in a spectral 
energy distribution (SED), our model fits are underconstrained.  Some parameters are
especially unconstrained.  E.{\ts}g., we can trade burst fraction against the absorption $A_V$ 
of the older population (modifying the burst extinction) and produce good fits that are dominated 
either by the starburst or by the older population.  But putting more light into the starburst forces us 
to increase its absorption.  And trying to force higher mass-to-light ratios by adding 
priors that favor older stars forces the fit to put more light into~less obscured, young
stars in order to fit the $V$- and $I$-band points.  Trying to increase $M/L_I$ by allowing higher extinctions
has the same effect.  The fitting procedure wants most of the light to be only moderately extincted. 
The result is that the extincted $M/L_I$ is constrained to be $\sim 0.2$ -- 1.  Favored mass-to-light ratios are 
$M/L_I = 0.42$ (median) to 0.61 (most probable) and
$M/L_K = 0.044$ (median) to 0.039 (most probable).

      Urged by the referee, we also tried three-component models.  The added, old population has
an age of 8 Gyr.  The results are similar.  Favoring young stars produced a $\chi^2 = 2.1$, acceptable fit 
with an extincted, total $M/L_I = 0.30$.  Forcing the intermediate-age population to contribute most of the light 
forced the extinction to be very low; $M/L_I = 0.63$ at $\chi^2 = 5.5$.  Forcing the old population to contribute 
significantly at $H$ and $K$ forced the young population to fit $V$~and~$I$.
That model has $M/L_I = 0.80$ but $\chi^2 = 5.9$.  As long as the SED observations control the population
mix, $M/L_I \simeq 0.2$ to 1 rather than $2 \pm 0.1$ as observed.

      This implies a weak detection of more dynamical mass than we can account for with stellar populations 
that fit our SED. From $M_{\rm nuc} = (76 \pm 16) \times 10^6$~$M_\odot$ and a stellar mass of $M_* =
16$ (8 -- 31) $\times 10^6$ $M_\odot$ from the Fig.~13 models or (12 -- 31)\ts$\times 10^6$ $M_\odot$ from the
three-component models, we can estimate that the nonluminous material has a mass of (20 -- 50) $\times 10^6$ $M_\odot$.

\subsubsection{Molecular Gas Mass in the NGC 6946 Nucleus}

      It turns out that the above, nonluminous mass is reasonably consistent with the molecular gas mass
in the nucleus.  There is a large literature on the gas content and starburst in the center of NGC 6946; we 
concentrate on results that help us to interpret our mass measurement.  In the optical, the center of the 
galaxy shows an H{\ts}II region spectrum but not a LINER or a Seyfert nucleus (Ho \etal 1995, 1997). 
 The nucleus plus pseudobulge contain both an $N$-band mid-infrared source 
(Telesco \etal 1993)
and an X-ray source
(Ptak \etal 1999;
Schlegel 1994;
Schlegel \etal 2000, 2003).
However, it satisfies X-ray--infrared correlations for starburst galaxies that are clearly separated 
from correlations for Seyferts (Krabbe \etal 2001).  This is one sign among many that a starburst is
in progress.

      An early study by Engelbracht \etal (1996) found $\sigma = 45 \pm 10$ km s$^{-1}$ and
$\sigma = 53$ km s$^{-1}$ from two independent analyses of the spectra of CO absorption bands at 2.3 
$\mu$m wavelength taken in a 2\farcs4 $\times$ 8$^{\prime\prime}$ aperture.  This is consistent
with our $\sigma$ measurement.  They fitted their flux and mass constraints with 
starburst models and favored a model with two instantaneous bursts, one that made $4 \times 10^6$
$M_\odot$ of stars 7 million years ago and a second burst that made $1.8 \times 10^7$
$M_\odot$ of stars 27 million years ago.  They concluded that ``the high rate of
star formation in the nucleus of NGC 6946 must be episodic in nature rather than continuous 
throughout the lifetime of the galaxy.''  

      The highest-resolution line observations are IRAM Plateau de Bure Interferometer CO measurements by 
Schinnerer \etal (2006, 2007) with resolutions $\sim 0\farcs58 \times 0\farcs48$ and $\sim 0\farcs35$, 
respectively.    Their highest-resolution (``inspector'') rotation curve in
Fig.~10 formally gives $M(r=0\farcs5) \sim 26 \times 10^6$ $M_\odot$.  This is
not very different from our $M(r_c)$.  The true mass is likely to be larger because beam-smearing 
affects $V$ and because the velocity dispersion of the gas is neglected.  The molecular gas reaches a 
maximum central velocity dispersion of 50 and 42 km s$^{-1}$ in
the $^{12}$CO(1--0) and $^{12}$CO(2--1) lines, respectively (Schinnerer \etal 2006).  These values 
are reassuringly consistent with Engelbracht \etal (1996) and with our results.

      Most importantly, Schinnerer \etal (2006) estimate that the mass of molecular gas interior to 
$r \simeq 1\farcs0 = 29$ pc is \hbox{$M_{H_2} \sim 17 \times 10^6$ $M_\odot$.}   Within measurement errors,
this is similar to our dynamical estimate of the central dark matter.  There is room for a BH whose mass 
is a few tens of millions of $M_\odot$ but no secure dynamical evidence that such a BH must be present. 
\vskip -15pt
\null

\subsubsection{Caveat $\Rightarrow$ No $M_\bullet$ Limit}

\vskip -2pt
      Many of the above papers conclude that the central starburst is essentially completely obscured.
      Our results do not require this conclusion.  The $H$- and $K$-band HST  images show the same 
nucleus as the $V$ and $I$ images even though the extinction is much smaller in the infrared.  The
stellar population models require that the optical light comes mostly from young stars.  Of course, some
stellar mass could be hidden from all photometry by putting it behind a completely opaque screen.
But that screen would be transparent to CO line measurements.  It is reassuring that the CO velocity 
dispersion of the central molecular cloud agrees with the $I$-band stellar velocity dispersion of the nucleus.
Since their linear sizes are similar, the implied dynamical masses are similar.  Nevertheless, the potential 
that some stellar mass is completely hidden makes it impossible for us to derive an $M_\bullet$ value or limit.
\vskip -15pt
\null

\subsubsection{Episodic Growth of the Nucleus and Pseudobulge}

\vskip -2pt
      Schinnerer \etal (2006, 2007) note that NGC 6946 contains prototypical examples of a nucleus and
pseudobulge that are caught in the act of growing by the internal secular evolution of isolated 
galaxy disks (Kormendy \& Kennicutt 2004).  With a stellar mass of $M_{\rm nuc} \gtrsim 2 \times 10^7$ $M_\odot$
and a molecular gas mass of $M_{H_2} \gtrsim 1.7 \times 10^7$ $M_\odot$ (Schinnerer \etal 2006), the stellar
mass will at least double when the present gas has turned into stars.

      Other nuclei are seen in earlier and later stages.  The blue nucleus of M{\ts}33 (\S\ts3.1) still has 
an A-type optical spectrum indicative of several past starbursts 
(van den Bergh 1976, 1991;
O'Connell 1983;
Schmidt \etal 1990;
Kormendy \& McClure 1993; 
Lauer \etal 1998; 
Gordon \etal 1999;
Long \etal 2002),
but it has no substantial molecular gas (Rosolowsky~et~al.~2007).   In NGC 4593, gas is accumulating near
the center but~not~yet starbursting. Kormendy~et~al.~(2006) suggest that the
``starburst events that contribute to pseudobulge growth can be episodic.''

\begin{figure*}[t]

\vskip 4.4truein

\includegraphics{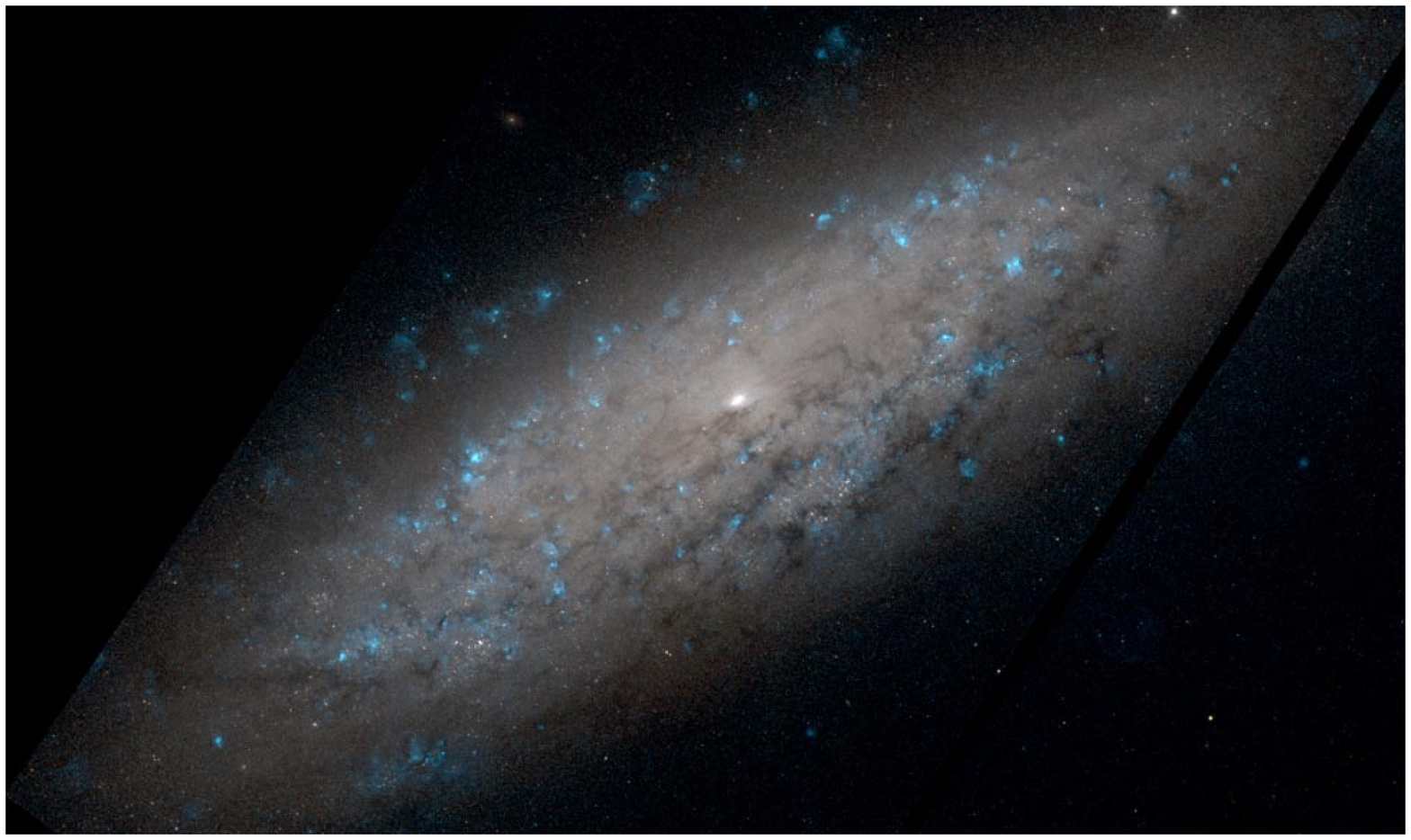}

\figcaption[]
{Color image of NGC 6503 taken with the {\it Hubble Space Telescope\/} Advanced Camera for Surveys.
Colors are bland because the wavelength range available is small.  Blue corresponds to the F650N filter 
(H$\alpha$), red to F814W ($I$ band) and green to their average.  Brightness here is proportional to the 
square root of the brightness in the galaxy.  North is up and east is at left.
Like NGC 5457 and NGC 6946, this is a pure-disk galaxy.  But NGC 6503 is smaller;
it has a flat outer rotation curve with $V_{\rm circ} \simeq 115$ km s$^{-1}$ 
compared with $V_{\rm circ} \simeq 210$ km s$^{-1}$ for the previous galaxies.  Like those galaxies, its
Hubble type is Scd.  And like them, a tiny, bright center visible in this image proves to be a pseudobulge that 
makes up 0.11\ts\% of the $I$-band light of the galaxy (see text).  The nucleus that we use to constrain 
$M_\bullet$ makes up only 0.040\ts\% of the $I$-band light of the galaxy.  It is completely invisible here 
but is illustrated in Figure 15.
\lineskip=-4pt \lineskiplimit=-4pt
}
\end{figure*}

\vfill\eject

\subsection{NGC 6503}

      NGC 6503 (Figures 14 and 15) is an Scd galaxy that is smaller than NGC 5457 and NGC 6946.  
It has a rising rotation curve over the inner $100^{\prime\prime}$, i.{\ts}e., roughly the radius range shown
in Figure 14, and then a well known, flat outer rotation curve with $V_{\rm circ} \simeq 115$ km s$^{-1}$
(van Moorsel \& Wells 1985; Begeman 1987; Begeman \etal 1991) out to $r \simeq 800^{\prime\prime}$.  This is
similar to $V_{\rm circ}$ in M{\ts}33.  NGC 6503 is another example of a pure-disk galaxy; it is not in 
the \S\ts4 sample because $V_{\rm circ} < 150$ km s$^{-1}$.

      Two HST archive images include the nucleus, an F814W image that defines our $I$ photometry bandpass 
and an F650N image that includes H$\alpha$ emission.  Color images of the galaxy and its nucleus
plus pseudobulge are constructed from these images in Figures 14 and 15.  The wavelength range is small, so 
colors look bland.  But absorption and star-formation regions are recognizable, and the figures serve to
emphasize how thoroughly this is a pure-disk galaxy.

      The tiny, bright center that is saturated in Figure 14 is resolved in Figure 15 into an elongated
structure that resembles a nuclear bar (see also Gonz\'alez-Delgado \etal 2008).  The disk-like or bar-like
morphology is sufficient to identify this as a pseudobulge.  It surrounds a distinct, 
high-surface-brightness nucleus.  NGC 6503's distance is only 5.27 Mpc (Karachentsev et al.~2003c;
Karachentsev \& Sharina 1997).  So the nucleus provides another opportunity to use 
ground-based spectroscopy to derive an $M_\bullet$ limit in a pure-disk galaxy.

\cl{\null}

\vskip 3.38truein
\vskip 0.12truein

\includegraphics{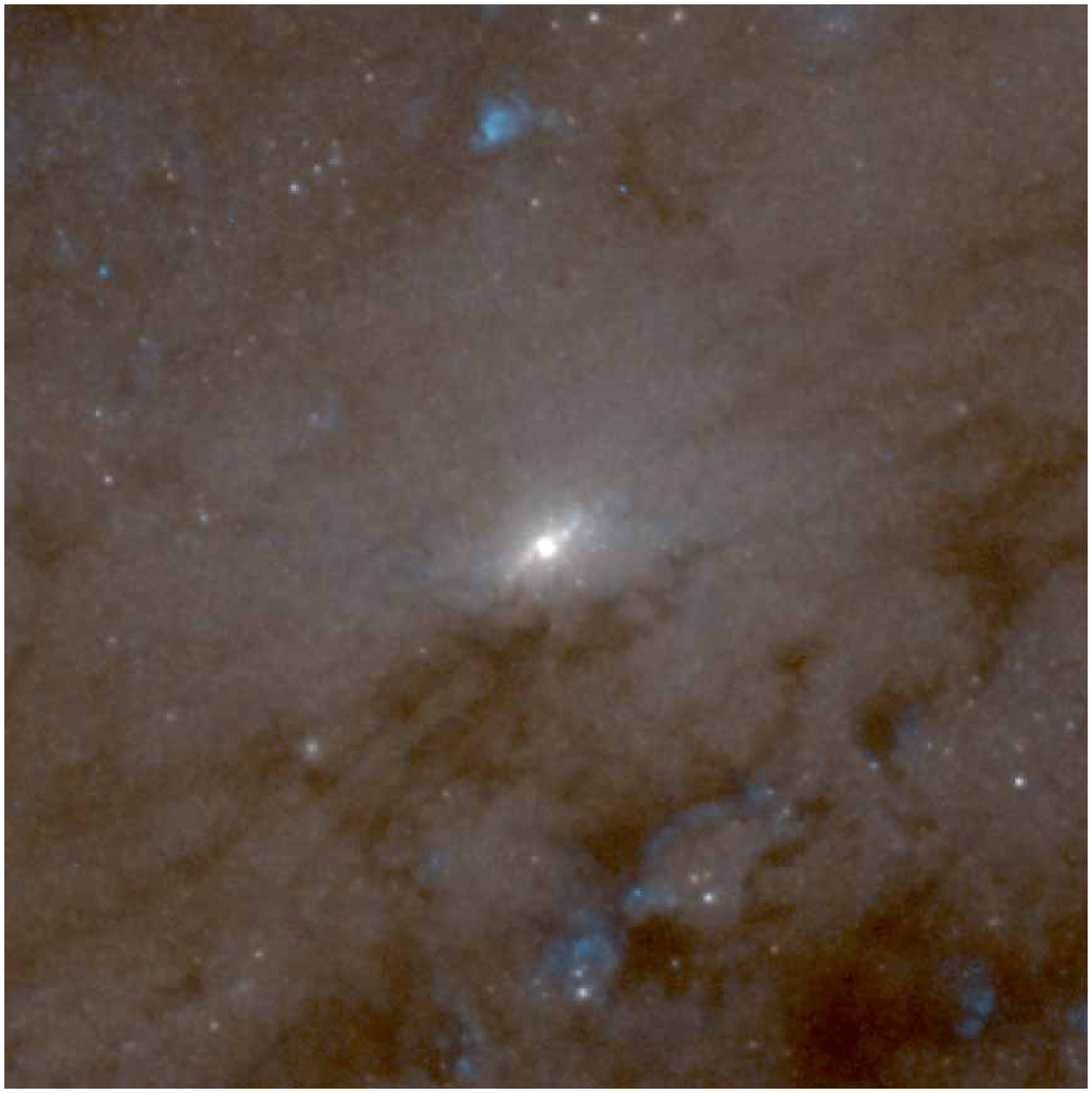}

\figcaption[]
{Color image of the central $20\farcs5 \times 20\farcs5$ of NGC 6503 made as in Figure 14
but with a different square-root stretch to show the central bar-like pseudobulge and
nuclear star cluster.  Both together are saturated in Figure 14.  
\lineskip=-4pt \lineskiplimit=-4pt
}

\vskip 17pt

\vfill\clearpage

\cl{\null}

\vskip 4.7 truein

\includegraphics{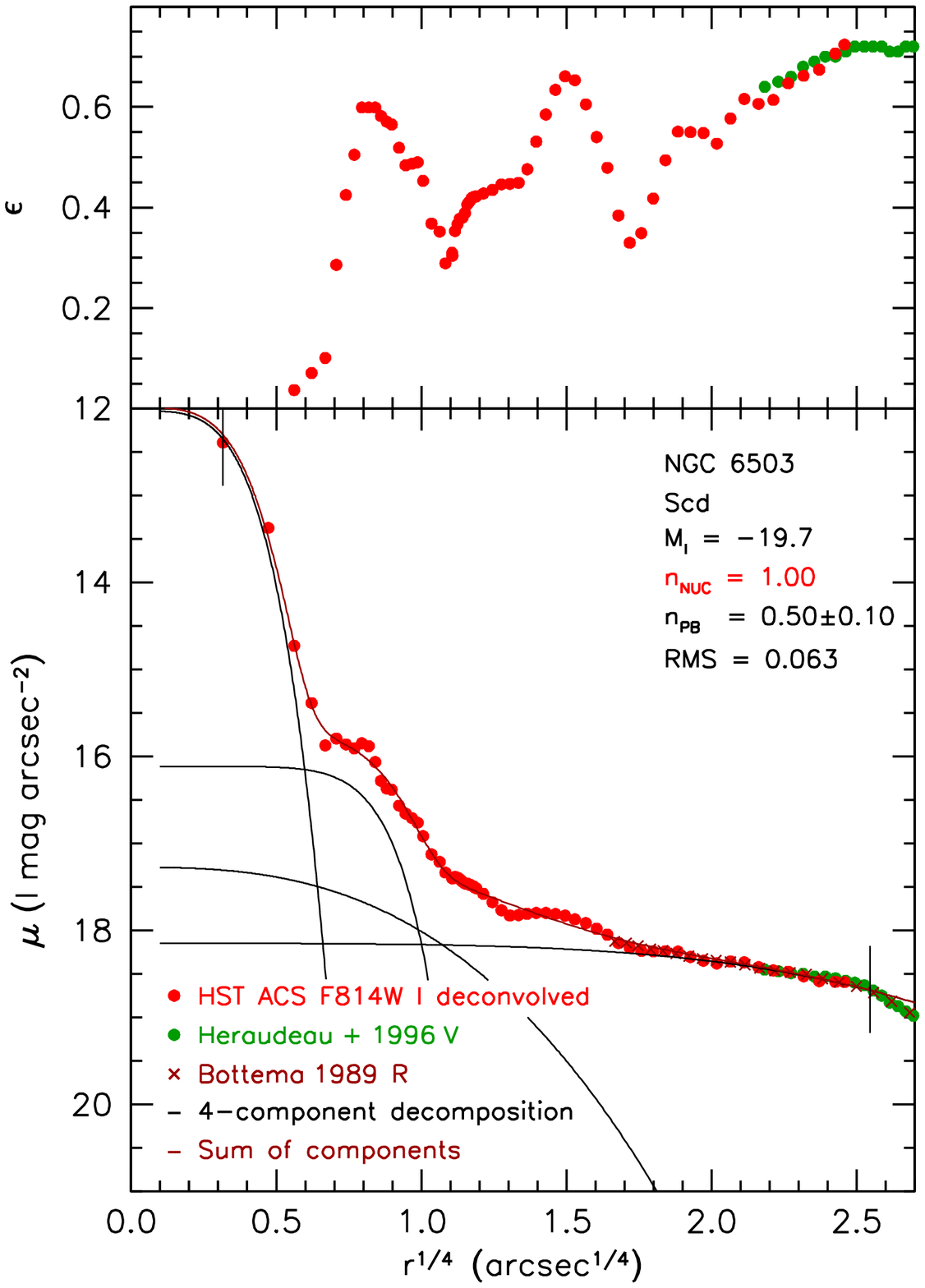}

 \figcaption[]
{VEGAmag $I$-band surface brightness and ellipticity profiles of NGC 6503.  Black 
lines show a decomposition into nuclear, pseudobulge, and disk components; their sum is 
shown in dark red.  The nuclear profile is exponential.  Two S\'ersic functions,
the inner of which is essentially Gaussian, are required to fit the pseudobulge;
they should not be interpreted as physically separate or distinct components.  The outer
exponential is fitted to the inner, flat part of the Freeman (1970) ``Type II'' profile
of the disk.  The fit range is shown by vertical dashes; the RMS
of the fit is 0.063 $I$ mag arcsec$^{-2}$.  The inner pseudobulge brightness shelf
has essentially the same flattening as the outer disk; it is the disk-like or bar-like
feature seen around the nucleus in Figure 15.  The ellipticity wiggles at $r^{1/4} \simeq
1.1$ to 1.9 are caused by patchy dust (see Figures 14 and 15).
\lineskip=-4pt \lineskiplimit=-4pt
}
      
\vskip 10pt

      The major-axis brightness profile of NGC 6503 is shown in Figure 16.
All of our results except our measurement of the total magnitude of the galaxy are 
based on the $I$-band profile derived from the HST ACS F814W image.  However, this
profile was extended to $r = 201^{\prime\prime}$ by averaging an $R$-band profile
from Bottema (1989) and a $V$-band profile from H\'eraudeau \etal (1996) both shifted
to the present zeropoint.  Over this wavelength range, color gradients in the galaxy
are small.  

      Figure 16 shows the central profile after 40 iterations of Lucy-Richardson
deconvolution.  However, all analysis was also carried out in parallel on the
undeconvolved profile.  We use both sets of results below.

      To estimate the total magnitude of the galaxy and to check our
zeropoint, we extended the observed profile by fitting an exponential to the outer,
steep profile that is just beginning to be visible at the largest radii shown in
Figure 16.  This is the outer exponential in the Freeman (1970) Type II profile of
the galaxy.  Integrating this extended profile and the ellipticity profile gives 
a total apparent magnitude~of~\hbox{$I_T = 8.96$}.  This compares well with $I_T = 8.94$ 
in H\'eraudeau \etal (1996) and with \hbox{$I_T = B_T - (B-V)_e - (V-I)_e = 8.93$ and 9.02} 
using magnitudes and colors from the main and integrated photometry tables in HyperLeda.  
Magnitudes and colors from the RC3~(NED) give $I_T = 9.06$.  Makarova (1999) gets $I_T = 9.20$.  
We adopt our total magnitude but again note that it is unrealistic to think that this is more 
accurate than $\pm 0.1$ mag since extrapolations of $\mu(r)$, $\epsilon(r)$, and color are required.  
The agreement in zeropoint and total magnitude of our results with published photometry is good.

      Figure 16 shows that the nucleus is tiny and dense compared to the $r \simeq 1^{\prime\prime}$ 
main part of the pseudobulge.  Its deprojected outer profile is much steeper than $I \propto r^{-2}$,
so we treat it as an independent self-gravitating cluster.  To estimate its mass and a limit on
$M_\bullet$, we need its effective radius, harmonic mean radius, and total magnitude with the small 
contribution from the rest of the galaxy removed. 

      A profile decomposition of NGC 6503 is shown in Figure~16.  The inner, shallow part of the
Freeman Type II disk profile is accurately exponential.  The pseudobulge is too complicated to be 
fitted by a single function, S\'ersic or otherwise.  We~fit~it with the sum of two S\'ersic functions,
an inner one for the nuclear bar and an outer one for the rest.  However, we interpret both as 
being parts of the same pseudobulge in the same way that, in any barred galaxy, the bar and the 
rest of the disk are parts of the same disk.  The secular evolution that makes pseudobulges is 
complicated and often involves starburst rings (Kormendy \& Kennicutt 2004); it is convenient that 
the results are often nearly-S\'ersic profiles with $n \lesssim 2$, but this is not guaranteed.  
Here, we need to fit the inner shelf in the pseudobulge profile well enough for a robust extrapolation
into the nucleus.  The decomposition in Fig.~16 serves this purpose. 
 
      Then the nucleus is exponential and $r_e = 0\farcs057 \pm 0\farcs011$.  Its profile falls off 
steeply, so we expect that the deconvolution ``rings'' and makes the profile slightly {\it too} steep.  
So we carried out the same analysis on the undeconvolved profile.  This is PSF-blurred, so
$r_e = 0\farcs088 \pm 0\farcs006$ overestimates the effective radius.  We therefore average these two 
values and adopt \hbox{$r_e = 0\farcs072 \pm 0\farcs016 = 1.8 \pm 0.4$ pc.}  If 
$\sigma = 40 \pm 2$ km s$^{-1}$, then $M_{1/2} = (2.7 \pm 0.6) \times 10^6$ $M_\odot$.  The total mass
of the nucleus, $M_{\rm nuc} = (5.5 \pm 1.3) \times 10^6$ $M_\odot$, is included in Table 1.

      The total magnitude of the nucleus given by the raw and deconvolved profiles is 
$I_{T,\rm nuc} = 17.54$ and $17.39$, respectively.  We adopt $I_{T,\rm nuc} = 17.47 \pm 0.07$; 
$M_{I,\rm nuc} = -11.2 \pm 0.07$, and $L_{1/2} = (0.65 \pm 0.04) \times 10^6$ $L_{I\odot}$.  
So $M_{1/2}/L_{1/2} = 4.2 \pm 1.0$.

      This value is too large to be easy to understand.  Of course, star formation histories and
internal absorptions that make $M/L_I = 4.2$ can be devised.  But the color of the nucleus is normal 
for an Scd galaxy: the central five $B - V$ measurements in HyperLeda range from 0.69 to 0.88 
and average $0.79 \pm 0.03$.  The aperture diameters are 1\farcs4 to 6\farcs9; that is, 
these are measurements of the nucleus and pseudobulge.  Correcting for Galactic reddening, 
$(B - V)_0 = 0.75 \pm 0.03$.  Bell \& de Jong (2001, Table 1) list the relationship between color 
and stellar population $M/L$ for a formation model that, while not unique, is suitable for NGC 6503.  
For the above color, it predicts that $M/L_I = 1.50 \pm 0.12$.

      Absent exotic star formation histories, two possibilities look plausible.  The stellar population 
may be as above and we may have weakly detected a BH of mass $M_\bullet \sim 1 \times 10^6$~$M_\odot$.
But a more conservative interpretation is more likely.  Integrating the light profiles of the components
shown in Figure 16 shows that $<$\ts10\ts\% of the light in our spectroscopic aperture comes from
the nucleus.  Our measurement of $\sigma = 40 \pm 2$ km s$^{-1}$ is a measurement of the pseudobulge.
The same is true of Ho \etal (2009) quoting Barth \etal (2002).  Many galactic nuclei have velocity 
dispersions of 20 -- 25 km s$^{-1}$; M{\ts}33 and NGC 5457 are two of them.  NGC 6503 may be another.
That is, the velocity dispersion may decrease from the pseudobulge into the nucleus and its stellar 
population $M/L_I$ may be entirely normal.

      The irony is that Bottema (1989) found $\sigma = 25 \pm 3$ km s$^{-1}$, even though we cannot
understand how he did it, because he got similar dispersions even at larger radii where~we~get~40~km~s$^{-1}$.  
Bottema's $\sigma$ gives $M_{1/2}/L_{1/2} = 1.66 \pm 0.4$.  Moreover, our $\sigma = 40 \pm 2$ km s$^{-1}$ 
gives a pseudobulge mass-to-light ratio of $(M/L)_I = 2.7$.  The first value is as expected and the 
second is more plausible than $M/L_I = 4$.  We clearly need a high-spatial-resolution measurement of $\sigma$ in
the {\it nucleus\/} of NGC 6503.

      The same uncertainty applies to constraints on $M_\bullet$.  For the undeconvolved and
deconvolved nuclear profiles, we measure $<$\null$1/r$\null$>^{-1} = 0\farcs090$ and 0\farcs053,
respectively.  We adopt the mean, $<$\null$1/r$\null$>^{-1} = 0\farcs071 \pm 0\farcs019 =
1.8 \pm 0.5$ pc.  It fortuitously equals $r_e$.  Then $M_\bullet \lesssim M_{\rm min} =
(2.0 \pm 0.6) (\sigma / 40~{\rm km~s}^{-1})^2 \times 10^6$ $M_\odot$.

      This limit is not restrictive in the context of an extrapolation of the 
\hbox{$M_\bullet$\ts--\ts$\sigma$} correlation (Ferrarese \& Merritt 2000; Gebhardt \etal 2000; 
Tremaine \etal 2002).  For $\sigma = 40$ km s$^{-1}$ and 25 km s$^{-1}$, it predicts
 $M_\bullet = 0.4 \times 10^6$ $M_\odot$ and $0.07 \times 10^6$ $M_\odot$, respectively.  
All allowed $M_\bullet$ are adequate to explain any low-level AGN activity in NGC 6503.  
It was classified as a Seyfert-LINER transition object (``T2/S2'') by Ho \etal (1997), and it 
contains a weak nuclear X-ray source (Panessa \etal 2006, 2007; Desroches \& Ho 2009). 
The latter papers suggest that the nucleus may be powered by young stars rather than an AGN.
\vskip -15pt
\cl{\null}

\section{How Can Hierarchical Clustering Make So Many Bulgeless, Pure-Disk Galaxies?}

      Hierarchical clustering in a cold dark matter universe (White \& Rees 1978) is
a remarkably successful theory of galaxy formation.  The remaining struggle is with baryonic
physics.  The most serious problem has been emphasized many times by observers (e.{\thinspace}g., 
Freeman 2000; 
Kormendy \& Kennicutt 2004; 
Kormendy \& Fisher 2005, 2008;
Kautsch \etal 2006; 
Carollo et al.~2007; 
Kormendy 2008;
Barazza \etal 2008;
Weinzirl \etal 2009;
Kautsch 2009),
by modelers
(Steinmetz \& Navarro 2002; 
Abadi et al.~2003;
Governato \etal 2004, 2010;  
Robertson \etal 2004;
Mayer \etal 2008,
Stewart \etal 2008, 2009;
Hopkins \etal 2009a, and
Croft \etal 2009
is a very incomplete list),
and by reviewers 
(e.{\ts}g., Burkert \& D'Onghia 2004;
Lake 2004;
Brooks 2010;
Peebles \& Nusser 2010).  
Given so much merger violence, how can hierarchical clustering make so many pure-disk galaxies with no
signs of merger-built bulges?  That is:

      How can dark matter halos grow (e.{\ts}g.)~to $V_{\rm circ} \sim 210$~km~s$^{-1}$ 
without letting the mergers that accomplished that growth destroy the fragile thin disks of
stars that predate the mergers (T\'oth \& Ostriker 1992) and without 
scrambling disks into recognizable classical bulges (Toomre 1977; Schweizer 1989)?
Minor mergers are not a problem; they do no damage.  But major mergers -- with range of mass ratios to be
determined -- scramble disks into classical bulges.~Can we explain pure disks?

      The problem gets much harder when we realize that many (we thought) small bulges are 
unlikely to be merger~remnants; rather, they are pseudobulges made mainly by secular evolution of 
isolated galaxy disks (e.{\ts}g., 
Kormendy \& Kennicutt~2004; 
Kormendy \& Fisher 2008;
Weinzirl \etal 2009).  
{\it From a galaxy formation point of view, galaxies that contain only pseudobulges are
pure-disk systems.}
The luminosity function of ellipticals is bounded at low luminosities
(Sandage \etal 1985a, b; Binggeli \etal 1988); the faintest ones resemble M{\ts}32
but are very rare (Kormendy \etal 2009).  Recognizing pseudobulges shows us that 
small classical bulges are rarer than we thought, too.  How much rarer is the subject of this section.

      The problem of bulgeless disks is least difficult for small galaxies.  They accrete gas in 
cold streams or as gas-rich dwarfs more than they suffer violent mergers 
(Maller \etal 2006;
Dekel \& Birnboim 2006;
Stewart \etal 2009;
Koda \etal 2009;
Brooks \etal 2009;
Hopkins \etal 2009b, 2010).
Energy feedback from supernovae is effective in counteracting gravity
(Dekel \& Silk 1986;
Robertson \etal 2004;
D'Onghia \etal 2006; 
Dutton 2009;
Governato \etal 2010). Attempts to explain pure disk galaxies have come closest
to success in explaining dwarf systems (Robertson \etal 2004; Governato \etal 2010).  So:

       {\it The pure-disk galaxies that most constrain our formation picture are the ones that 
live in the highest-mass dark halos.}  We know of no Sc or later-type galaxy that has a classical~bulge 
(Kormendy \& Kennicutt 2004).  In this section, we inventory classical and pseudo bulges in the nearby 
universe and conclude that the solution to the problem of giant bulgeless galaxies is not to hope that 
they are rare enough so they can be explained as the tail of a distribution of formation histories that 
included a few fortuitously mergerless galaxies.

      Consider first the Local Group.  Only our Galaxy has had an uncertain bulge classification.  
Its boxy shape 
(Maihara \etal 1978; 
Weiland \etal 1994; 
Dwek \etal 1995) 
implies that the high-latitude structure is a pseudobulge -- the part of the disk 
that heated itself vertically when it formed the Galactic bar
(Combes \& Sanders 1981; 
Combes \etal 1990; 
Raha \etal 1991;
Athanassoula \& Misiriotis 2002;
Athanassoula 2005).
Particularly compelling is the observation of a perspective effect -- the near side of the 
thick bar looks taller than the far side, so the pseudobulge is not just boxy, it is a 
parallelogram (Blitz \& Spergel 1991).  Further evidence is provided by the observation 
that the rotation velocity is almost independent of height above the disk plane, as in 
other boxy bulges and as in $n$-body models of edge-on bars (Howard \etal 2009; Shen \etal 2010).
Further, the low velocity dispersion of the bulge merges seamlessly with that of the disk 
(Lewis \& Freeman 1989).  Finally, the complicated central $\sigma$ profile derived by 
Tremaine et al.~(2002) also implies a pseudobulge.  
Only the old, $\alpha$-element-enhanced stellar population is suggestive of a classical 
bulge.  But these stars could have formed before the bar structure (Freeman 2008).
Kormendy \& Kennicutt (2004) discuss caveats.  Like Freeman (2008), 
we conclude that there is no photometric or dynamical evidence for a classical bulge. 
    
      Our Galaxy provides an additional important conclusion.  Its disk stars are as old
as 9 -- 10 Gyr (Oswalt \etal 1995; Winget \& Kepler 2008).  Unless our Galaxy is
unusual, this suggests: {\it The~solution to the problem of forming giant, pure-disk galaxies 
is not to use some physical process like energy feedback to delay star formation until 
recently and thereby to give the halo time to grow without forming a classical bulge.}

 Then the Local Group contains one tiny elliptical, M{\thinspace}32, and one big classical bulge, 
in M{\thinspace}31.  In the most massive three galaxies, there is only one classical bulge.

      Looking beyond the Local Group, the nearest giant Sc-Scd galaxies include the well known
objects M{\ts}101, NGC 6946, and IC 342.  All have outer rotation velocities 
$V_{\rm circ} \simeq 200$ km s$^{-1}$.  All have extraordinarily tiny pseudobulges and 
no sign of classical bulges (\S\ts3).  To further test whether such galaxies 
could be rare enough to have formed as the quiescent tail of a distribution of merger histories, 
we inventory similar giant galaxies in the nearby universe.  This section expands on 
Kormendy \& Fisher (2008) to provide better statistics.

      The problem of pure disk galaxies proves to depend  on environment{\ts}--{\ts}it is a
puzzle in the field but not in rich~clusters.  Also, we need detailed observations to classify
(pseudo)bulges.  These considerations motivate us to restrict ourselves to a nearby volume that
contains small groups of galaxies like the Local Group but not any denser environments that 
approach the conditions in the Virgo cluster.  M{\ts}101 is the most distant bulgeless disk 
discussed in \S\ts3, at $D = 7$ Mpc.  We look for all
giant galaxies with $D \leq 8$ Mpc.  As our cutoff for giant galaxies, we will be
conservative and choose $V_{\rm circ} > 150$ km s$^{-1}$ or central
$\sigma \sim V_{\rm circ}/\sqrt{2} > 106$ km s$^{-1}$.
We use Tully (1988), HyperLeda, and NED to construct a master list of nearby galaxies and then use
individual papers that provide accurate measures of $D$, $V_{\rm circ}$, and $\sigma$
to cull a sample that satisfies the above criteria. 

\cl{\null} 

\vskip 7.2truein

\includegraphics{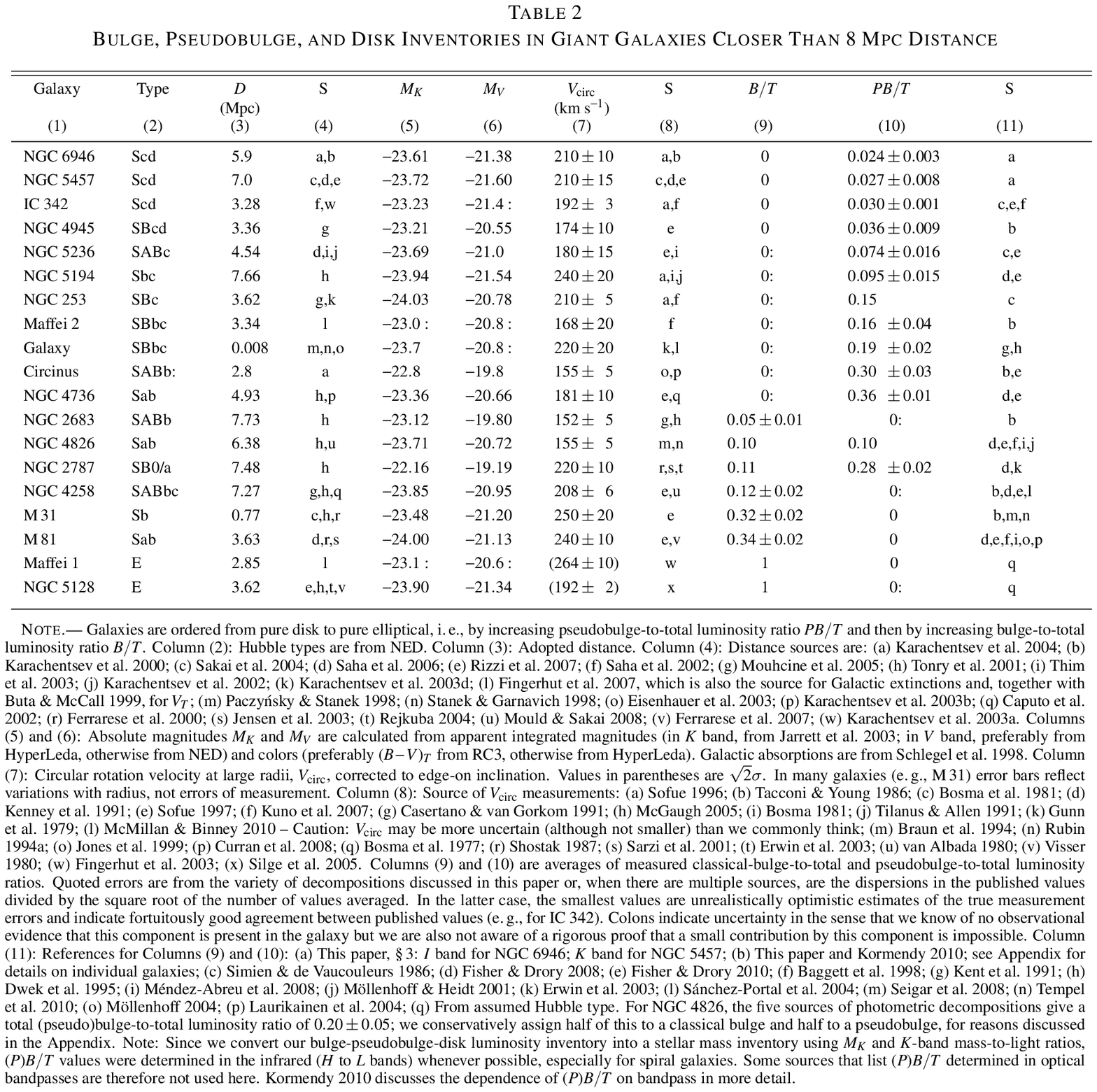}

      Table 2 lists the resulting 19 galaxies in order from pure disk to pure elliptical.
Distances are a complicated problem; we use averages (Column 3) of the most 
accurate determinations that we could find in the sources in Column (4).  Column (5) gives the 
$K$-band absolute magnitude of the galaxy from the total magnitude in the 2MASS Large Galaxy Atlas 
(Jarrett \etal 2003).  Column (6) is the $V$-band total absolute magnitude.  Column (7) gives
the outer rotation velocity $V_{\rm circ}$ from sources in Column (8).  For the two ellipticals, 
we use $V_{\rm circ} = \sqrt{2} \sigma$, where $\sigma$ is an approximate velocity dispersion.
Finally, classical-bulge-to-total and pseudobulge-to-total ratios $B/T$ and $PB/T$, respectively, 
are listed in Columns (9) and (10).  We averaged the values given by the sources listed in Column (11).  
Bulge classifications are discussed in the Appendix.

\vfill\eject

      M{\thinspace}101, NGC 6946, and IC 342 are well-known examples of giant, pure-disk
galaxies, but they are not unique even in our sample.  Four of the 19 galaxies have $PB/T \lesssim$\ts4\ts\%.  
No classical bulge can be hidden in these
galaxies -- not even one as small as M{\ts}32.  In M{\ts}101 and in NGC 6946, we find 
nuclear star clusters that make up 0.03\ts\% and 0.1\ts\% of the light of the galaxy; these are
as faint as or fainter than the smallest ellipticals known.  But they are nuclei -- they are too 
small and dense to be bulges.  {\it We emphasize: we do not have the freedom to postulate classical bulges 
which have arbitrary properties (such as low surface brightnesses) that make them easy to hide.
Classical bulges and ellipticals satisfy well defined fundamental plane correlations} 
(Kormendy \etal 2009 and Kormendy 2009 show these to the faintest luminosities).   {\it Objects that 
satisfy these correlations cannot be hidden in the above galaxies.}  So $B/T = 0$ in 4/19 
of the giant galaxies in our sample.

      Of the rest, 7 more are dominated by pseudobulges and show no signs of classical bulges.
The pseudobulge classifications are robust.  There is no sign of a multi-component bulge structure.  
Many of these objects have long been discussed as prototypical examples of pseudobulge formation by 
secular evolution (e.{\ts}g., NGC 4736, see the Appendix).  So 11 of the 19 
giant galaxies in our sample either cannot have a classical bulge or have dominant pseudobulges and 
show no sign of a classical bulge.  

      Four galaxies are listed in Table 2 as having tiny classical bulges ($B/T \lesssim 0.1$).
Except in NGC 4258, the identification of these as classical bulges is uncertain.  E.{\ts}g., Erwin \etal 
(2003) decompose the complicated inner light profile of NGC 2787 into two components that they
interpret as coexisting classical and pseudo bulges with $B/T \ll PB/T$.  It is not clear that the 
smaller of these components is a classical bulge.  But we do expect that classical and pseudo bulges 
coexist in some galaxies (Kormendy 1993; Erwin \etal 2003; Kormendy \& Kennicutt 2004).  In Table 2 and
in the Appendix, we err on the side of caution in identifying small classical bulges.  However, note that in 
NGC 2683 -- NGC 4258, $B/T = 0.05$ to 0.12.  This is still small compared to the classical
bulges that are made in simulations of hierarchical clustering (Abadi \etal 2003).

      Finally, substantial merger remnants are not absent from our sample.  Maffei 1 and 
NGC 5128 = Centaurus A are ellipticals.  They are sometimes classified as peculiar S0s, but we
assign $B/T \equiv 1$ to both.  NGC 5128 is the most massive classical bulge in our sample.  
Two other galaxies have classical bulges with $B/T \simeq 1/3$ and no sign of pseudobulges, 
M{\thinspace}31 and M{\thinspace}81.  

      {\it We conclude that bulgeless galaxies do not form the rare tail of the distribution of 
galaxy formation histories; they are \hbox{58\ts--\ts74\ts\%} of our sample.  Almost all of the classical
bulges that we do identify -- some with substantial uncertainty -- are smaller than those normally 
made in simulations of galaxy formation.  In field environments, the 
problem of forming giant, pure-disk galaxies in a hierarchically clustering universe is acute.}

      Finally, we estimate the stellar mass in disks, pseudobulges, and classical bulges $+$
ellipticals summed over the 19 galaxies in our sample.  The calculation is approximate,
e.{\ts}g., because we do not have  dynamical stellar mass measurements and because
many of the galaxies have large and somewhat uncertain Galactic obscurations.  We estimate
a stellar population, $K$-band mass-to-light ratio, $\log{M/L_K} = -0.692 + 0.652(B-V)_0$,
from the dereddened $B-V$ color, following Bell \& de Jong (2001).  Classical bulges are
redder than their associated disks; we use the correspondingly higher $M/L_K$
ratios.  For pseudobulges, we use the disk $M/L_K$ (``Bulges are more like their disks
than they are like each other.'' -- Wyse \etal 1997; see also Peletier \& Balcells 1996; Gadotti
\& dos Anjos 2001).
We assume that the $B/T$ and
$PB/T$ values in Table 2 apply at $K$ band and apply them to $M_K$ to get the luminosity
of each component.  The result is that the total stellar masses in the 
galaxies, in their pseudobulges, and in classical bulges $+$ ellipticals are
$\Sigma M_{\rm total}       = 6.0 \times 10^{11}$ $M_\odot$;
$\Sigma M_{\rm pseudobulge} = 4.7 \times 10^{10}$ $M_\odot$;
$\Sigma M_{\rm bulge}       = 1.34 \times 10^{11}$ $M_\odot$.
That is, $22 \pm 4$\ts\% of the mass is in bulges and ellipticals,
         $8 \pm 4$\ts\% is in pseudobulges, and so
         $78 \pm 4$\ts\% is in pseudobulges plus disks, i.{\ts}e., not in major merger remnants.
In the above, the high-bulge-mass-fraction error bar is derived by assigning half of all pseudobulge
mass to classical bulges; this is certainly too conservative, because it is inconsistent with the
properties observed for the biggest pseudobulges.  The low-bulge-mass-fraction error bar is 
similarly derived by assigning 1/2 of the classical bulge mass (not including ellipticals)
to the pseudobulges; this also is inconsistent with observations. 

\lineskip=-4pt \lineskiplimit=-4pt
 
      Our conclusions are robust to uncertainties in assumptions.  For example, 
if we use the same mass-to-light ratio for all stellar populations, then  
$17^{-3}_{+4}$\ts\% of the mass is in bulges and ellipticals,
$8^{+3}_{-4}$\ts\% is in pseudobulges, and 
$83^{+3}_{-4}$\ts\% is in pseudobulges plus disks.
The stellar mass results are even robust to any uncertainty in the distinction between classical and 
pseudo bulges, because the ratio of stellar mass in both together is $(\Sigma B + \Sigma PB)/\Sigma T \simeq
0.25$ to 0.30 for the above possible choices of mass-to-light ratios.  That is, the total mass in
bulges is small because most $(P)B/T$ values in the field are small.  Note also that 
our procedure underestimates disk and pseudobulge masses significantly, because
we do not inventory cold gas and because we do not correct for internal absorption,
which is large in some disks but small in bulges and ellipticals.  

      We conclude that, in the nearby field, most stellar mass and most baryonic mass is in disks; 
in fact, in pure disks.  The ratio of pseudobulge-to-bulge stellar mass is $\Sigma PB/\Sigma B =
0.41^{+0.31}_{-0.24}$; that is, significant but not dominant.  However, the importance of pseudobulges is
not in their total mass but rather in the fact that they are not merger remnants.  From the point of view 
of galaxy formation by hierarchical clustering, their mass should be included in the disk inventory.
So only 1/5 of the stellar mass in giant galaxies in our 8-Mpc-radius, field volume 
is in probable remnants of major mergers.  And this is distributed in no more than 8 but possibly 
as few as 5 of the 19 giant galaxies in our sample.   Pure disk galaxies are 
the dominant population among our giant galaxies in extreme field 
environments.\footnote{Our 
results generally agree with published studies, but quantitative comparison is difficult:
(1) Pseudobulges have not generally been identified; most of the ones discussed here
could not easily be classified far away.  So studies of large (e.{\ts}g., SDSS) samples find 
that pure disks are common -- and, indeed, more common in the field than in clusters (Kautsch \etal 2009) --
but they probably underestimate the fraction of bulgeless galaxies.  (2) Many studies combine their statistics 
over a variety of environments; they find that late-type galaxies dominate strongly at low 
redshifts (e.{\ts}g., Nair \& Abraham 2010).  We concentrate on the extreme field in part to emphasize the 
stark contrast with Virgo.  Jogee \etal (2009) provide an up-to-date discussion of
how observed merger rates over the past 7 Gyr agree with theory.  Engineering 
consistency between our theoretical picture and our observations of zero- and high-redshift galaxies 
is a growth industry that is still in its early stages.}

      In contrast, in the Virgo cluster, about 2/3 of the stellar mass is in
elliptical galaxies and some additional mass is in classical bulges (Kormendy \etal 2009).  So
the above statistics are a strong function of environment.

      We therefore restate the theme of this section: What is special about 
galaxy formation in low-density, Local-Group-like environments that allows the majority of
galaxies with halo $V_{\rm circ} > 150$ km s$^{-1}$ to form with no sign of a major merger?

\acknowledgments

      We thank the anonymous referee for a very helpful report that led to important improvements
in this paper.  In particular, the referee and Scott Tremaine both suggested the use of the Virial 
theorem minimum mass estimator $M_{\rm min}$ as an $M_\bullet$ limit.  We also thank Scott Tremaine for
elaborating on an argument implicit in Tremaine \& Ostriker (1982) that a nuclear star cluster 
can be treated as a dynamically isolated system ``if there is a well-defined radial range in which
the surface brightness falls off faster than $1/r$.''
We are grateful to Tod Lauer for providing the Lucy deconvolution program and to Luis Ho, Jim Peebles, 
and Jean Turner for helpful conversations and/or comments on the paper.
\null
We also thank David Fisher for providing (pseudo)bulge classifications and $(P)B/T$ luminosity ratios
from Fisher \& Drory (2008, 2010) and for providing the McDonald Observatory 0.8 m telescope 
$I$-band brightness profile of NGC 6946.

      The Hobby-Eberly Telescope (HET) is a joint project of the University of Texas at Austin, 
Pennsylvania~State~University, Stanford University, Ludwig-Maximilians-Universit\"at Munich, and 
Georg-August-Universit\"at G\"ottingen. The HET is named in honor of its principal benefactors, 
William P.~Hobby and Robert E.~Eberly.

      The NGC{\ts}3077 spectrum (Appendix) was obtained with the Marcario Low Resolution 
Spectrograph (LRS) and the HET.  LRS is named for Mike Marcario of High Lonesome Optics; he made 
optics for the instrument but died before its completion.  LRS is a project of the HET partnership
and the Instituto de Astronom\'\i a de la Universidad Nacional Aut\'onoma de M\'exico.

      This publication makes extensive use of data products from the Two Micron All Sky Survey (Skrutskie 
\etal 2006), which is a joint project of the University of Massachusetts and the Infrared Processing and 
Analysis Center/California Institute of Technology, funded by the National Aeronautics and Space Administration
and the National Science Foundation. This paper would have been completely impractical without extensive use 
of the NASA/IPAC Extragalactic Database (NED), which is operated by Caltech and JPL under contract with NASA, 
of the HyperLeda database (Paturel \etal 2003): http://leda.univ-lyon1.fr, and of NASA's Astrophysics Data 
System bibliographic services.

      The wide-field color image of NGC 6946 (Figure 7) was taken by Vincenzo Testa and Cristian DeSantis with 
the Large Binocular Telescope (LBT).  We thank LBT Director Richard Green for permission to use it here.  The LBT
Observatory is an international collaboration; its partners are the University of Arizona on behalf of the Arizona 
university system; Istituto Nazionale di Astrofisica, Italy; LBT Beteiligungsgesellschaft, Germany, representing 
the Max Planck Society, the Astrophysical Institute Potsdam, and Heidelberg University; the Ohio State University;
and the Research Corporation, on behalf of the University of Notre Dame, University of Minnesota, and University of 
Virginia.  

      Finally, we are most grateful to the National Science Foundation for supporting this work under 
grant AST-0607490.

\vskip 5pt
{\it Facilities:} HET(HRS, LRS), HST(WFPC2, NICMOS)

\section*{Appendix}

\section*{BULGE VERSUS PSEUDOBULGE CLASSIFICATIONS IN TABLE 2}

      This Appendix discusses the bulge and pseudobulge classifications
in Table 2.  It is far from an exhaustive review; many of these galaxies have been
studied in great detail.  We provide enough information for a robust classification.

      {\it M{\ts}101 and NGC 6946} are discussed in \S\ts3.  They satisfy three of the
pseudobulge classification criteria in Kormendy \& Kennicutt (2004, hereafter KK04):
they have overall S\'ersic indices $n < 2$, they contain small-scale structure
that cannot be formed in a hot stellar system, and star formation is vigorously in progress.
In fact, in NGC 6946, the mass of molecular gas in the nuclear star cluster is very similar
to its stellar mass, showing that growth of the nucleus and, at larger radii, the growth 
of the pseudobulge are still very much in progress.

      {\it IC 342} is closely similar to the above galaxies. 
Fisher \& Drory (2010) find that $n < 2$.
A strong central concentration of molecular gas feeds vigorous star formation 
(Becklin \etal 1980;
Turner \& Ho 1983;
B\"oker \etal 1999; 
Meier \etal 2000; 
Helfer \etal 2003, and references therein).

      {\it NGC 4945's} pseudobulge is best fitted with $n \simeq 1.3$,
based on our decomposition of the 2MASS $K_s$ profile.  Here and for all decompositions 
in this paper, the different flattenings of the bulge and disk 
are taken into account in measuring $PB/T$.
A strong central concentration of molecular gas is associated with vigorous
star formation 
(Dahlem \etal 1993; 
Henkel \& Mauersberger 1993;
Wang \etal 2004),
particularly in a 100 pc nuclear ring (Marconi \etal 2000)
like the starburst rings seen in many other barred and oval galaxies that are
actively growing pseudobulges (see KK04 for a review).  The starburst is 
powerful enough to drive a polar wind of x-ray-emitting gas (Strickland \etal 2004).

      {\it NGC 5236} has a powerful nuclear starburst 
(Turner \& Ho 1994; 
Harris \etal 2001; 
Bresolin \& Kennicutt 2002; 
D\'\i az \etal 2006) 
with multiple density concentrations that are comparable in mass to giant molecular clouds 
(Thatte \etal 2000; 
Bresolin \& Kennicutt 2002; 
Rodrigues \etal 2009).  
Fisher \& Drory (2010) find that $n \ll 2$.  
The whole center of the galaxy is being re-engineered on a timescale of $10^7$ yr 
(Rodrigues \etal 2009).

      {\it NGC 5194 = M{\ts}51} shows strong central star formation (e.{\ts}g., 
Turner \& Ho 1994;
Calzetti \etal 2005)
associated with a central peak in molecular gas emission (Helfer \etal 2003).
It also has $n \simeq 0.5$ (Fisher \& Drory 2008, 2010) and a nuclear bar
(Men\'endez-Delmestre \etal 2007).

      {\it NGC 253} has an extraordinarily powerful nuclear starburst (e.{\ts}g., 
Rieke \etal 1980;
Engelbracht \etal 1998;
Ott et al.\ts2005b;
Mart\'\i n \etal 2006)
that drives a polar wind of x-ray-emitting gas (e.{\ts}g., Strickland \etal 2004).
As in other, similar starbursts, it is associated with a dense and massive central
concentration of molecular gas (e.{\ts}g., Peng \etal 1996).

      {\it Maffei 2} has a pseudobulge, based on the observation that molecular gas 
(e.{\ts}g., Kuno \etal 2007, 2008) feeds a nuclear starburst (e.{\ts}g., 
Turner \& Ho 1994;
Tsai \etal 2006;
Meier \etal 2008).
We constructed a composite profile by measuring an HST NICMOS NIC3 F190N image and grafting
its profile onto the center of the 2MASS $K_s$ profile.  The central arcsec is heavily
obscured even in the infrared.  Extinction and star formation both render the S\'ersic 
index uncertain; depending on assumptions about whether to include the obscured part of
the profile in the decomposition fit or not, S\'ersic indices from $2.5 \pm 1$ to $3.4 \pm 0.5$ 
fit the data reasonably well.  The derived $PB/T = 0.16 \pm 0.04$ is more robust; the quoted 
error estimate takes the above uncertainties into account.  Our value is measured in $K$ band.
For comparison, Buta \& McCall (1999) got 0.22 in $I$ band using an $r^{1/4}$ law for the pseudobulge.

      {\it Our Galaxy} is discussed in \S\ts4.  

\pretolerance=15000  \tolerance=15000

      {\it Circinus} is discussed in detail in Kormendy (2010).  The galaxy has a pseudobulge 
based on three classification criteria.  The weakest one is S\'ersic index.  Kormendy (2010) 
constructs a $K$-band composite profile from the 2MASS data at large radii and from HST NICMOS data
near the center.  The best-fit S\'ersic-exponential decomposition has $n = 1.7 \pm 0.3$, which
is formally but not significantly less than 2.  A stronger argument is provided by the observation
of a nuclear disk -- a shelf in the brightness distribution that has almost the same
flattening as the outer disk.  Most compelling is the observation of a strong central concentration 
of molecular gas and star formation
(Marconi et~al.\ts1994;
Oliva \etal 1995; 
Maiolino \etal 1998;
Elmouttie \etal 1998;
Jones \etal 1999;
Wilson \etal 2000;
Curran \etal 1998, 2008; 
Greenhill \etal 2003; 
Mueller S\'anchez \etal 2006).

      {\it NGC 4736} is the ``poster child'' for pseudobulges.  It satisfies five classification
criteria in KK04.
It has a nuclear bar (e.{\ts}g., Kormendy 1993; M\"ollenhoff \etal 1995) which implies that 
a pseudobulge dominates the light even at small radii.
Spiral structure reaches in to the nuclear bar essentially undiluted by a classical bulge 
(Chincarini \& Walker 1967; Kormendy 1993; Fisher \etal 2009).
Especially important is the observation that the pseudobulge has a large ratio of
rotation velocity to velocity dispersion (Kormendy 1993; KK04).  The pseudobulge has a
complicated light profile (cf.~the pseudobulge in NGC 6946: Fig.~9\ts--\ts11), but the
main part has a S\'ersic function profile with $n \simeq 1.4 \pm 0.2$ (Fisher \& Drory 2008; 2010).
Finally, star formation in central molecular gas (Regan \etal 2001; Helfer \etal 2003) 
is modest now (Turner \& Ho 1994) but was more vigorous in the past (Pritchet 1977; Walker \etal 1988);
in addition, vigorous star formation is under way now in a molecular gas ring farther out in the 
pseudobulge (Wong \& Blitz 2000; Bendo \etal 2007).  This is consistent with the general picture in 
which star formation -- often in rings -- builds pseudobulges from the inside outward as the gradual
increase in central mass concentration shifts to larger radii the annulus at which infalling gas 
stalls and makes stars (KK04).

      {\it NGC 2683} is a difficult case, because the center of this almost-edge-on galaxy is
obscured by dust in the optical.  However, the 2MASS $K_s$-band outer profile and an
HST NICMOS NIC3 image taken with the F160W filter and calibrated to $K_s$ yield a composite
profile that clearly shows a tiny central bulge.  Is it classical or pseudo?  We cannot~be~sure,
because a bulge that is comparable in size to the thickness of the disk is not always classifiable
using the KK04 criteria.  The range of plausible decompositions gives $n = 2.5^{+0.6}_{-0.3}$.
Its structural parameters satisfy the fundamental plane correlations for small ellipticals and
classical bulges (Kormendy \etal 2009; Kormendy 2009).  Both results favor but do not guarantee
a classical bulge.  It could be a pseudobulge, as is the case for the bright, tiny center of the
similar, edge-on, ``boxy bulge'' galaxy NGC 4565 (Kormendy \& Barentine 2010).  But we err on the 
side of caution and call the bulge in NGC 2683 classical.  It is important to note that this tiny 
bulge with $B/T = 0.05 \pm 0.01$ is not the boxy bulge seen at larger radii and confidently 
identified as an edge-on bar via the observation of ``figure 8'' structure in the emission lines
of ionized gas 
(Rubin 1993;
Merrifield \& Kuijken 1999;
Funes \etal 2002;
Kuzio de Naray \etal 2009).

      {\it NGC 4826} is tricky, because the KK04 bulge \hbox{classification} criteria send a 
mixed message.  Among three \hbox{pseudobulge} characteristics, the most important is that the 
(pseudo)bulge has a relatively high ratio of rotation velocity to velocity dispersion.  This puts
it near other dynamically classified pseudobulges and above the ``oblate line'' that describes 
rotating, isotropic oblate spheroids in the  $V/\sigma$ -- $\epsilon$ diagram (Kormendy 1993).  
This is a disky property (see KK04 for a review).  For its luminosity, NGC 4826 also has a low, 
pseudobulge-like velocity dispersion (Kormendy 1993).  A somewhat weaker argument  is that 
it shows small-scale, mostly spiral structure all the way to the center (Lauer \etal 1995; 
{\tt http://heritage.stsci.edu/2004/04/big.html}).  The problem with interpreting this is
that it could be caused by the prominent dust disk.  The dust is associated with strong 
and centrally concentrated molecular gas emission 
(Regan \etal 2001; 
Helfer \etal 2003;
Garc\'\i a-Burillo \etal 2003), 
but the star formation rate is not particularly high (Turner \& Ho 1994).  All this is
suggestive of a pseudobulge.  On the other hand, the S\'ersic index of the bulge is variously 
derived to be $\sim 1.8$ (M\"ollenhoff \& Heidt 2001) to $\sim 3.6$ (Fisher \& Drory 2008;
M\'endez-Abreu \etal 2008), and the apparent axial ratio of the bulge is considerably rounder
than that of the disk (see all three of the above papers).  These properties favor a classical 
bulge interpretation, although they are not conclusive.  A complication is the observation of
counterrotating gas at large radii
(Braun \etal 1992, 1994;
Rubin 1994a, b),
although its mass is small and Rix \etal (1995) conclude that ``NGC 4826 has not undergone
a merger with another galaxy of significant size since the formation of its stellar disk.''
Plausibly, they argue that ``any [prograde-orbiting] gas \dots that is
likely to have existed originally in NGC 4826 \dots would have suffered inelastic collisions
with the [accreted] retrograde disk and would have gradually lost angular momentum and 
spiraled into the center of the galaxy.  This mechanism offers an elegant explanation for the 
abnormally high gas surface density in the center of NGC 4826 (Braun \etal 1994)'' and 
perhaps also for the dust disk.  We conclude with some confidence that the recent
minor accretion event -- while intrinsically interesting -- does not affect our 
classification of the (pseudo)bulge.  The weight of the evidence favors a pseudobulge.
However, in this paper more than most, it is exceedingly important that we not overestimate 
the importance of pseudobulges.  Moreover, it is clear that classical and pseudo bulges must 
co-exist in some galaxies 
(Kormendy 1993;
Erwin \etal 2003; 
KK04),
the best candidates are galaxies in which classification criteria send a mixed message.
We are therefore conservative and assign half of $(P)B/T$ to a classical bulge and half
to a pseudobulge.

      {\it NGC 2787} satisfies at least three pseudobulge classification criteria:
high ratio of rotation to random velocities, a nuclear disk structure, and $n \sim 1$ to 2
(Erwin \etal 2003; Kormendy \& Fisher 2008; Fisher \& Drory 2008).  Erwin \etal (2003)
decompose the profile into what they interpret as classical bulge and pseudobulge parts.
The complicated central profile is not in doubt, but all of the bulge may be pseudo.
To be conservative, we follow Erwin's decomposition in Table 2.

      {\it NGC 4258} contains a classical bulge with $n > 2$ and $V/\sigma$ value
consistent with the ``oblate line'' in the $V/\sigma$ -- $\epsilon$
diagram (Fisher \& Drory 2008, 2010; Siopis \etal 2009).  Molecular gas 
is observed (Helfer \etal 2003), but the emission drops in the center as it does in 
other classical bulges (Regan \etal 2001). 

      {\it M{\ts}31} contains a classical bulge with $n \simeq 2.5$ (Kormendy \& Bender 1999)
and rotation that is slightly below the oblate line in the $V/\sigma$ -- $\epsilon$ diagram 
(Kormendy \& Illingworth 1982).

      {\it M{\ts}81} contains a classical bulge with $n \simeq 3.8 \pm 0.1$ 
(Fisher \& Drory 2008, 2010)
and rotation that is consistent with the oblate line in the $V/\sigma$ -- $\epsilon$ diagram 
(Kormendy \& Illingworth 1982).
Like other classical bulges, M{\ts}81 has a central minimum in molecular gas emission 
(Helfer \etal 2003)
and a low central star formation rate (Turner \& Ho 1994).

      {\it Maffei 1 and NGC 5128:} We adopt elliptical galaxy classifications
for these two galaxies (for Maffei 1, see Buta \& McCall 1999, 2003).  We neglect the 
light of the small, late-type galaxy that is in the process of being inhaled by NGC~5128.
The absolute magnitudes of both galaxies are somewhat uncertain: the intrinsic colors implied 
by $M_K$ and $M_V$ listed in Table 2 are $(V - K)_0 = 2.5$ for Maffei 1 and 2.56 for NGC 5128; 
these values are bluer than normal colors  $(V - K)_0 = 3.0$ for old stellar populations.  We use
only the $K$-band magnitudes consistently adopted from the 2MASS Large Galaxy Atlas.

      Two galaxies that were included in Kormendy \& Fisher (2008) are omitted here
because $V_{\rm circ} < 150$ km s$^{-1}$:

      {\it NGC 3077} is usually classified as a Type II irregular (Sandage 1961) or
equivalently as an I0 galaxy (de Vaucouleurs \etal 1991) because of patchy dust near
its center.  However, it is participating in a spectacular, three-way gravitational
interaction with M{\ts}82 and M{\ts}81 (Yun \etal 1994), and its H{\ts}I connection with the latter 
galaxy makes it likely that it has accreted cold gas during the interaction.  The 
central dust and prominent star formation (e.{\ts}g., Ott \etal 2003, 2005a; Harris \etal 2004)
therefore are likely to be recent additions to what previously was probably a more 
normal, early-type galaxy.  If it was an elliptical, then it is particularly important
that we not bias our results by excluding it unfairly.
      Kormendy \& Fisher (2008) included the galaxy to be safe but were not certain
that it was big enough to make their sample cut.  We have now checked this by obtaining
spectra with the 9.2 m Hobby-Eberly Telescope and LRS Spectrogpaph (Hill \etal 1998).
The slit width was 1\farcs0 and the instrumental velocity dispersion was $\sigma_{\rm instr} =
119$ km s$^{-1}$ near the Mg b lines ($\lambda \sim 5175$ \AA).  The K0 III standard star was HD172401.  
Our signal-to-noise ratios were very high, and the absorption lines in NGC 3077 are very 
obvious.  However, we completely failed to resolve their line widths.  We conclude that
$\sigma \ll 119$ km s$^{-1}$ in NGC 3077.  This is consistent with the estimate that the
central {\it escape} velocity from the galaxy is $\sim 110$ km s$^{-1}$ (Ott \etal 2003).
Therefore NGC 3077 is too small to be included in our sample.

      {\it NGC 5195}, the companion of M{\ts}51, was also included in Kormendy \&
Fisher (2008).  However, Kohno \etal (2002) find from CO observations that the maximum rotation 
velocity is ``$160$ km s$^{-1}$ at $r \sim 50$ pc in the plane of the galaxy'' but that there is 
a ``steep rise of rotation velocity toward the center'' to the above value from smaller rotation
velocities at larger radii (their Figure 6).  We therefore omit the galaxy.  However, we
note that the central concentration of molecular gas and star formation -- possibly fed 
by the interaction with M{\ts}51 -- is most consistent with a pseudobulge and (ii) that the 
pseudobulge-to-total luminosity ratio is small (Smith \etal 1990 estimate that $PB/T \sim 0.06$ in $K$ band).  
If we are incorrect in omitting NGC 5195, then we underestimate the importance of pseudobulges 
in our mass inventory and therefore overestimate the importance of classical bulges.  However, 
any error introduced is small, because the pseudobulge of NGC 5195 is small.

\end{document}